\documentclass[prc]{revtex4-2} 

\usepackage{graphicx}
\usepackage{multirow, array}
\usepackage{color}
\usepackage{xcolor}

\usepackage{url}
\usepackage{physics}
\usepackage{float} 
\usepackage{caption}
\usepackage{subcaption}
\usepackage{isotope}
\usepackage{booktabs}
\usepackage{hyperref}

\begin{document}
\captionsetup{format=plain, font=footnotesize, labelfont=bf, margin=2cm}

\title[Benchmarking electromagnetic observables in PHF]{Benchmarking electromagnetic observables in {angular-momentum} projected Hartree-Fock} 

\author{Raul Bernal-Gonzalez}
\author{Calvin W. Johnson}
\address{San Diego State University,
5500 Campanile Drive, San Diego, CA 92182-1233}


\begin{abstract}
We benchmark electromagnetic transitions and moments (electric quadrupole and magnetic dipole) in angular-momentum projected-after-variation Hartree-Fock calculations against full 
configuration-interaction diagonalization results in a shell model basis.  For such a simple approximation we find 
reasonably good agreement, including for many odd-$A$ and odd-odd nuclides. 
As previous work on excitation spectra found, 
results are frequently improved in cases with shape coexistence. Here we considered select cases from 
the $sd$- and $pf$-valences spaces.
While electric quadrupole moments and transitions are, as one might expect, frequently (though not always) well reproduced, especially in even-even nuclides, magnetic dipole moments and transitions are overall better 
than expected.
This continues the benchmarking of  projected Hartree-Fock as a simple yet effective alternative to full 
configuration-interaction as well as an underlying foundation for many other  many-body methods. 
\end{abstract}


\maketitle

\section{Introduction}

Atomic nuclei are surprisingly complex, inspiring 
the development of a number of different calculational approaches~\cite{ring2004nuclear}. 
Arguably the most fundamental are mean-field methods,  such as Hartree-Fock (HF), where independent particles/quasi-particles move in an average field generated by the rest of the system. 
Other many-body methods explicitly or implicitly start from a mean-field picture. These include configuration-interaction (CI)  methods~\cite{bg77,towner1977shell,lawson1980theory,br88,ca05}, as well as generator-coordinate methods~\cite{ring2004nuclear,reinhard1987generator,robledo2018mean,klupfel2008systematics} and the 
Monte Carlo shell model~\cite{PhysRevLett.77.3315,otsuka2001monte}.  
The latter two methods construct mean-field states deformed by external fields, and conserved quantities such as angular momentum and fixed particle number conservation must be restored. 
More recently, coupled cluster calculation using a symmetry-restored reference state has been developed and demonstrates significant improvement~\cite{PhysRevC.105.064311,PhysRevC.110.L011302}. 
Symmetry-projected mean-field theory is also used in quantum chemistry \cite{jimenez2012projected}. 
 This background highlights the important technical task of projection of broken symmetries ~\cite{ring2004nuclear,jimenez2012projected,sheikh2021symmetry}; 
benchmarking the underlying base methodology helps to demonstrate the sound foundations of more sophisticated approaches. 

A previous paper~\cite{lauber2021benchmarking} benchmarked angular-momentum projected (after variation) Hartree-Fock (PHF) calculations against full configuration interaction (FCI) shell-model calculations, with both sets of calculations carried out in the same model space and using the same input shell-model Hamiltonian matrix elements, and found surprisingly good agreement for excitation spectra, at least in medium-light nuclides where pairing is not strong. Such spectra were significantly improved by the use of multiple minima when they occurred.

This paper extends the project of \cite{lauber2021benchmarking} by 
benchmarking electromagnetic observables in PHF calculations against FCI results. 
In the next section, we present our methods; we  focus on electric quadrupole (E2) and magnetic dipole (M1) moments and transitions. (In  Appendix~\ref{sec:details}, we provide   details for computing general one-body density matrices in PHF and multi-reference extensions.) In Section~\ref{results}, we present our results. 
Although it is not surprising to find that quadrupole observables are approximately reproduced for even-even nuclides, whose structures are known to be dominated by quadrupole deformation, we are gratified to find similar approximate agreement 
for odd-$A$ and odd-odd nuclides. Somewhat surprising is the reasonable agreement for magnetic dipole observables. 
Of course, the agreement is far from perfect.  Nonetheless, our results demonstrate that the roots of success in various methods--generator coordinate (including the related discrete non-orthogonal shell model~\cite{nkck-ctvd}), Monte Carlo shell model~\cite{otsuka2001monte}, coupled clusters~\cite{PhysRevC.105.064311,PhysRevC.110.L011302}, and so on--are heavily enabled by a good starting point already in PHF.

\section{Methods}

\label{methods}

In order to compare the projected Hartree-Fock calculations against full configuration 
interaction results, we carry out both sets of calculations in the same shell-model space using the 
same interaction input.

\subsection{Configuration-interaction}

We benchmark against full configuration-interaction (FCI)~\cite{bg77,towner1977shell,lawson1980theory,br88,ca05}, expanding the wave function in a shell-model basis $\{ | \alpha \rangle \}$:
\begin{equation}
| \Psi \rangle  = \sum_\alpha c_\alpha | \alpha \rangle.
\end{equation}
For our basis we use the occupation representation of antisymmetrized products of single-particle states, or Slater determinants: 
if $\hat{a}^\dagger_i$ is the creation operator for the $i$th single-particle state, then the occupation representation of 
an $A$-body Slater determinant is
\begin{equation}
\hat{a}^\dagger_1 \hat{a}^\dagger_2 \hat{a}^\dagger_3 \ldots \hat{a}^\dagger_A | 0 \rangle,
\label{SD}
\end{equation}
where $| 0 \rangle$ is the fermionic vacuum, or, equivalently, a frozen core. 
In practice, because the single-particle states are fixed, these states can be represented compactly by bit strings.

Because both total angular momentum $\hat{J}^2$ and the $z$-component $\hat{J}_z$ commute with 
our Hamiltonians, we choose many-body basis states with fixed eigenvalues of the latter, labeled as $M$. This is known as an \textit{M-scheme basis}.   Other than fixing the $z$-component of angular momentum $J_z$ and parity $\pi$, for FCI 
we take all possible Slater determinants. 
The  {\tt BIGSTICK} code  \cite{Johnson20132761,johnson2018bigstick} efficiently computes matrix elements of the Hamiltonian in this basis, $\langle \alpha | \hat{H} | \beta \rangle$, 
and then uses the Lanczos algorithm to find low-lying eigenstates \cite{ca05,Lanczos}.  Although there is no explicit constraint on total angular momentum $J$, states with good $J$ naturally emerge except in the case of accidental degeneracies. 
It is these FCI results that we use as a benchmark, against which we compare PHF.

\subsection{Angular-momentum projected Hartree-Fock and finding minima}

Details of our angular-momentum projected Hartree-Fock calculations, including on the efficacy of gradient descent, excitation spectra and separation energies, and so on, are given in \cite{lauber2021benchmarking}. 
Here we survey the most important points. We work in the same shell-model spaces as FCI, so that the single particle states are expanded in an occupation basis. More specifically, we minimize  
$\langle \hat{H} \rangle $ for an arbitrary Slater determinant $| \Psi \rangle$, by   defining a single-particle basis 
via an $N_s \times N_s$ unitary transformation, where $N_s$ is the number of single-particle states,
\begin{equation}
\hat{c}^\dagger_a = \sum_{i=1}^{N_s} U_{ia} \hat{a}^\dagger_i.
\label{Utransform}
\end{equation}
Aside from having good particle number, 
we do not impose any constraints, such as axial symmetry, on our Slater determinants.
We only assume   $U_{ia}$ is real, which is equivalent to an $RT$ symmetry, which is invariance under rotation by $\pi$ radians about the $y$-axis, followed 
by time reversal $T$. Implicitly then we choose for our trial wave function 
(note the subtle but importance difference with Eq.~(\ref{SD}))
 \begin{equation} 
 | \Psi \rangle = \hat{c}^\dagger_1 \hat{c}^\dagger_2 \ldots \hat{c}^\dagger_A | 0 \rangle.\label{Slater}
 \end{equation}
The Slater determinant can be represented as a rectangular matrix ${\Psi}$, which for $N_p$ particles is given by $N_p$ columns, each of length $N_s$,  of the matrix $\mathbf{U}$ in 
Eq.~(\ref{Utransform}). 
We use separate proton and neutron Slater determinants.
One can compute $\langle \hat{H} \rangle$ for any ${\Psi}$ and then vary the elements of $\mathbf{U}$  to minimize  \cite{SHERPA}.  

We routinely use gradient descent to robustly find local minima~\cite{lauber2021benchmarking}. 
By using randomly chosen starting points, we can effectively find both global and local minima.


In general, 
 an arbitrary state $| \Psi \rangle$, including Slater determinants and those found through Hartree-Fock minimization, are not eigenstates of angular momentum. Instead, they can be expanded 
as sum of states with good angular momentum.
In the standard approach one accomplishes this by an integral  \cite{ring2004nuclear,sheikh2021symmetry}, but we perform the projection by solving a set of linear algebra equations \cite{PHF1, PHF2}.  Some details about finding states of good angular momentum are given in Appendix \ref{sec:details}. 


\begin{figure}[h!]
    \centering
    \includegraphics[width=\textwidth]{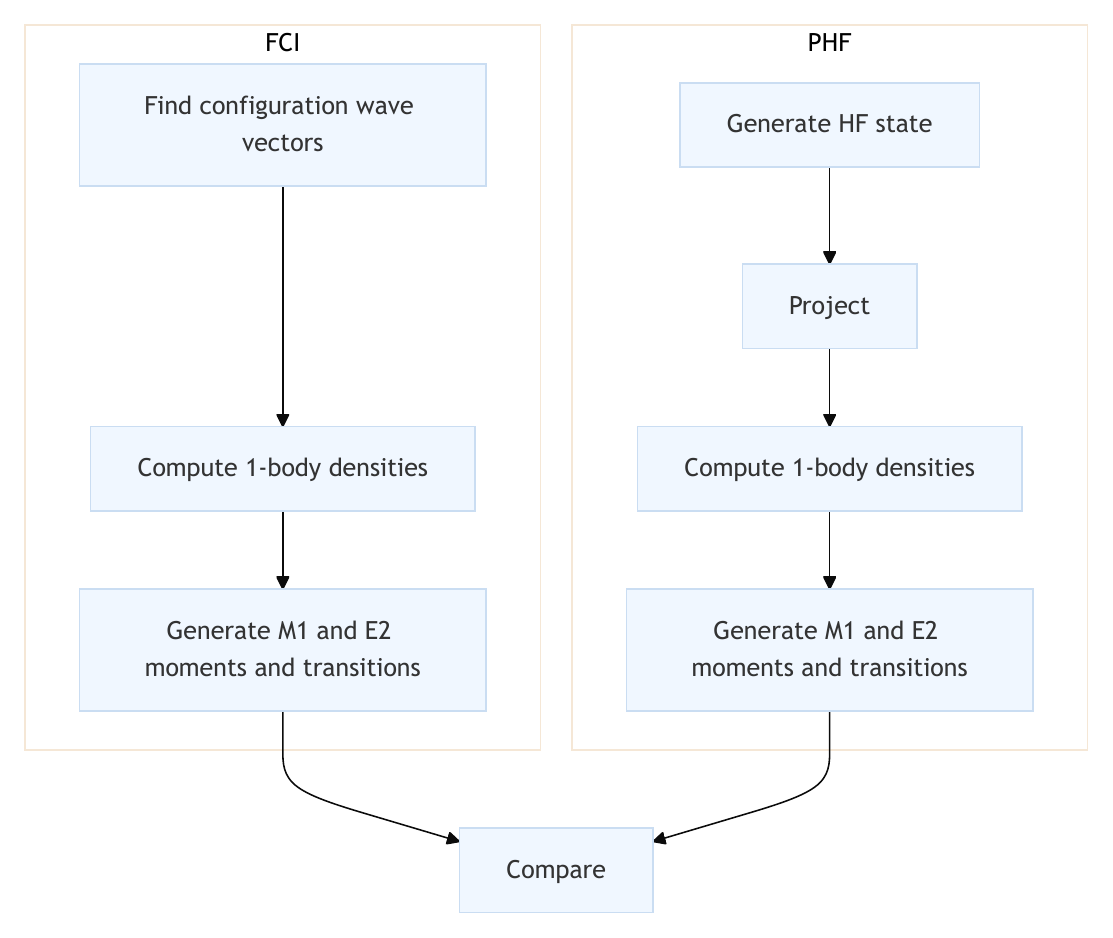}
    \caption{Schematic of our work flow. The one-body densities were computed from each method, then the electromagnetic transition and moments were calculated. The same code calculates the M1 and E2 moments and transitions from the densities.  
}
    \label{fig:workflow}
\end{figure}



\subsection{Electromagnetic observables and density matrices}

In this work we focus on matrix elements of one-body operators, 
the electric quadrupole (E2) and magnetic dipole (M1) operators.
The electric quadrupole operator is
\begin{equation}
\hat{Q}_{2,m} =    \sum_{i = 1}^A q_i \,  r_i^2 \, Y_{2,m}(\theta_i, \phi_i),
\end{equation}
where 
$\theta_i, \phi_i$ are the spherical coordinates for the $i$th nucleon 
and $q_i$ is the effective charge for the $i$th nucleon. 
It is important to emphasize that our proximal goal here is not specifically to  accurately reproduce  experimental values, but rather to compare a computationally cheap approximate method to a full valence space calculation. 
Therefore we used the effective charges $e_p = 1.5 \, e$ and $e_n = 0.5 \, e$, which  generally reproduce experimental results for our model spaces~\cite{PhysRevC.78.064302,PhysRevC.73.061305}, and set our oscillator parameter using $b \approx 1.0A^{1/6}$ fm.

The magnetic dipole operator is
\begin{equation}
    \hat{\mu}_{1,m} = \sqrt{\frac{3}{4\pi}} \sum_{i=1}^A g_{l,i} \vec{l}_{m,i}
+ g_{s,i}\vec{s}_{m,i},
\end{equation}
where $ \vec{l}_i$ and $\vec{s}_i$ are the orbital angular momentum and 
spin, respectively, of the $i$th nucleon, and $g_{l,i}, g_{s,i}$ are the 
respective g-factors.  We used the bare values $g_l$ = 1 $e$ and 0 for protons and neutrons, respective, and $g_s = 5.5857 \, \mu_N$ and
$-3.2863 \, \mu_N$, in units of the nuclear magneton, again for protons and neutrons, respectively.

For an arbitrary one-body operator $\hat{\cal O}_{K,m}$ of angular momentum rank $K$,
we compute reduced matrix elements using
\begin{equation}
    \langle f || \hat{\cal O} || i \rangle = \sum_{a,b}
    \langle a || \hat{\cal O} || b \rangle \rho^{fi}_K(ab).
    \label{eq:operatorme}
\end{equation}
 Here $\langle f || \hat{\cal O}_K || i \rangle$ 
is the reduced matrix element between initial and final many-body wave functions 
$| i \rangle$ and $| f \rangle$,  respectively, and $\langle a || \hat{\cal O}_K || b \rangle $
is the reduced matrix element between single-particle orbitals labeled 
by $a,b$. We use the convention of Edmonds~\cite{edmonds1996angular} for reduced 
matrix elements.  Finally, the one-body density matrix is
\begin{equation}
    \rho^{fi}_K(ab) = \frac{1}{\sqrt{2K +1 }}
    \langle f || \left [ \hat{a}_a^\dagger \otimes \tilde{a}_b \right ]_K 
    || i \rangle.
\end{equation}

The transition strength, or $B$-value, is the average over initial states and sum over final states of $|\langle f || \hat{\cal O} || i \rangle |^2$, which is
\begin{equation}
    B({\cal O}) = \frac{1 }{2J_i +1}|\langle f || \hat{\cal O} || i \rangle |^2.
\end{equation}

Static moments are defined in the stretched state,  with $z$-component $M = J$ ~\cite{bg77}, so that 
\begin{eqnarray}
    {\cal M}(\sigma \lambda) & = &  \langle J, M =J | \hat{\cal M}(\sigma \lambda) | J, M = J \rangle \\
    \nonumber &= & \frac{(J J, \lambda 0 | J J) }{\sqrt{2J+1}} \langle f || \hat{\cal M} (\sigma \lambda) || i \rangle;
\end{eqnarray}
furthermore, the moment operator $\hat{\cal M}$ is defined 
in Cartersian operators, not spherical tensors, so that 
$\hat{\cal M}(M1) = \sqrt{\frac{4\pi}{3}} \hat{\mu}_1$
and
$\hat{\cal M}(E2) = \sqrt{\frac{16\pi}{5}} \hat{Q}_2$.

The workflow for our calculations is sketched in Fig.~\ref{fig:workflow}. 
Not only are the valence spaces and interactions the same for both types of calculations, FCI and PHF, but the format of density matrices is the same. Hence our treatment of both approaches are highly parallel for ease of comparison.

\section{Results}

\label{results}

Here we present our results.  While some previous work has compared PHF and related generator-coordinate method calculations against FCI~\cite{PhysRev.159.885,baye1983electromagnetic,PhysRevC.65.024304,PhysRevC.98.054311,PhysRevC.103.064302}, this study is 
the widest systematic such benchmarking, emphasizing a variety of different nuclides 
including odd-A and odd-odd nuclides. We compare full configuration-interaction (FCI) results against the angular-momentum projected (after variation) Hartree-Fock (PHF) results.

\begin{figure}[h!]
    \centering
    \begin{subfigure}{0.49\linewidth}
        \includegraphics[width=\linewidth]{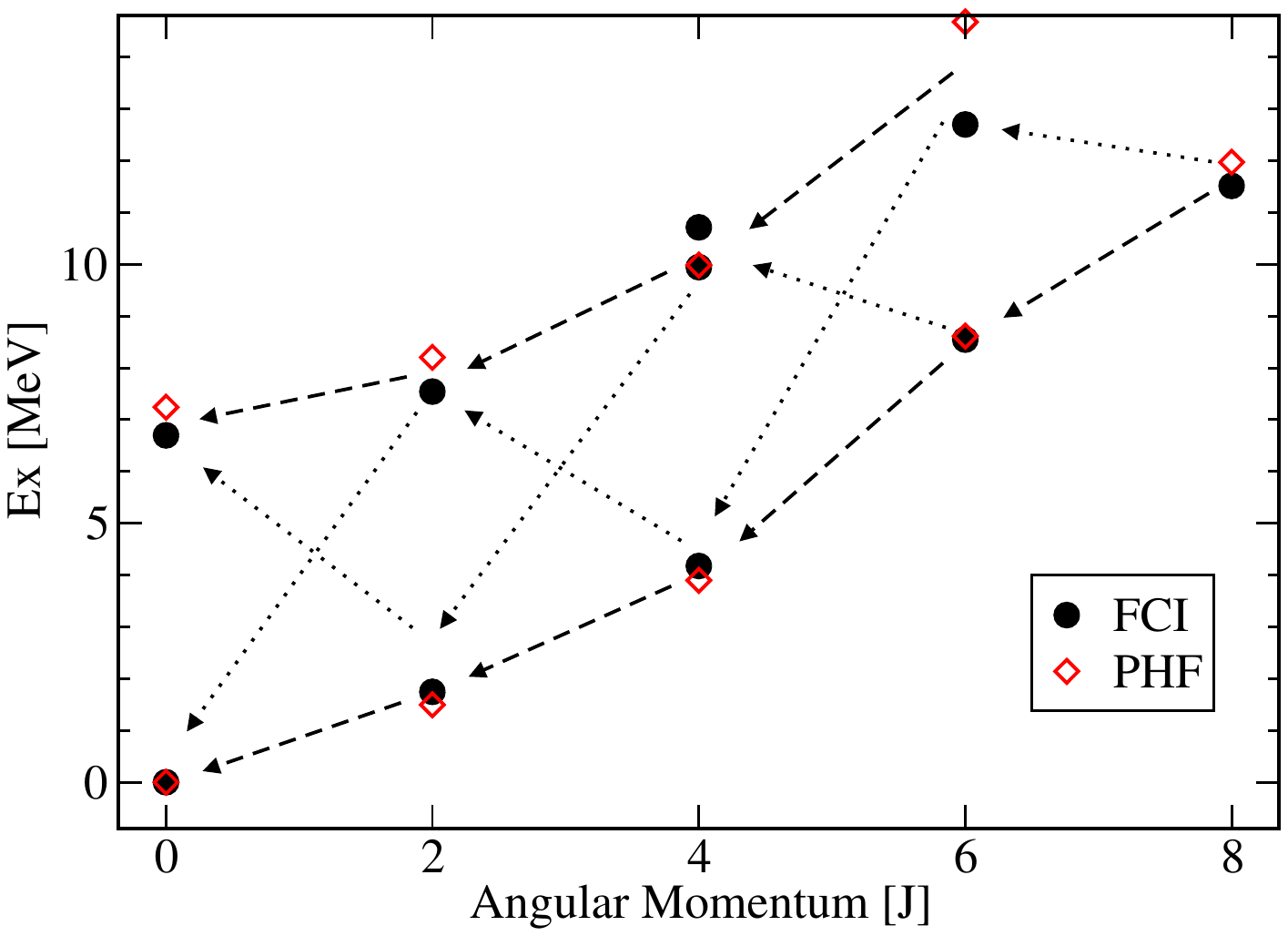}
        \caption{$^{20}$Ne}
    \end{subfigure}
    \begin{subfigure}{0.49\linewidth}
        \includegraphics[width=\linewidth]{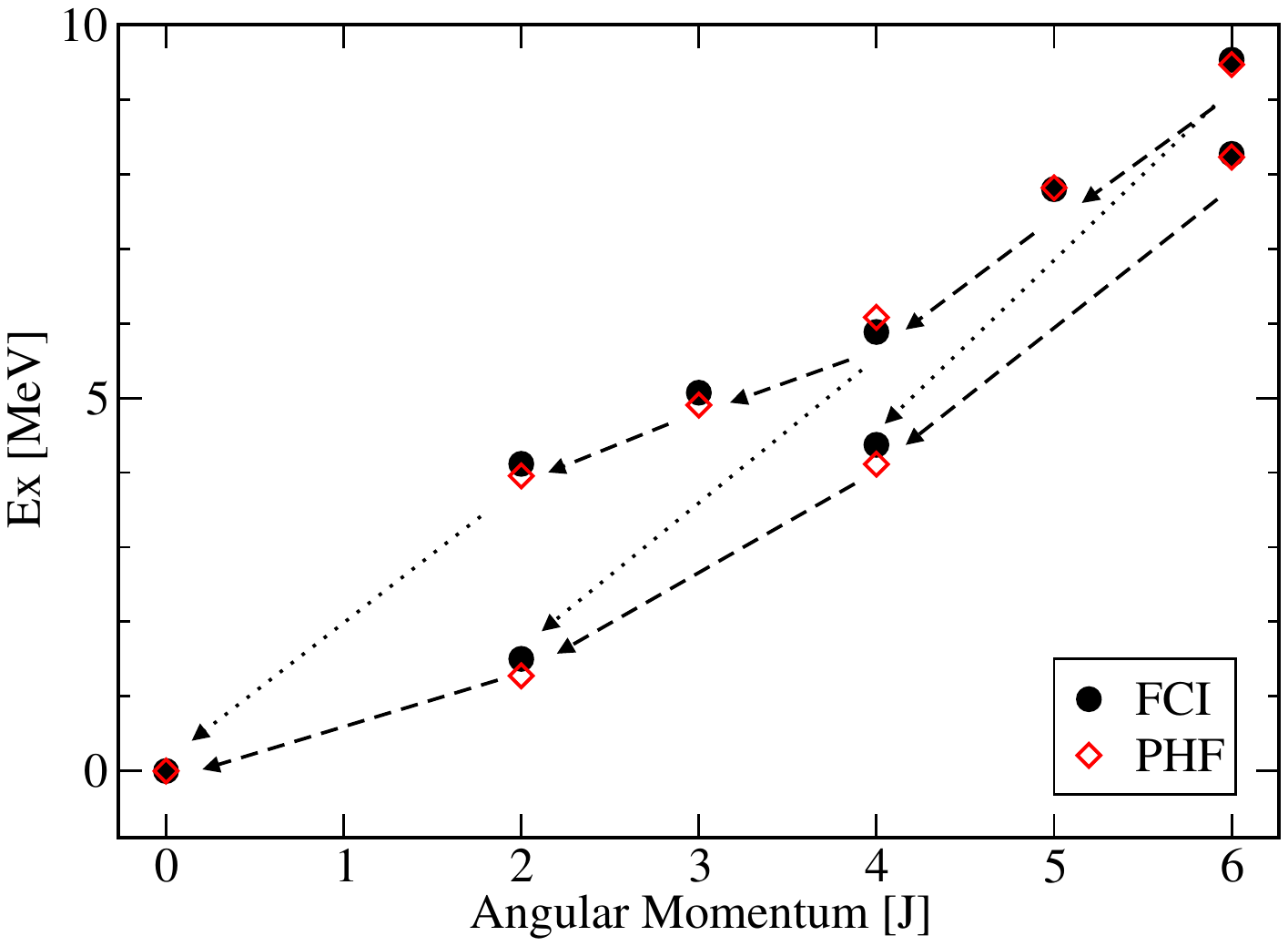}
        \caption{$^{24}$Mg}
    \end{subfigure}
    \vfill
     \begin{subfigure}{0.49\linewidth}
        \includegraphics[width=\linewidth]{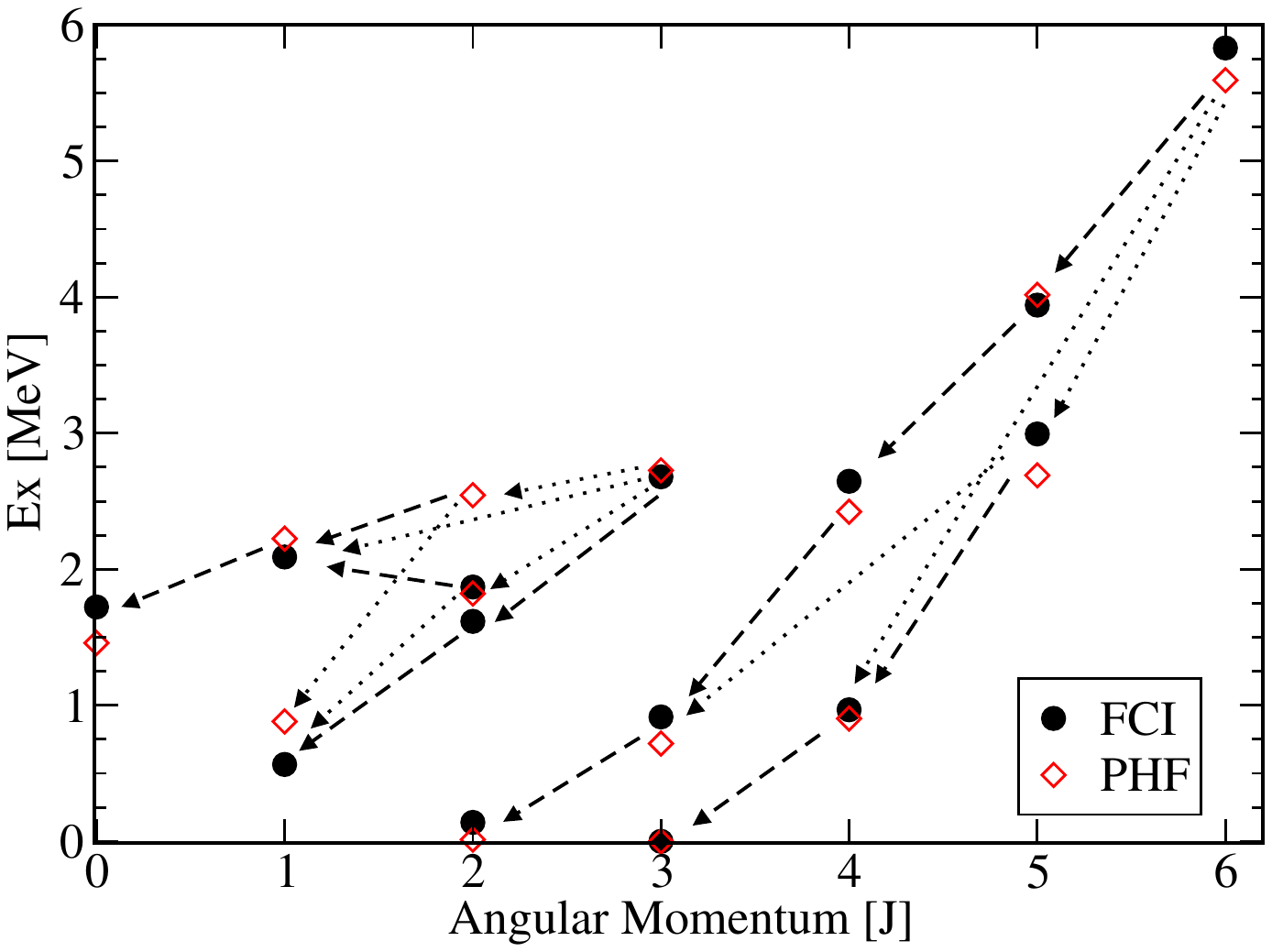}
        \caption{$^{30}$Al}
    \end{subfigure}
     \begin{subfigure}{0.49\linewidth}
        \includegraphics[width=\linewidth]{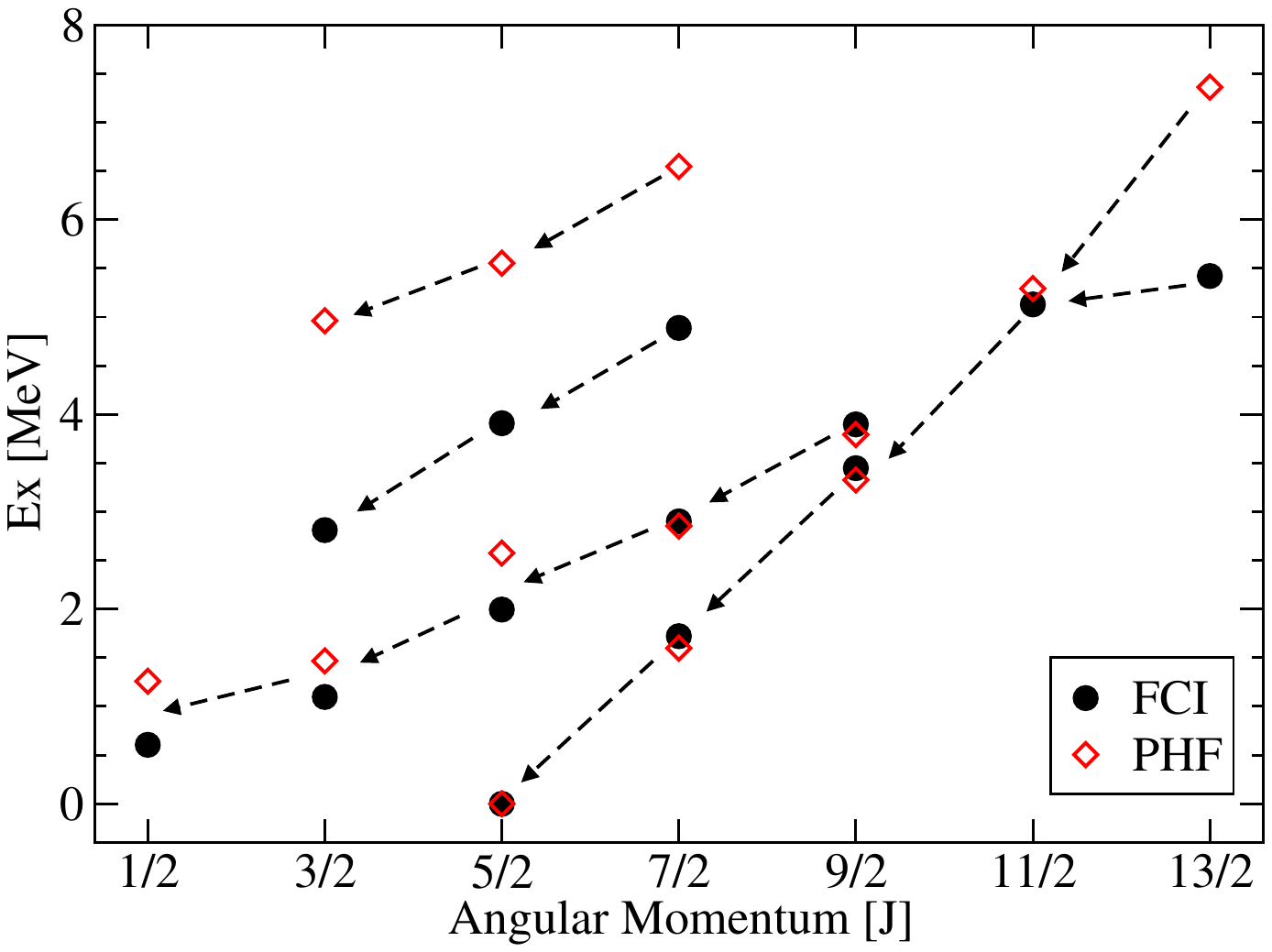}
        \caption{$^{25}$Mg}
    \end{subfigure}
    \caption{Excitation energy plots for four examples in the $sd$-shell, comparing full configuration-interaction (FCI) (filled circles) and angular-momentum projected Hartree-Fock (PHF) (open diamonds). The intra-band transitions are traced in dashed lines and the inter-band transitions are traced in dotted lines.}
    \label{fig:sdenergies}
\end{figure}

\subsection{Examples from the $sd$-shell}



We chose four test cases in the $sd$-shell, with a model space comprised of a frozen $^{16}$O core with valence $1s_{1/2}$-$0d_{3/2}$-$0d_{5/2}$ orbitals, with the universal $sd$ interaction version $B$ (USDB)  \cite{PhysRevC.74.034315}.

We plot selected results here: two even-even nuclei ($^{20}$Ne and $^{24}$Mg), one odd-odd nuclei ($^{30}$Al) and one odd-A nuclei ($^{25}$Mg).
shown in Figure \ref{fig:sdenergies}.  
Appendix \ref{data} has tables for $^{20}$Ne and  $^{24,25}$Mg, as well as additional 
 tables of results for $^{32}$Si and $^{34}$S, not plotted here.
 \newpage


\emph{Even-even.} 

$^{20}$Ne has two HF minima, a prolate minimum at -36.40 MeV, and a second oblate local minimum at -31.83 MeV. In general, both the yrast (ground state) band 
and yrare (excited state) band excitation energies are well reproduced by PHF, as 
plotted in Fig.~\ref{fig:sdenergies}(a); for reference, Table~\ref{tab:ne20energy} provides the absolute and excitation energies.

\begin{figure}[H]
    \centering
    \begin{subfigure}{.49\linewidth}
        \includegraphics[width=\linewidth]{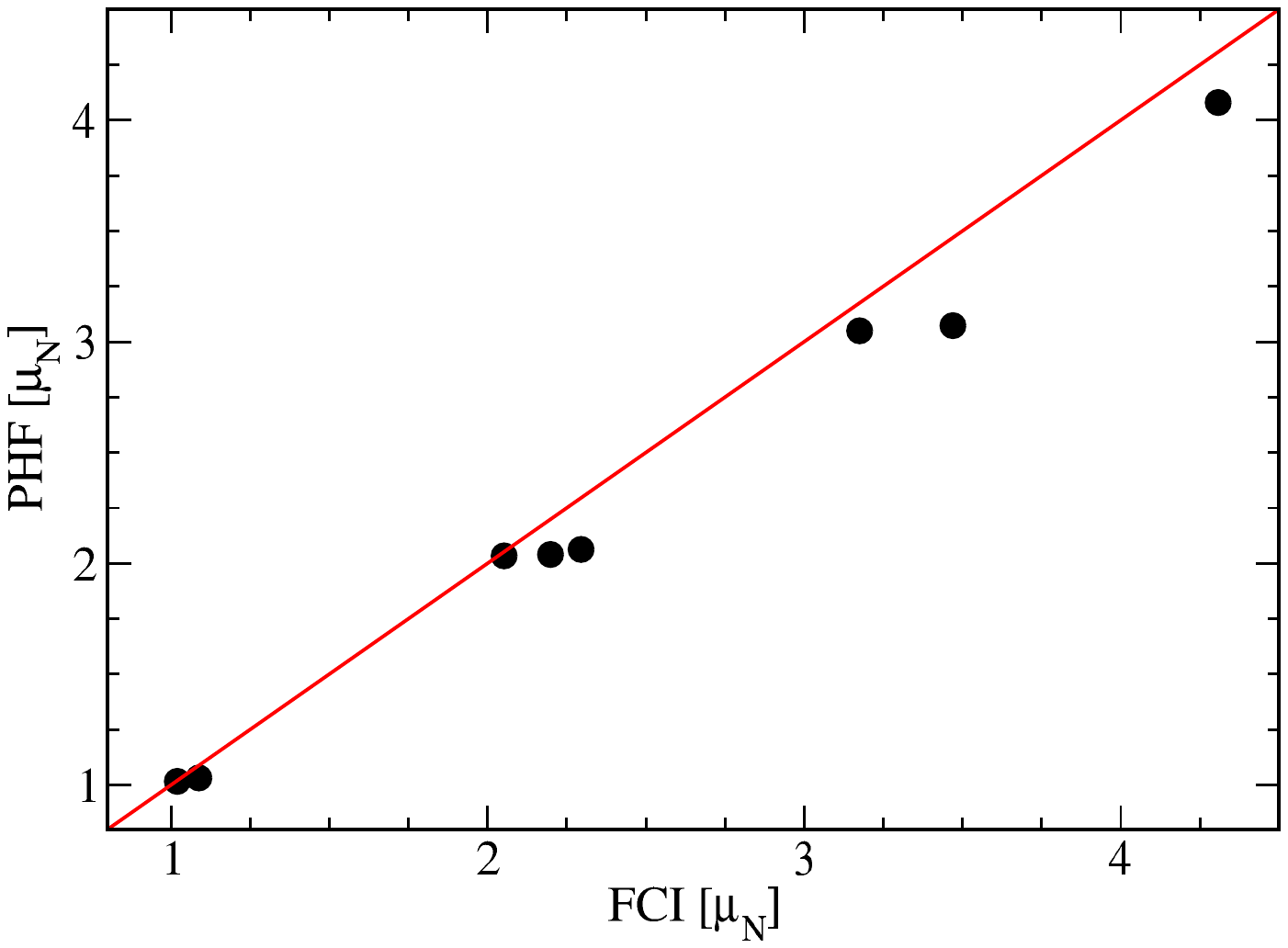}
        \caption{M1 moments for $^{20}$Ne.}
    \end{subfigure}
    \begin{subfigure}{.49\linewidth}
        \includegraphics[width=\linewidth]{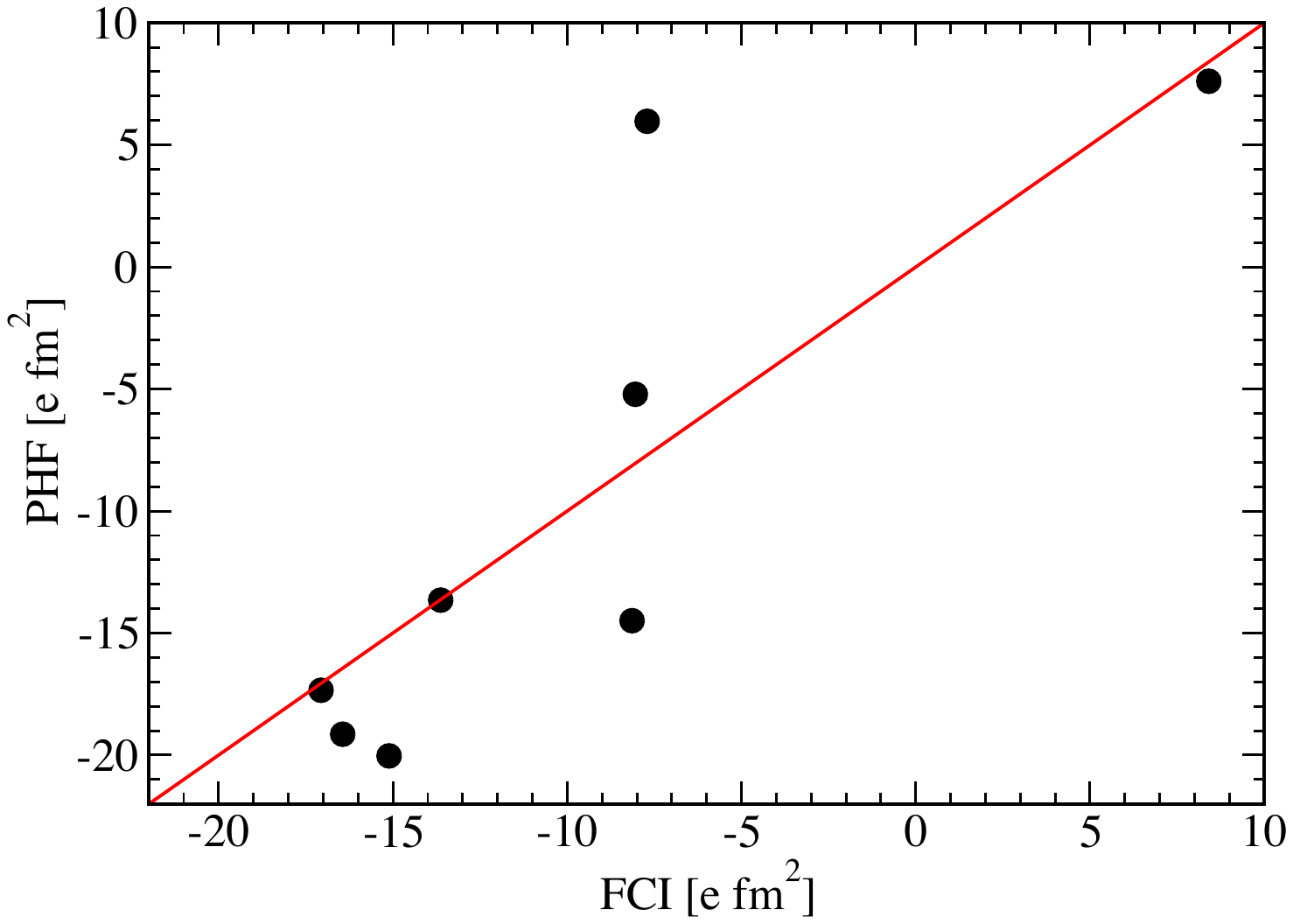}
        \caption{E2 moments for $^{20}$Ne.}
    \end{subfigure}
    \caption{Scatter plot of M1 and E2 moments for $^{20}$Ne. Moments calculated from FCI densities are plotted on the x-axis and from PHF on the y-axis. The $y=x$ line is shown for comparison.}
    \label{fig:ne20moms}
\end{figure}

\begin{figure}[H]
    \centering
    \begin{subfigure}{.49\linewidth}
        \includegraphics[width=\linewidth]{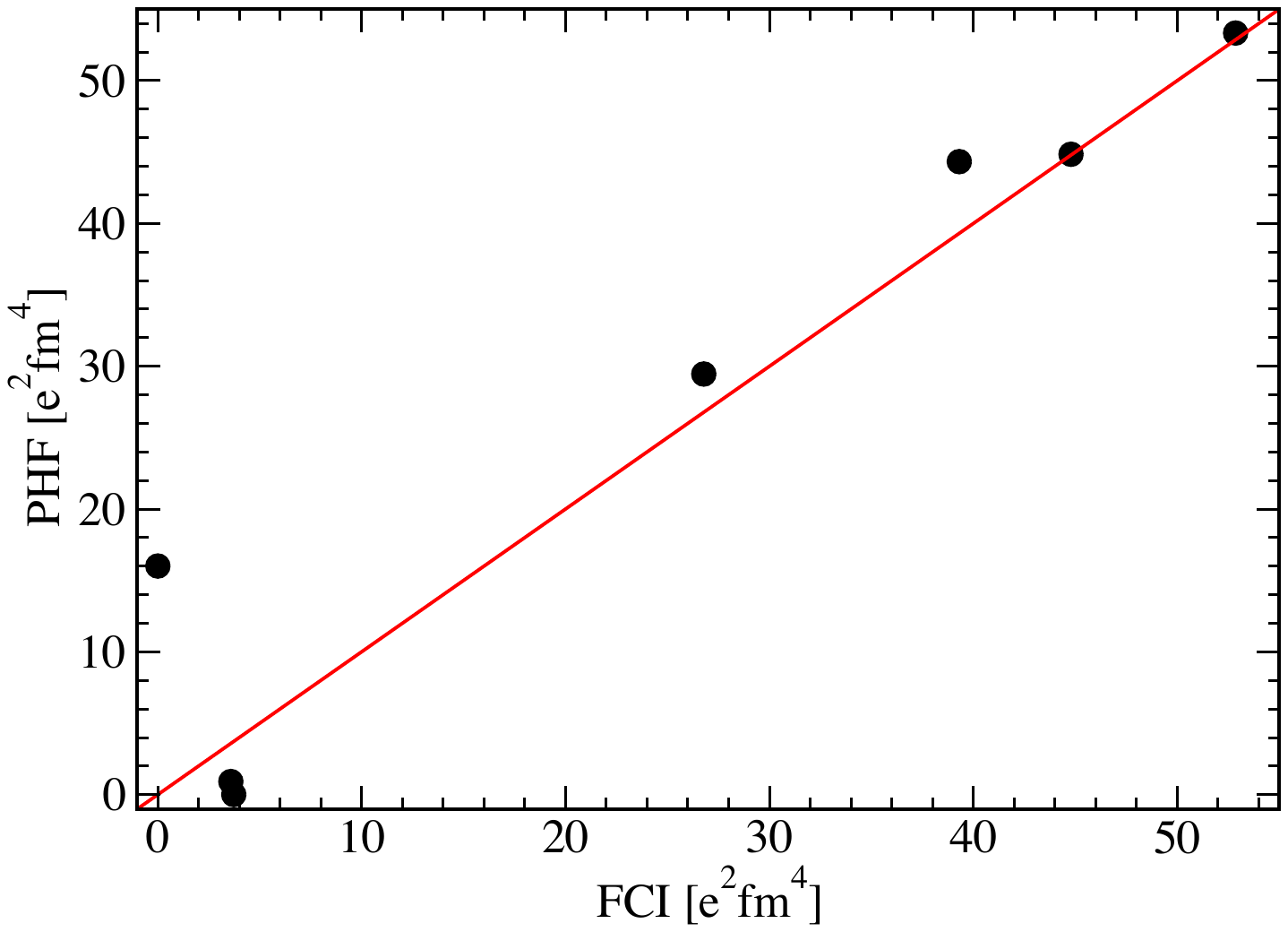}
        \caption{Intra-band B(E2) strengths $^{20}$Ne.}
    \end{subfigure}
    \begin{subfigure}{.49\linewidth}
        \includegraphics[width=\linewidth]{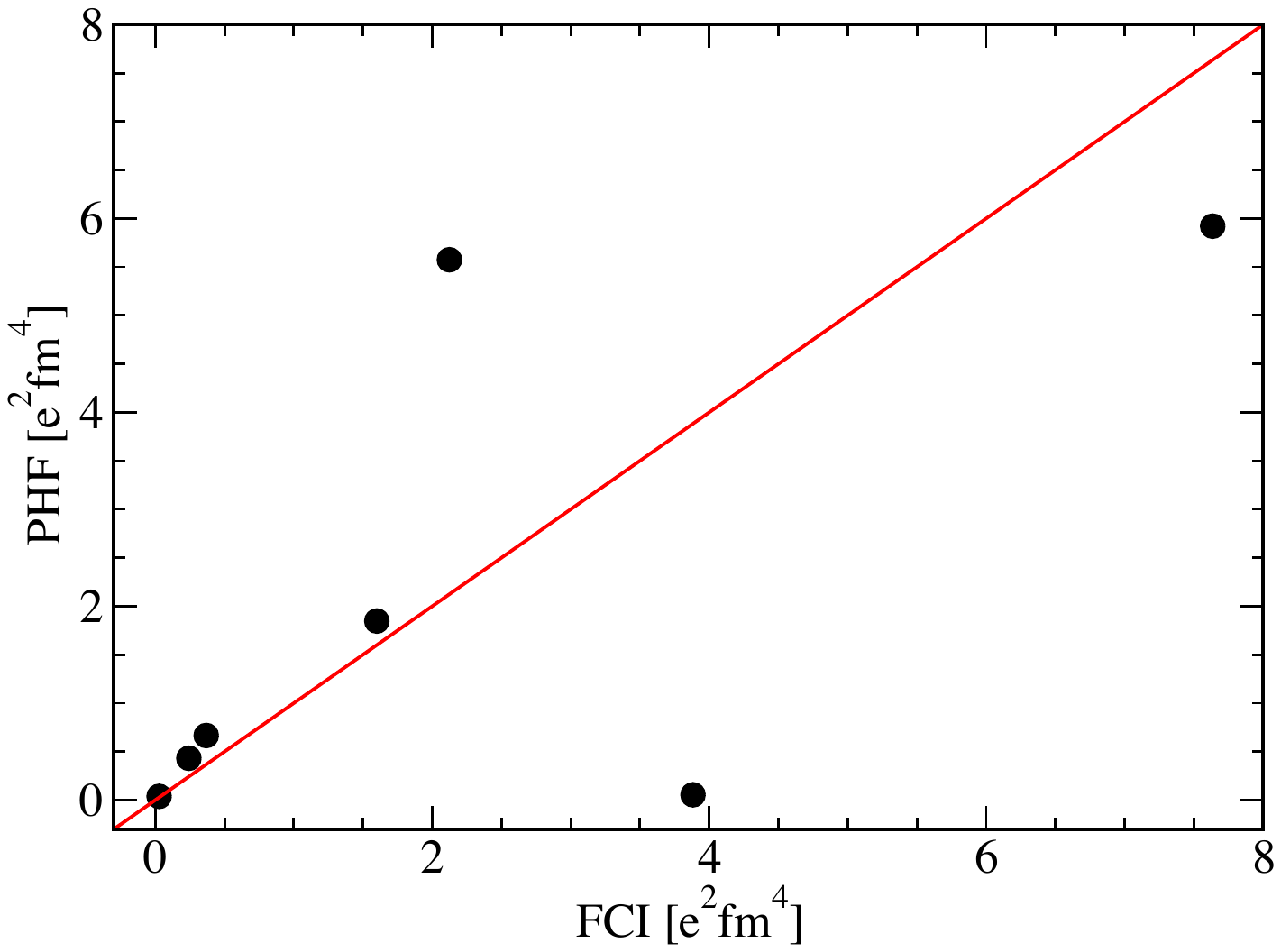}
        \caption{Inter-band B(E2) strengths $^{20}$Ne.}
    \end{subfigure}
    \caption{Scatter plot of B(E2) transition strengths for $^{20}$Ne. Transitions calculated from FCI densities are plotted on the x-axis and from PHF on the y-axis. The $y=x$ line is shown for comparison.}
    \label{fig:ne20trans}
\end{figure}

The scatter plots in Fig.~\ref{fig:ne20moms}(a) and (b) compare the FCI and PHF 
static spectroscopic M1 and E2 moments (the values are also given in Table~\ref{tab:ne20energy}). While the overall trend of both sets of moments is good, 
somewhat surprisingly the M1 moments have better agreement. Why this is so is not clear.
The agreement for intra-band B(E2) values, 
compared in a scatter plot in Fig.~\ref{fig:ne20trans}(a), 
is overall better than the inter-band B(E2) values compared in Fig.~\ref{fig:ne20trans}(b), 
which one might expect. Numerical B(E2) values can be found in Table~\ref{tab:ne20e2}.
 Because the PHF results 
had only even $J$ states, due to axial symmetry, we do not plot any M1 transitions 
for $^{20}$Ne.


\begin{figure}[H]
    \centering
    \begin{subfigure}{.49\linewidth}
        \includegraphics[width=\linewidth]{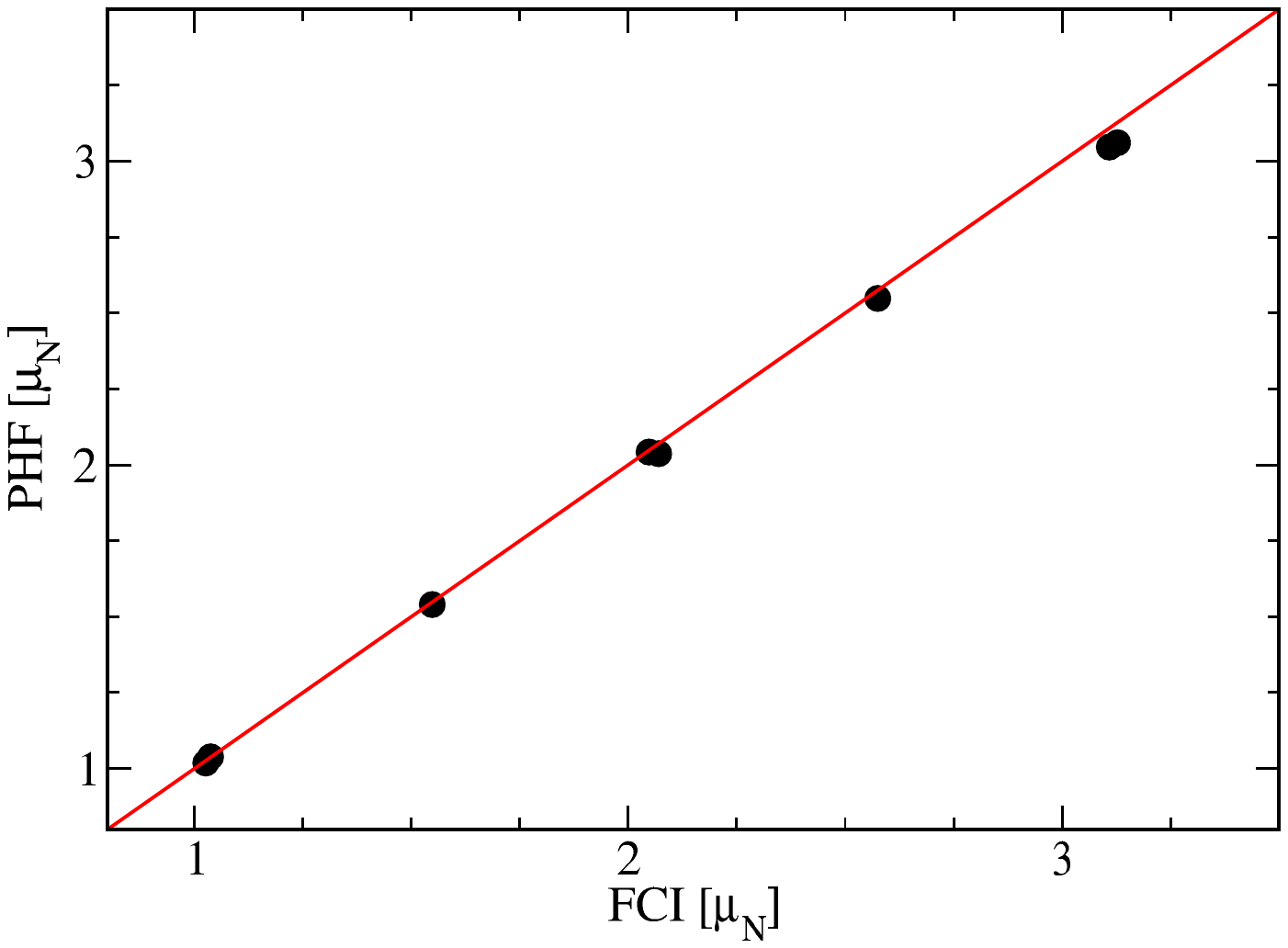}
        \caption{M1 moments for $^{24}$Mg.}
    \end{subfigure}
    \begin{subfigure}{.49\linewidth}
        \includegraphics[width=\linewidth]{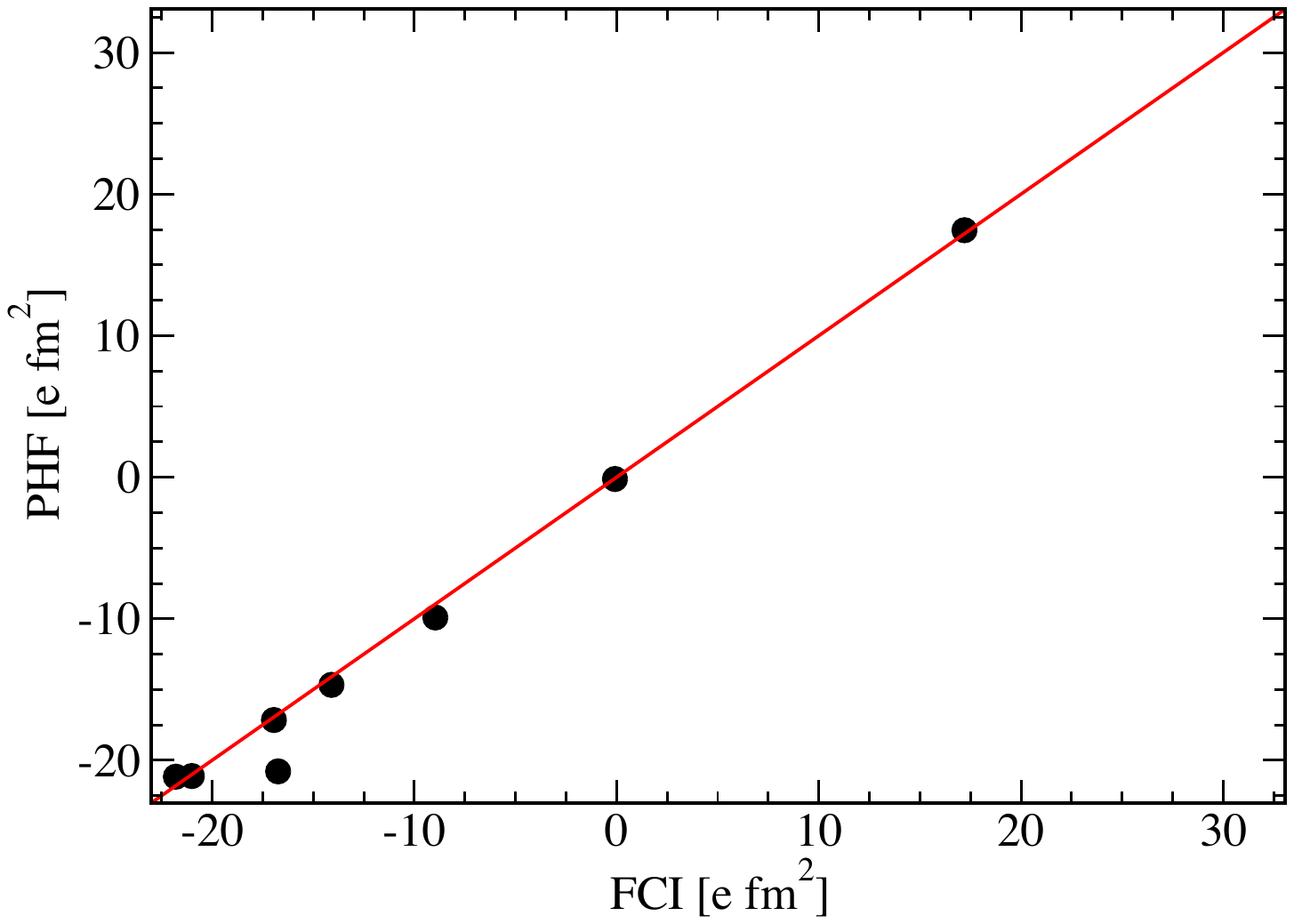}
        \caption{E2 moments for $^{24}$Mg.}
    \end{subfigure}
    \caption{Scatter plots of M1 and E2 moments for $^{24}$Mg. Moments calculated from FCI densities are plotted on the x-axis and from PHF on the y-axis. The $y=x$ line is shown for comparison.}
    \label{fig:mg24mom}
\end{figure}

\begin{figure}[H]
    \centering
    \begin{subfigure}{.49\linewidth}
        \includegraphics[width=\linewidth]{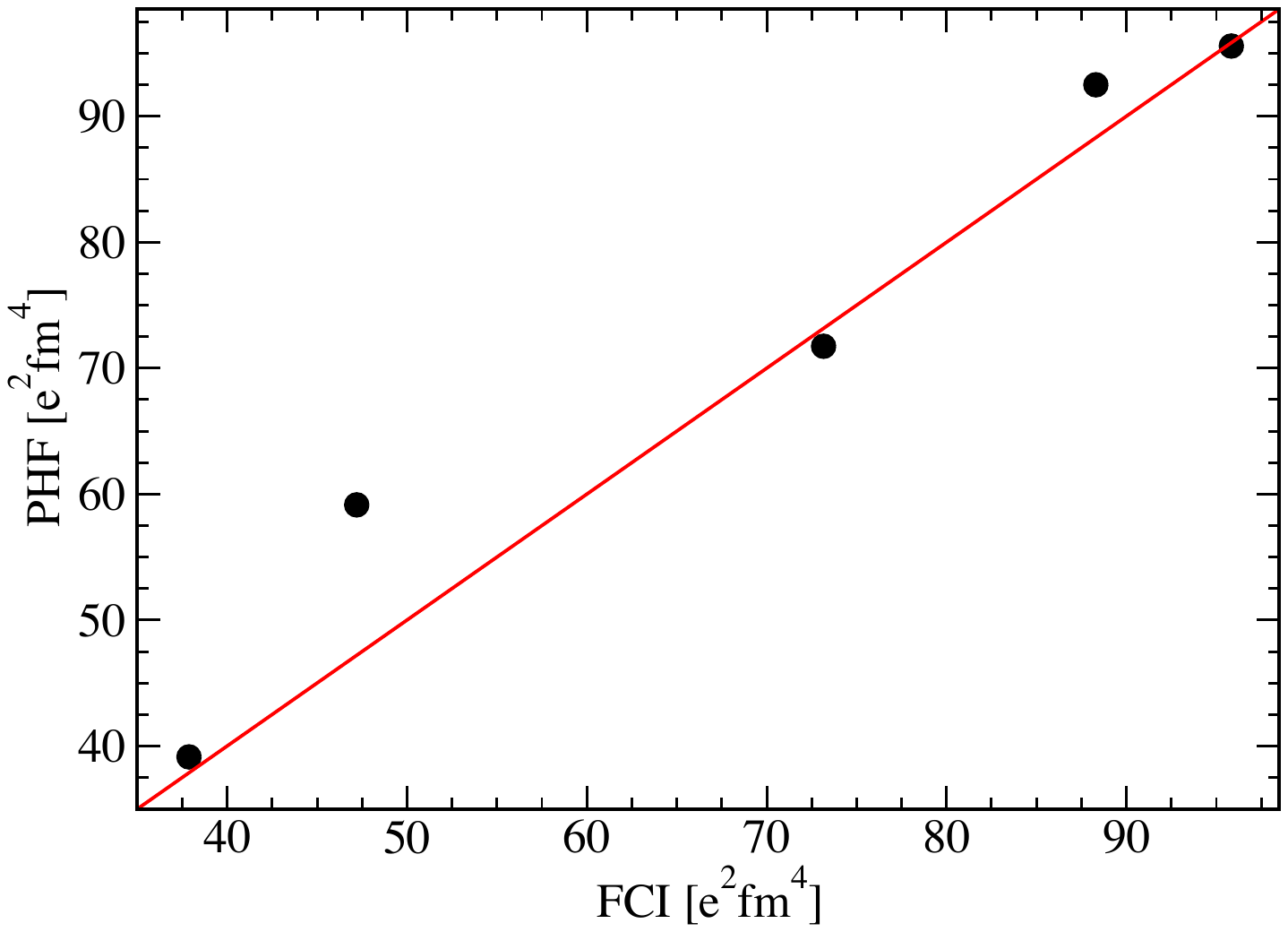}
        \caption{Intra-band B(E2) strengths for $^{24}$Mg.}
    \end{subfigure}
    \begin{subfigure}{.49\linewidth}
        \includegraphics[width=\linewidth]{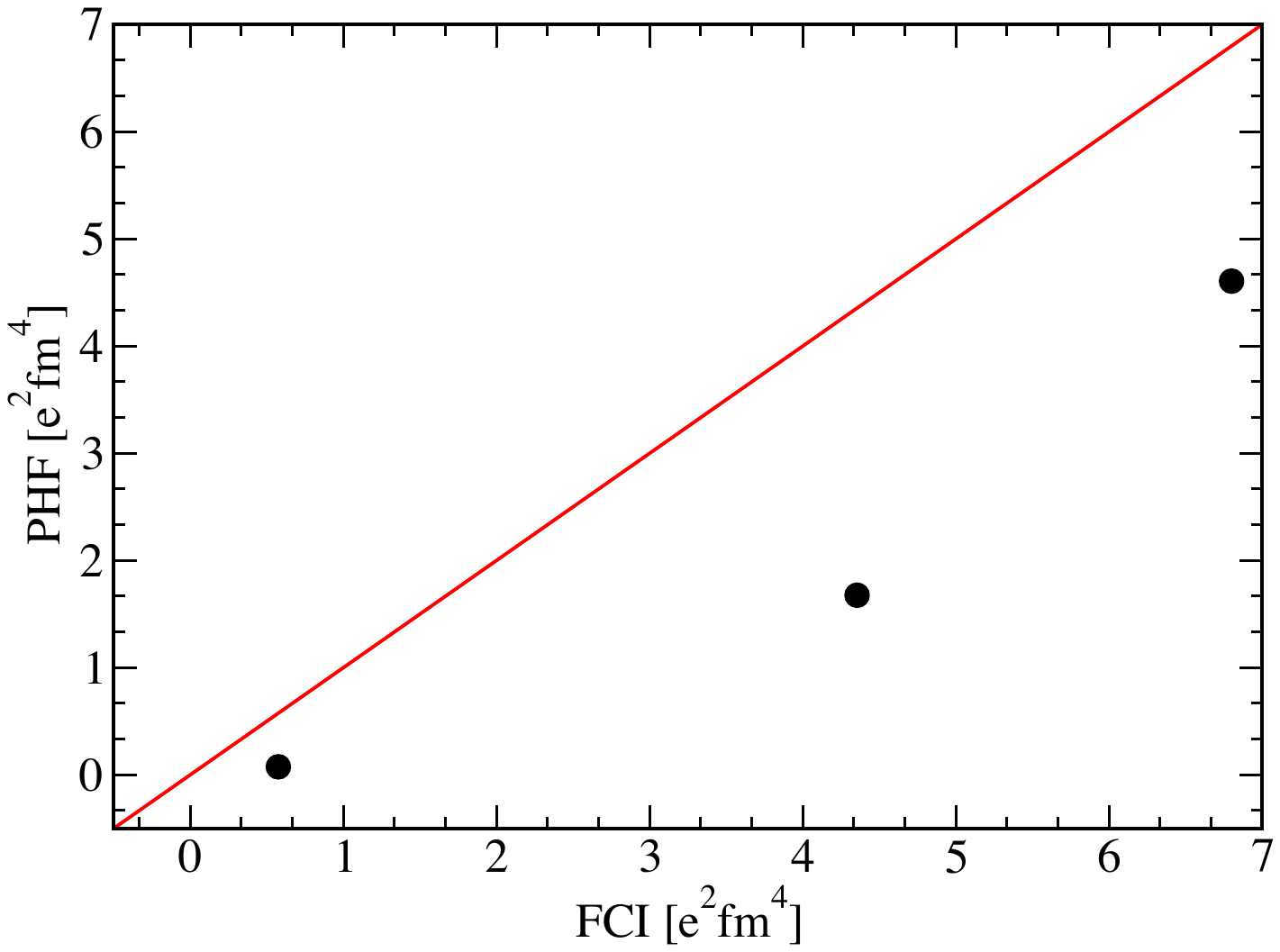}
        \caption{Inter-band B(E2) strengths for $^{24}$Mg.}
    \end{subfigure}
     \caption{Scatter plots of B(E2) transition strengths $^{24}$Mg. Transitions calculated from FCI densities are plotted on the x-axis and from PHF on the y-axis. The $y=x$ line is shown for comparison.}
    \label{fig:mg24trans}
\end{figure}

  $^{24}$Mg has a only a prolate, weakly triaxial ($\gamma = 11.94^\circ$) HF minima at -80.97 MeV.  The low-lying FCI and PHF excitation spectra agree reasonably well, as shown in
    Fig.~\ref{fig:sdenergies}(b), not only the ground state band but also an excited gamma band;  Table~\ref{tab:mg24energy} gives the 
    absolute and excitation energies. The static spectroscopic M1 and E2 moments, 
found also in Table~\ref{tab:mg24energy}, are shown in a scatter plot in 
Fig.~\ref{fig:mg24mom}(a) and (b), respectively; the FCI and PHF values are in 
remarkable agreement.
Table~\ref{tab:mg24e2} 
gives the B(E2) values, plotted in Fig~\ref{fig:mg24trans}(a) and (b) for 
intra-band and inter-band transitions, respectively. Just as for $^{20}$Ne, the intra-band 
B(E2) transition strengths agree better than the admittedly small number of inter-band 
transitinos.

\emph{Odd-odd.} $^{30}$Al has a triaxial ($\gamma = 34^\circ$) HF minimum at -137.85 MeV.
A single PHF minimum fails to provide the correct $3^+$ ground state for $^{30}$Al but the addition of a second minimum nearby lowers the $3^{+}_1$ excitation energy by roughly 800 keV, nearly 
degenerate with the $2^+_1$ state, as shown in Fig.~\ref{fig:sdenergies}(c). 
Adding the second minimum also puts the $4^+_1$ and $3^+_2$ very close in energy, as in the FCI spectrum. 

\begin{figure}[H]
    \centering
    \begin{subfigure}{.49\linewidth}
        \includegraphics[width=\linewidth]{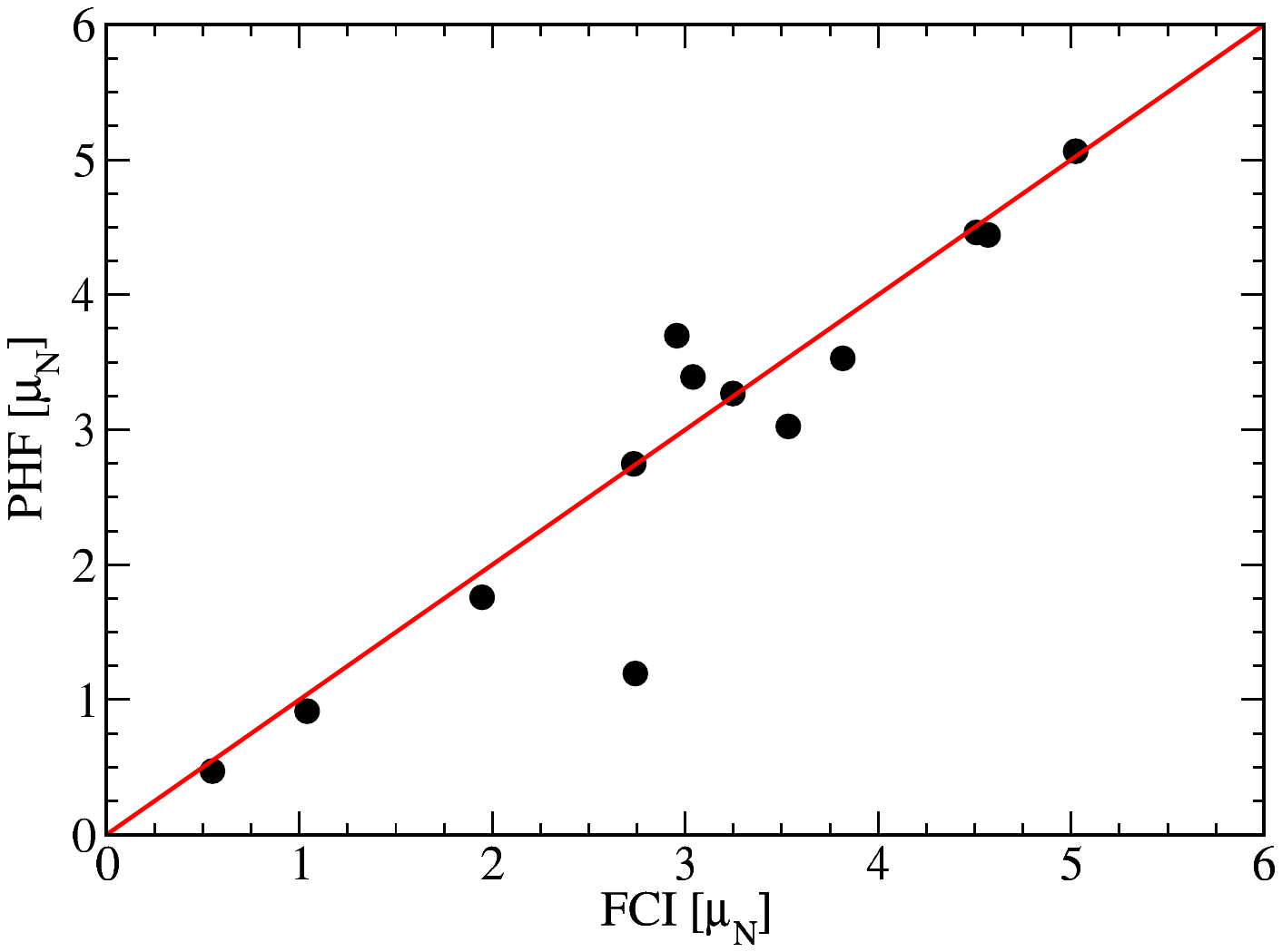}
        \caption{M1 moments for $^{30}$Al.}
    \end{subfigure}
    \begin{subfigure}{.49\linewidth}
        \includegraphics[width=\linewidth]{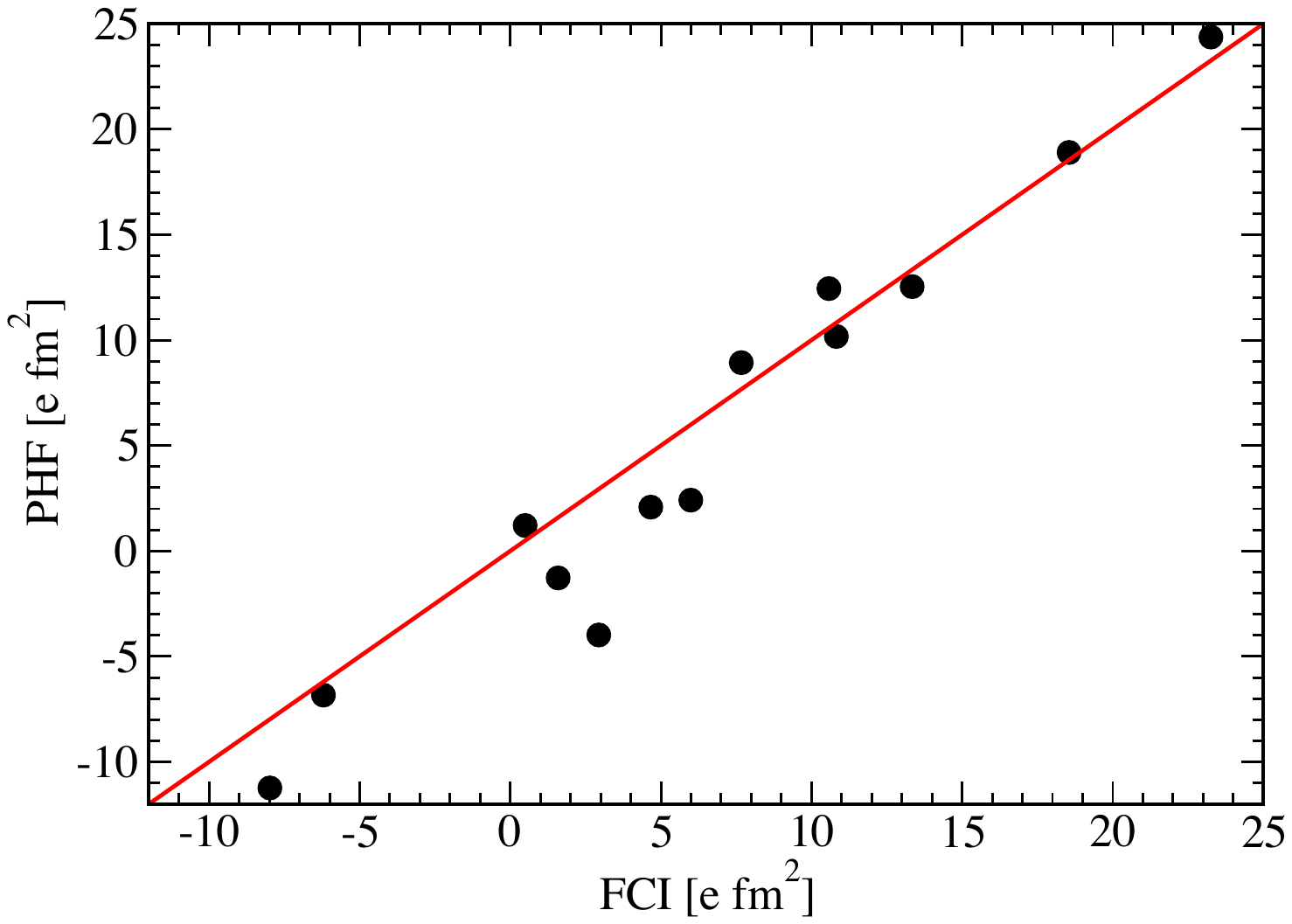}
        \caption{E2 moments $^{30}$Al.}
    \end{subfigure}
     \caption{Scatter plot of M1 and E2 moments for $^{30}$Al. Moments calculated from FCI densities are plotted on the x-axis and from PHF on the y-axis. The $y=x$ line is shown for comparison.}
    \label{fig:al30mom}
\end{figure}

\begin{figure}[H]
    \centering
         \begin{subfigure}{.49\linewidth}
        \includegraphics[width=\linewidth]{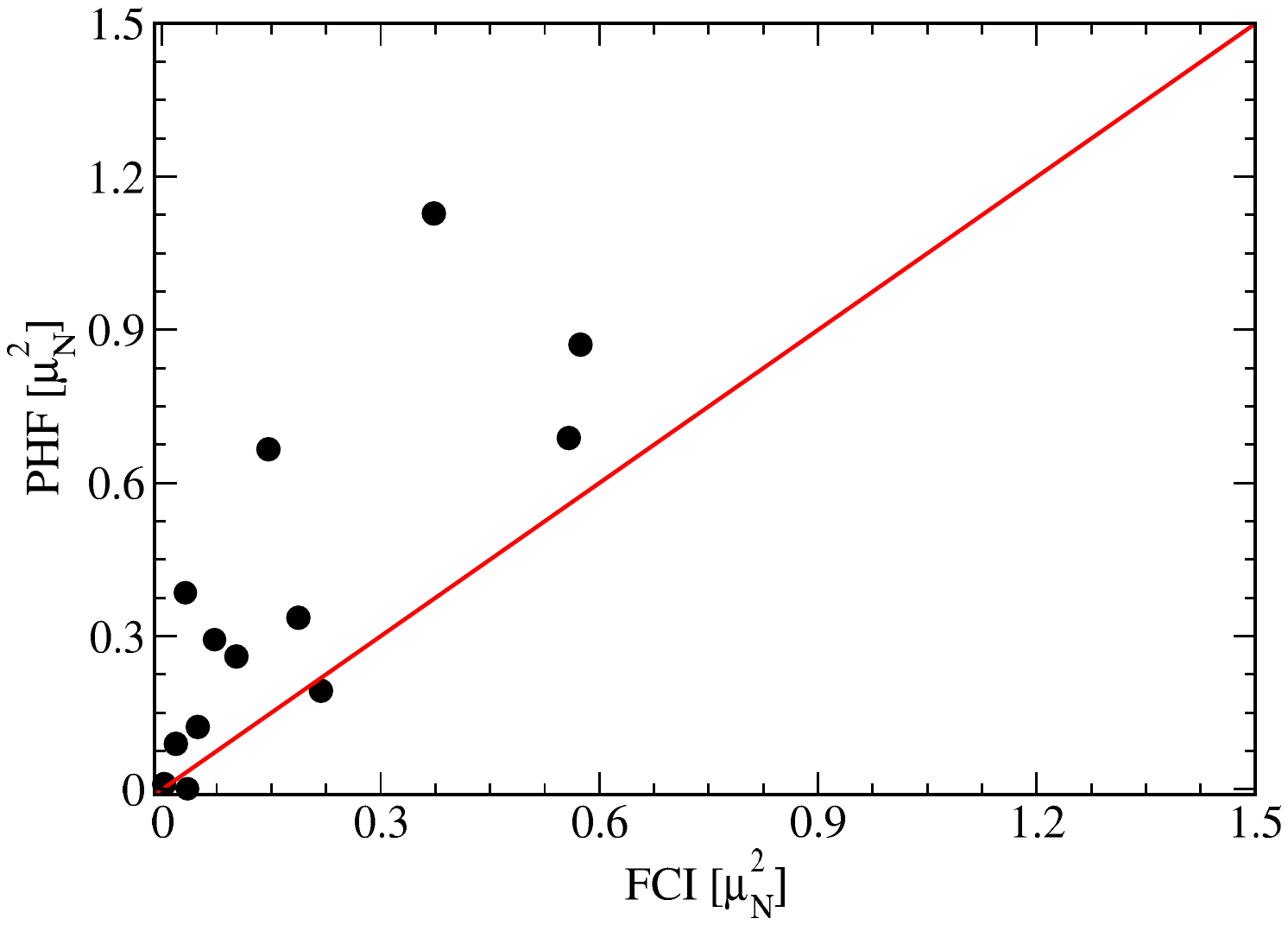}
        \caption{B(M1) strengths $^{30}$Al.}
    \end{subfigure}
    \begin{subfigure}{.49\linewidth}
        \includegraphics[width=\linewidth]{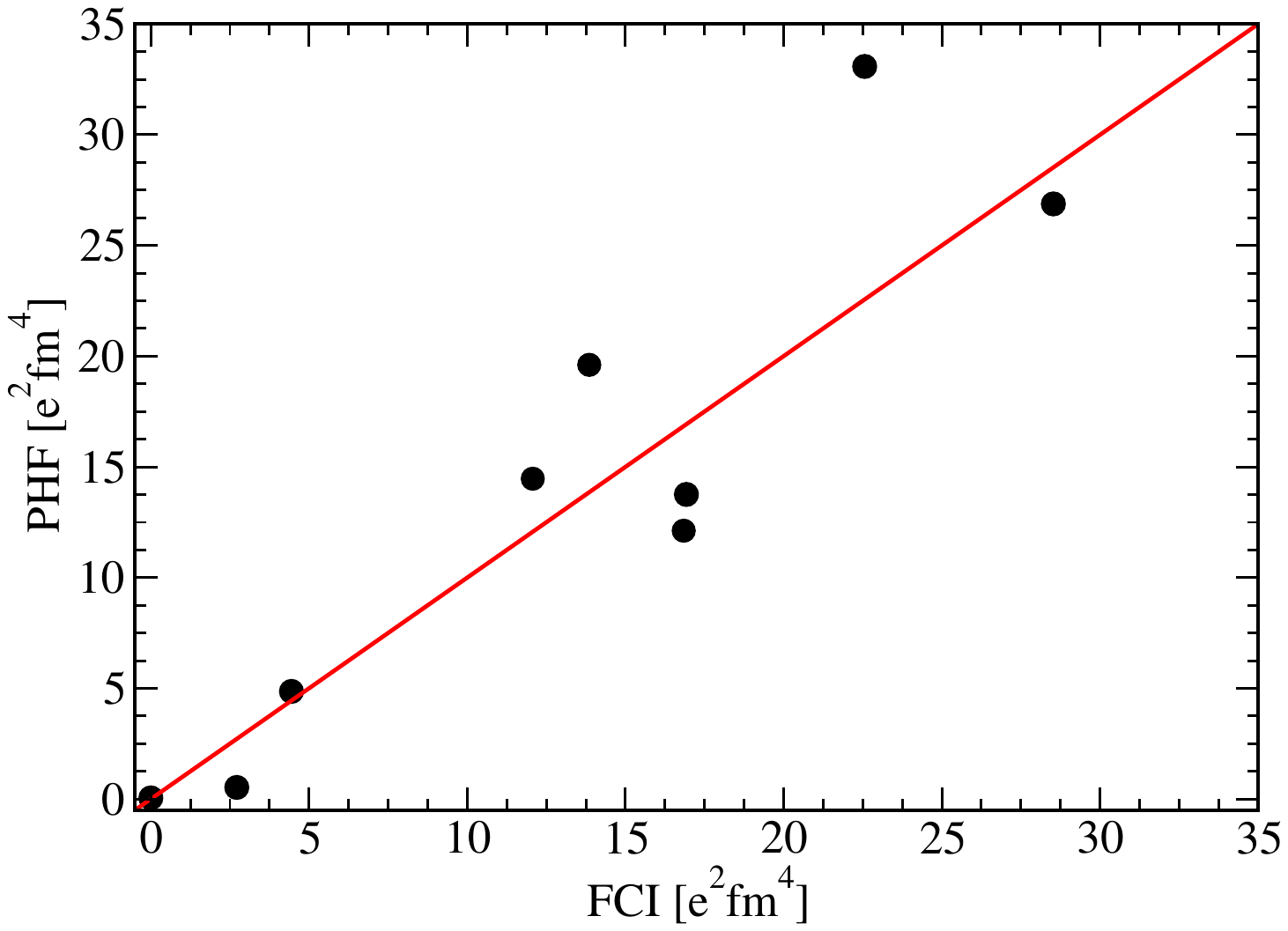}
        \caption{B(E2) strengths $^{30}$Al.}
    \end{subfigure}
     \caption{Scatter plot of selected B(E2) transition strengths for $^{30}$Al. Transitions calculated from FCI densities are plotted on the x-axis and from PHF on the y-axis. The $y=x$ line is shown for comparison.}
    \label{fig:al30trans}
\end{figure}

The structure of odd-odd nuclides are not expected to be simply captured by deformation, 
so it is somewhat surprising that the PHF values for static M1 and E2 moments, plotted in Fig.~\ref{fig:al30mom}(a) and (b), respectively, and B(E2) transition strengths, plotted in Fig.~\ref{fig:al30trans}(b), agree surprisingly well with the benchmark FCI values. 
The B(M1) transition strengths, however, have poorer agreement, as plotted in Fig.~\ref{fig:al30trans} (a).

\begin{figure}[H]
    \centering
    \begin{subfigure}{.49\linewidth}
        \includegraphics[width=\linewidth]{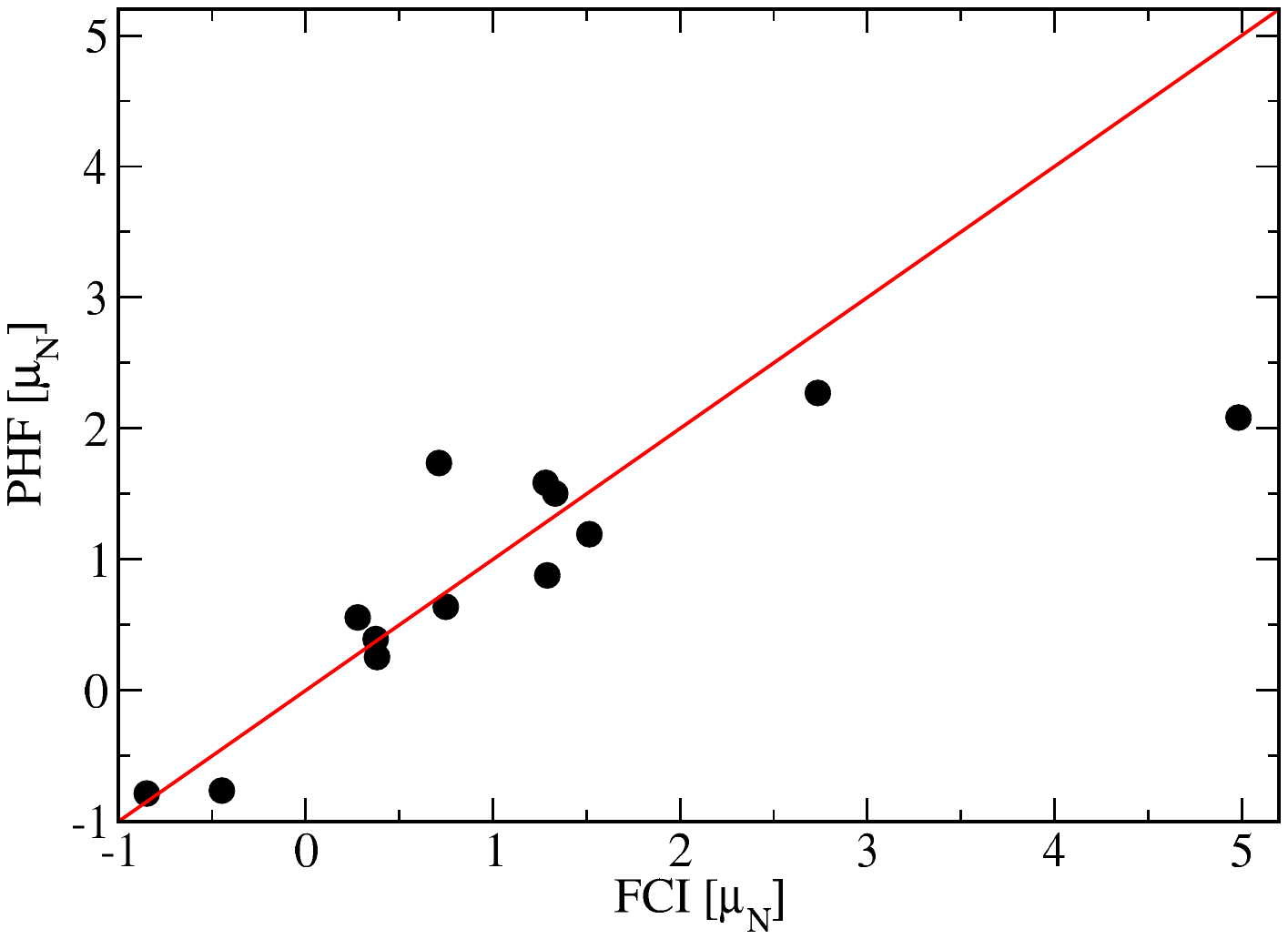}
        \caption{M1 moments for $^{25}$Mg.}
    \end{subfigure}
    \begin{subfigure}{.49\linewidth}
        \includegraphics[width=\linewidth]{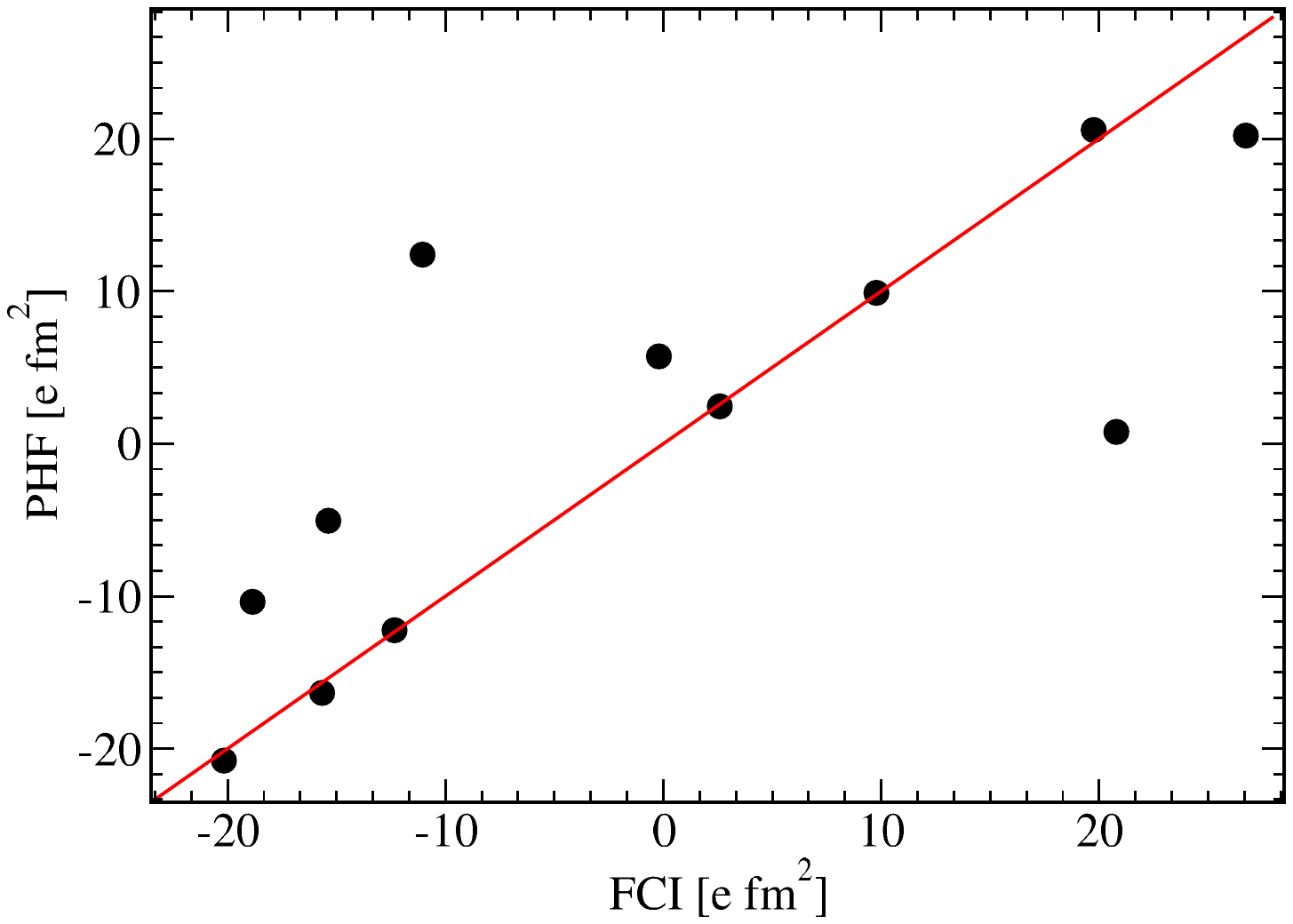}
        \caption{E2 moments for $^{25}$Mg.}
    \end{subfigure}
     \caption{Scatter plot of M1 and E2 moments for $^{25}$Mg. Moments calculated from FCI densities are plotted on the x-axis and from PHF on the y-axis. The $y=x$ line is shown for comparison.}
    \label{fig:mg25mom}
\end{figure}

\begin{figure}[H]
    \centering

    \begin{subfigure}{0.49\linewidth}
        \includegraphics[width=\linewidth]{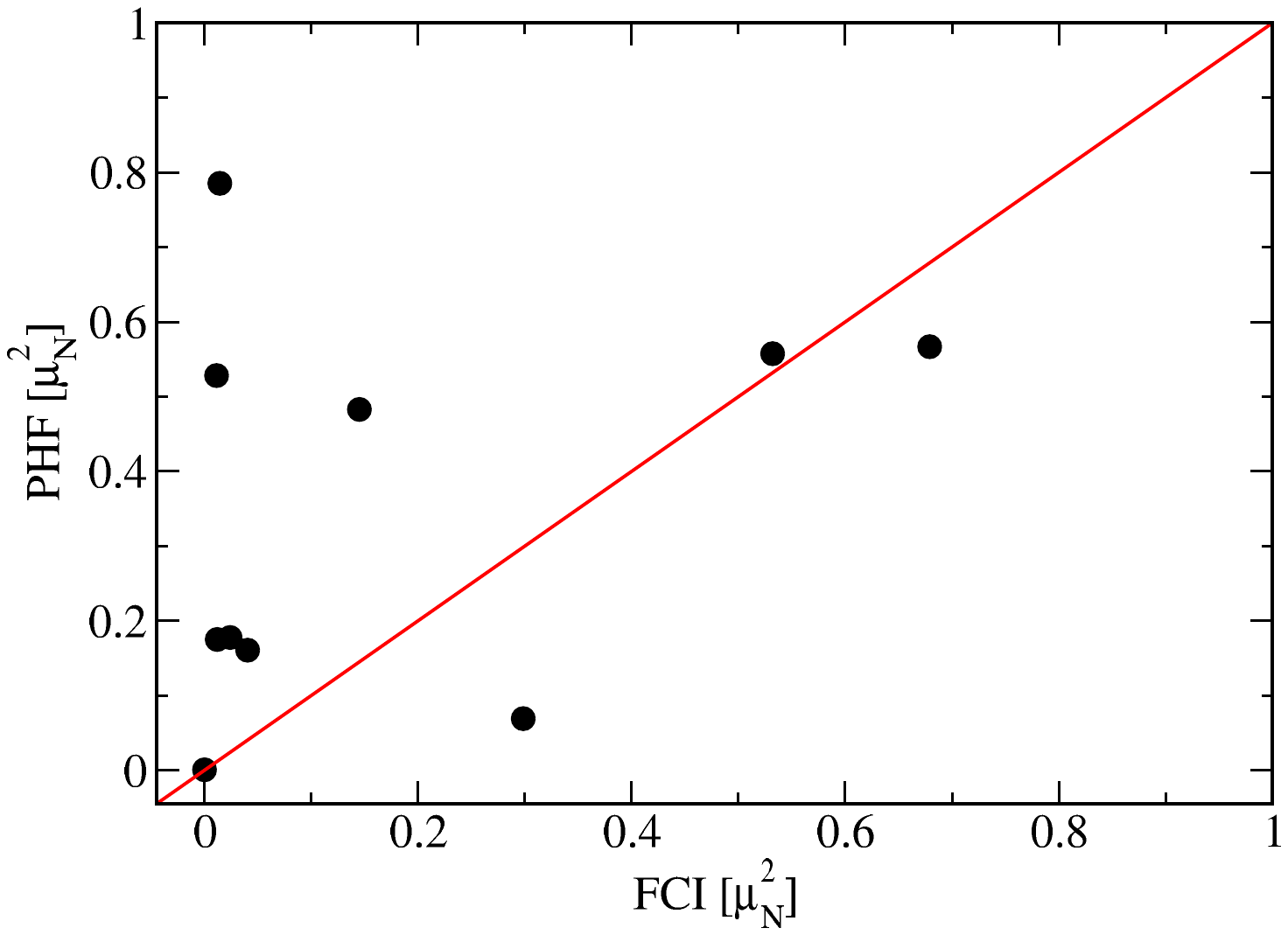}
        \caption{B(M1) strengths for $^{25}$Mg.}
    \end{subfigure}
        \begin{subfigure}{0.49\linewidth}
        \includegraphics[width=\linewidth]{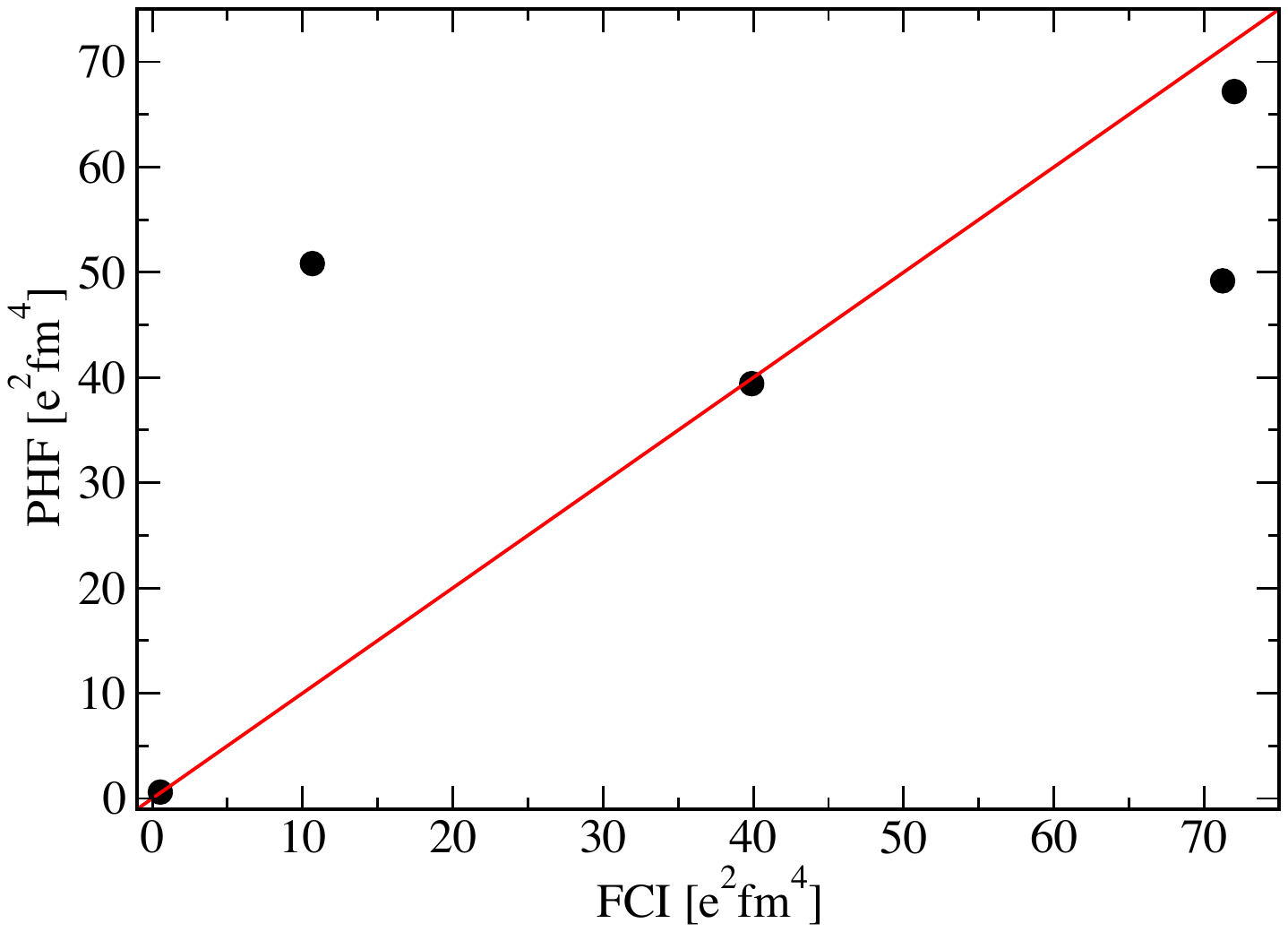}
    \caption{B(E2) strengths for $^{25}$Mg.}
    \end{subfigure}
    \caption{Scatter plot of (a) B(M1) transition strengths and (b) B(E2) transition strengths  for $^{25}$Mg. Transitions calculated from FCI densities are plotted on the x-axis and from PHF on the y-axis. The $y=x$ line is shown for comparison.}
\label{fig:mg25trans}
\end{figure}

\emph{Odd-A.} 
$^{25}$Mg has a single triaxial ($\gamma = 19.08^\circ$) HF minima at -89.11 MeV.
PHF reproduces the order of low-lying states found in the FCI calculations, as plotted in 
Fig.~\ref{fig:sdenergies}(d), and given in Table~\ref{tab:mg25energy}. Many, though not all, 
of the M1 and E2 moments agree well between PHF and FCI calculations, as plotted in Fig.~\ref{fig:mg25mom}(a) and (b), respectively, and values found in Table~\ref{tab:mg25energy}.
The transition B-values, plotted in Fig.~\ref{fig:mg25trans}, with values found for B(M1) in 
Table~\ref{tab:mg25m1} and for B(E2) in Table~\ref{tab:mg25e2}, agree less well.


\subsection{Examples from the $pf$-shell}
\label{pfexample}

We also computed examples in  
the $pf$ shell, which assumes a frozen $^{40}$Ca core with valence $1p_{1/2}$-$1p_{3/2}$-$0f_{5/2}$-$0f_{7/2}$ orbitals, and used a modified 
$G$-matrix interaction (GX1A) interaction \cite{PhysRevC.65.061301,PhysRevC.69.034335,honma2005shell}.

\begin{figure}
    \centering
    \begin{subfigure}{0.45\linewidth}
        \includegraphics[width=\linewidth]{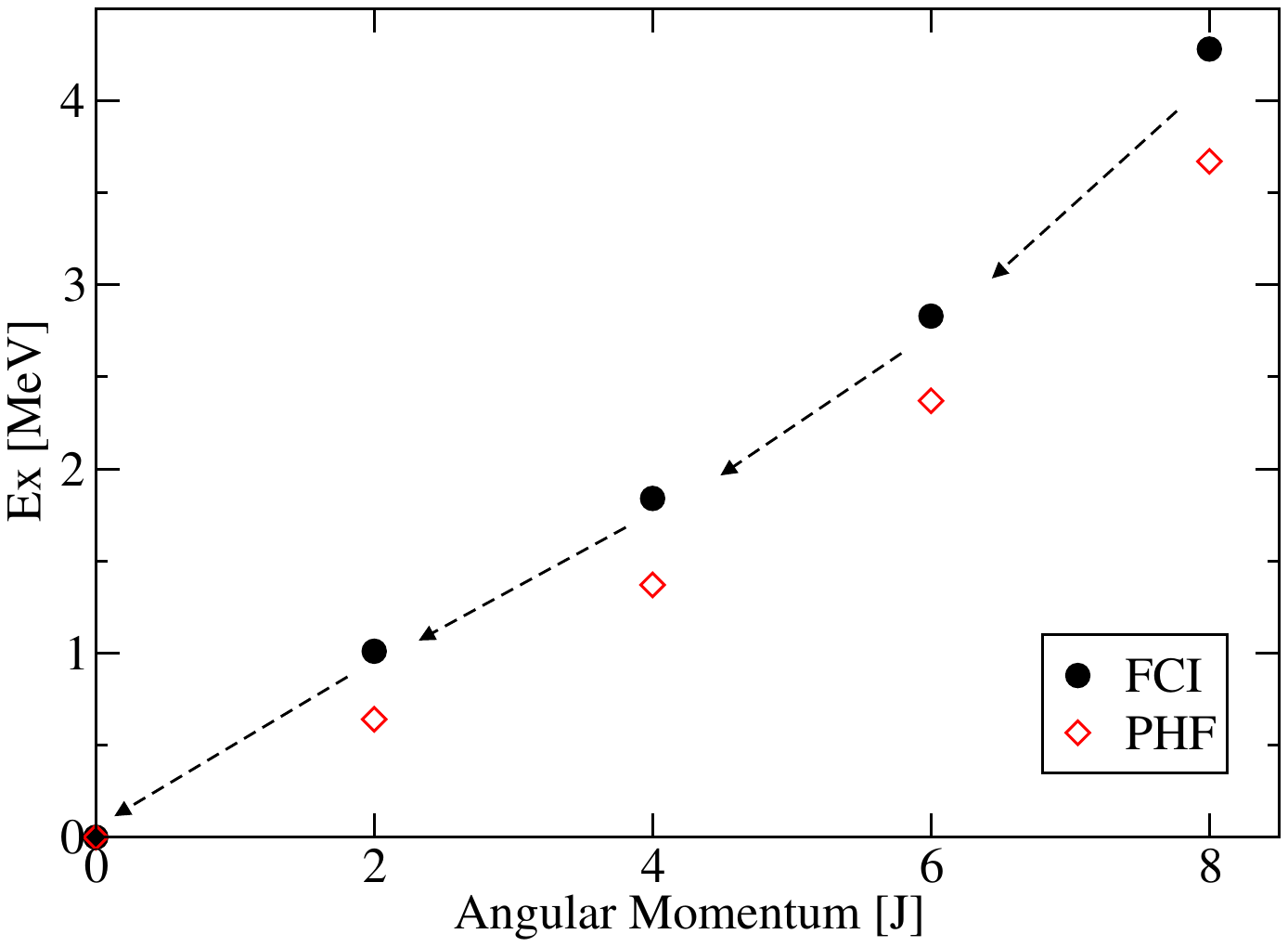}
        \caption{$^{46}$Ti}
        \end{subfigure}
          \begin{subfigure}{0.45\linewidth}
        \includegraphics[width=\linewidth]{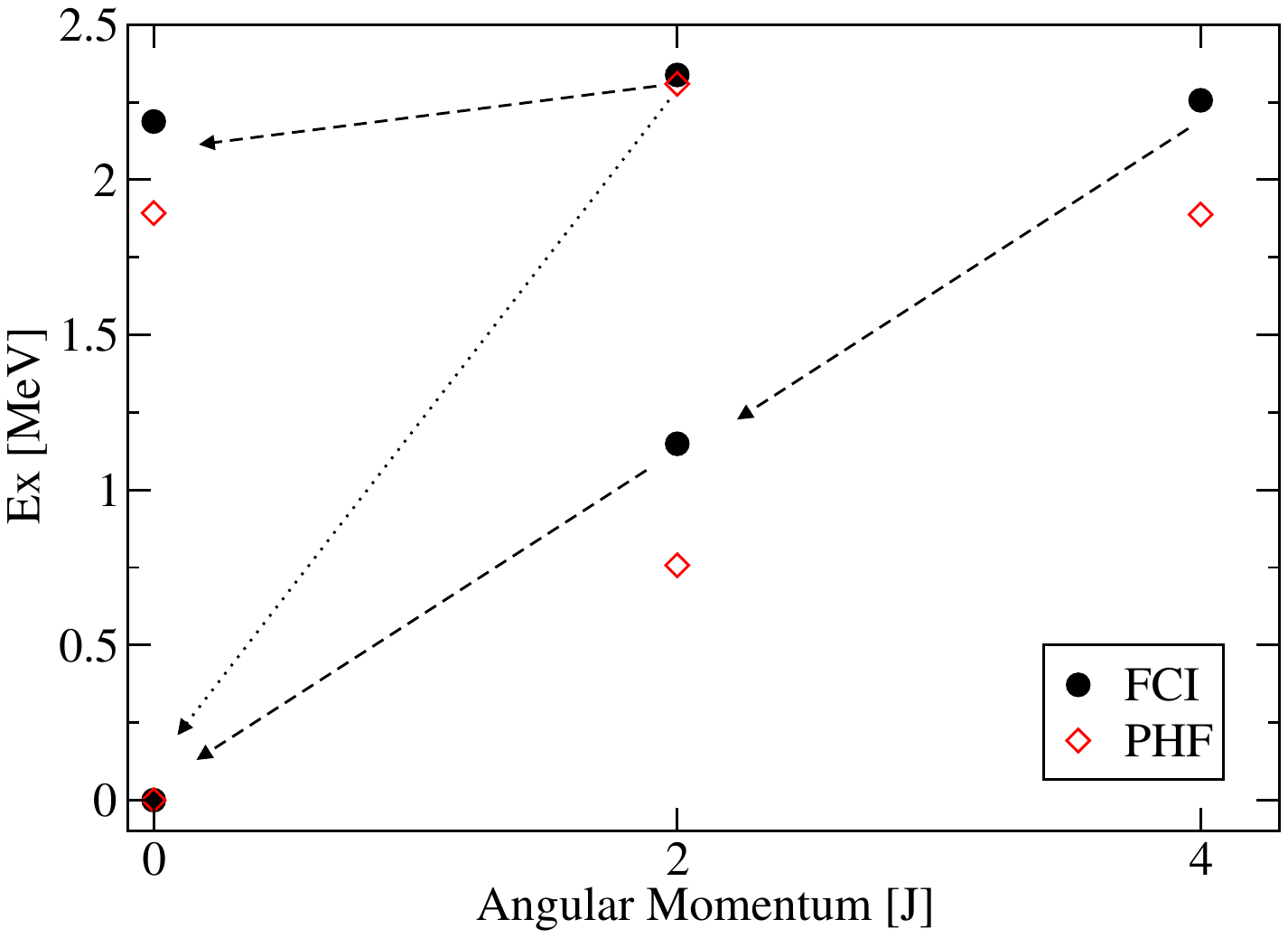}
        \caption{$^{62}$Ni}
    \end{subfigure}  
    \vfill
    \begin{subfigure}{0.45\linewidth}
        \includegraphics[width=\linewidth]{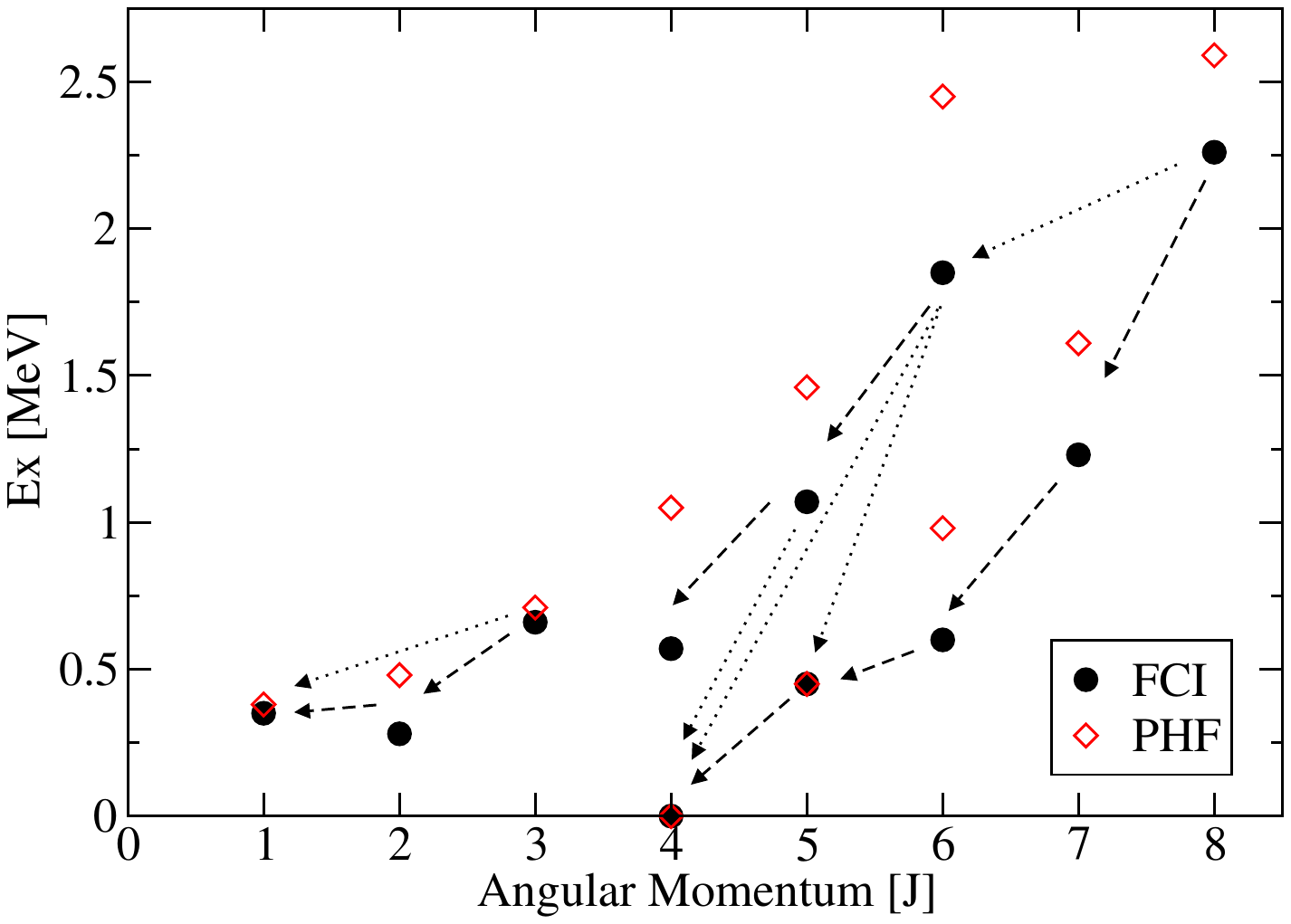}
        \caption{$^{48}$V}
    \end{subfigure}
    \begin{subfigure}{0.45\linewidth}
        \includegraphics[width=\linewidth]{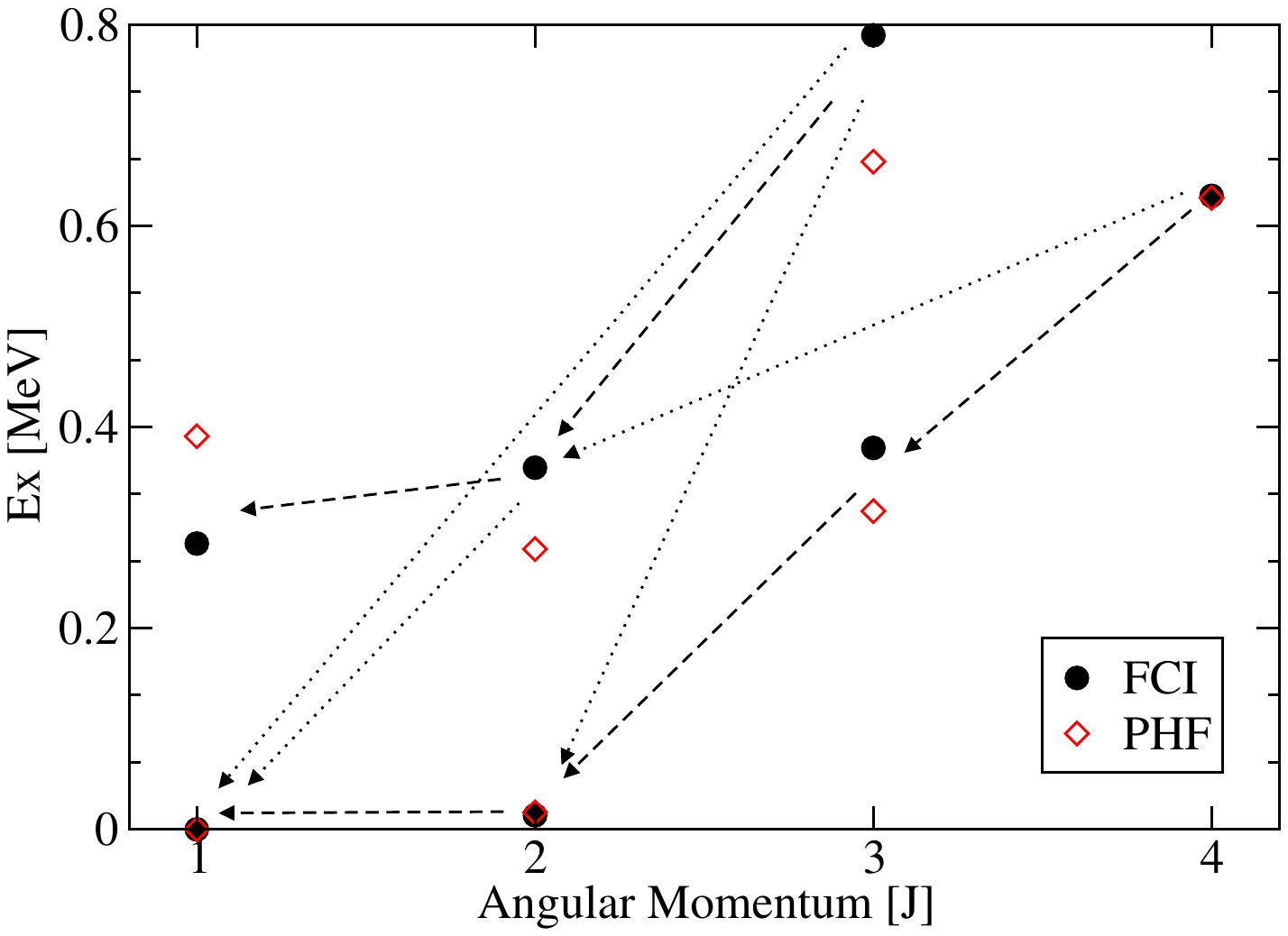}
        \caption{$^{64}$Cu}
    \end{subfigure}
    \vfill
    \begin{subfigure}{0.45\linewidth}
        \includegraphics[width=\linewidth]{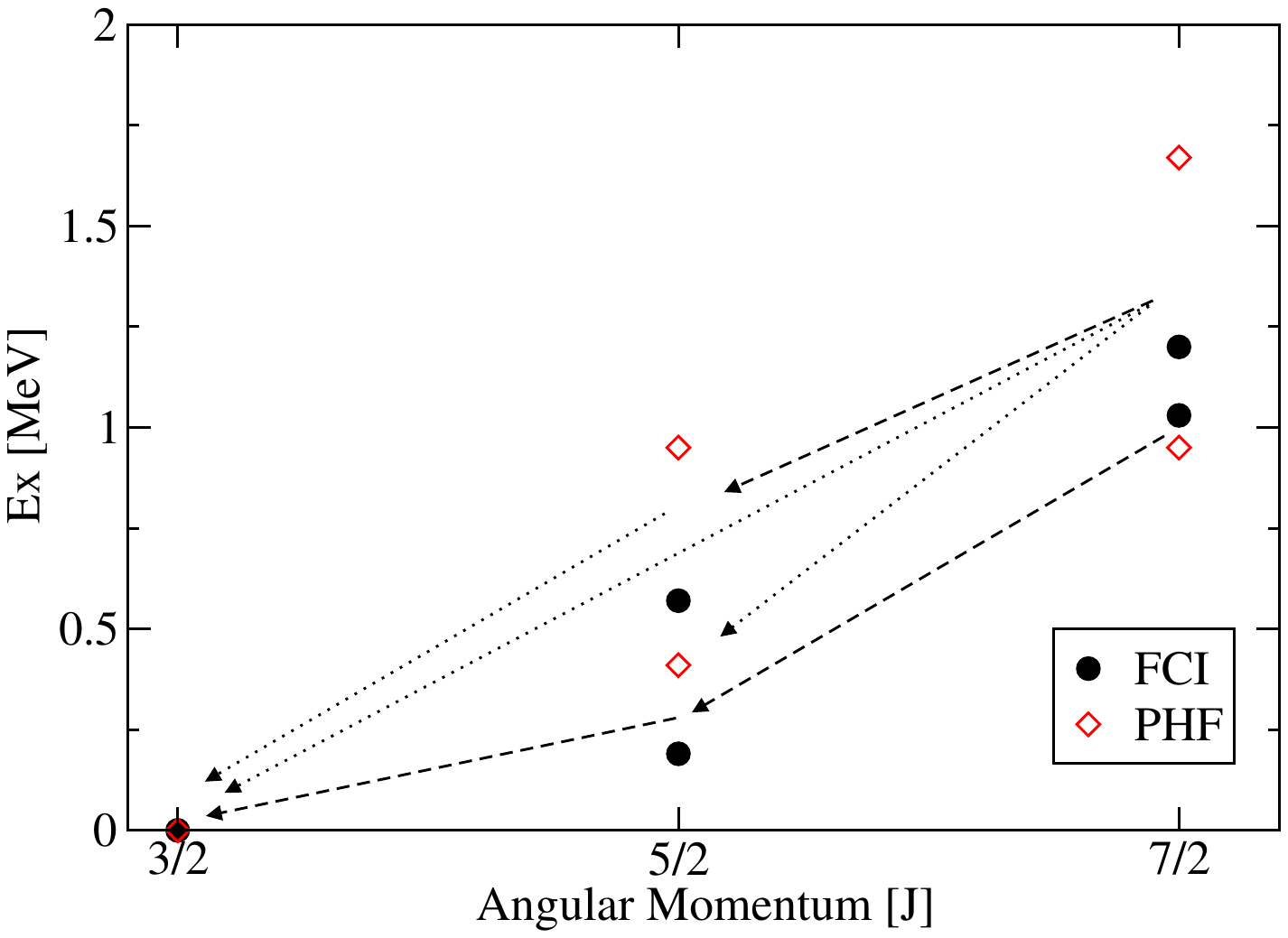}
        \caption{$^{63}$Zn}
    \end{subfigure}
    \caption{Excitation energy plots for five examples in the $pf$-shell, comparing FCI (filled circles) and PHF (open diamonds) results. The dashed and dotted lines are used to represent different paths from the same state.} 
    \label{fig:pfenergies}
\end{figure}

Five benchmark cases for the $pf$-shell are shown in Figure \ref{fig:pfenergies} with two even-even nuclei ($^{46}$Ti and $^{62}$Ni), two odd-odd nuclei ($^{48}$V and $^{64}$Cu) and one odd-A nucleus ($^{63}$Zn). 

\begin{figure}[H]
   \begin{subfigure}{0.45\linewidth}
       \includegraphics[width=\linewidth]{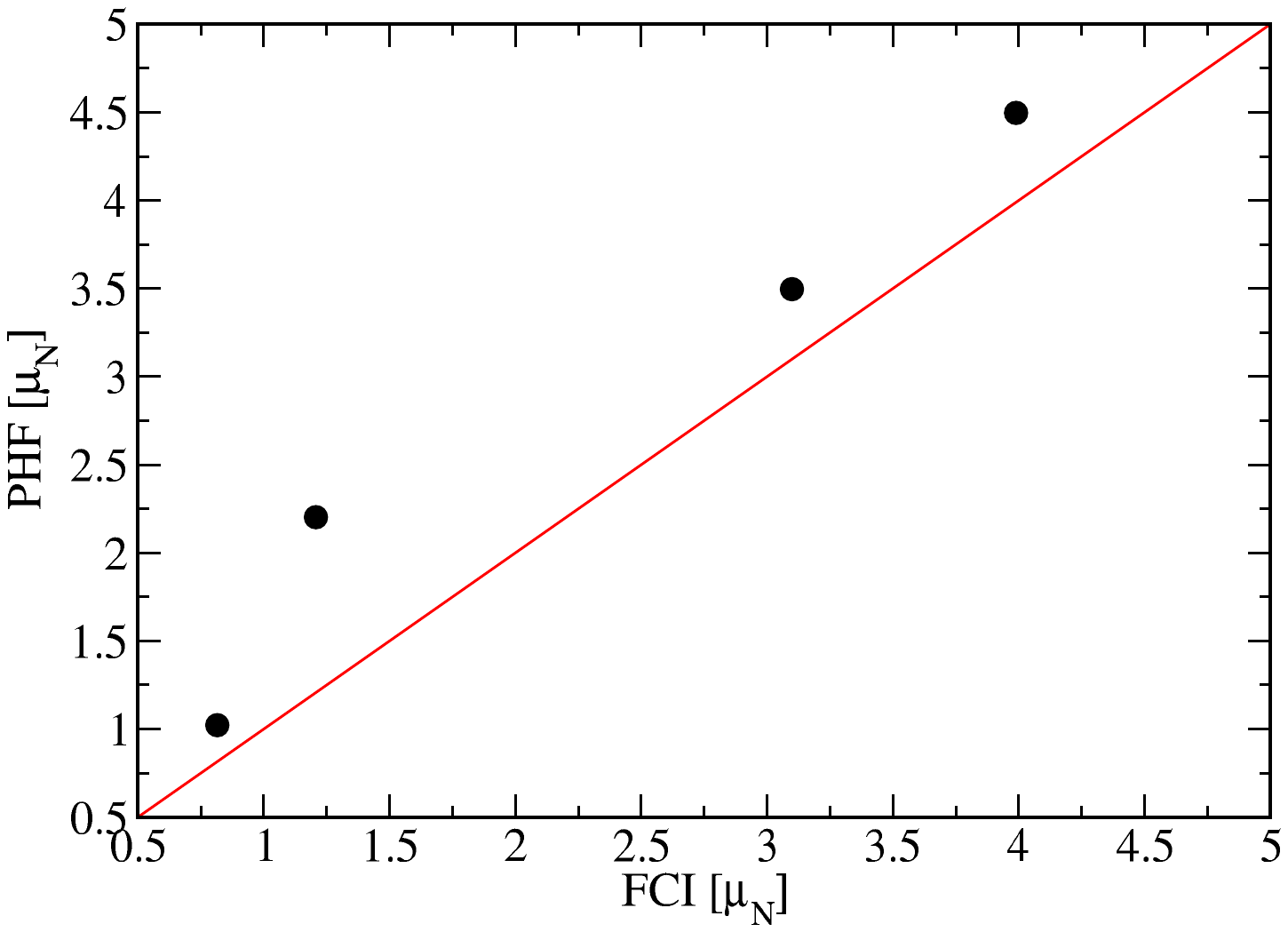}
       \caption{M1 moments for $^{46}$Ti.}
   \end{subfigure}
   \begin{subfigure}{0.45\linewidth}
       \includegraphics[width=\linewidth]{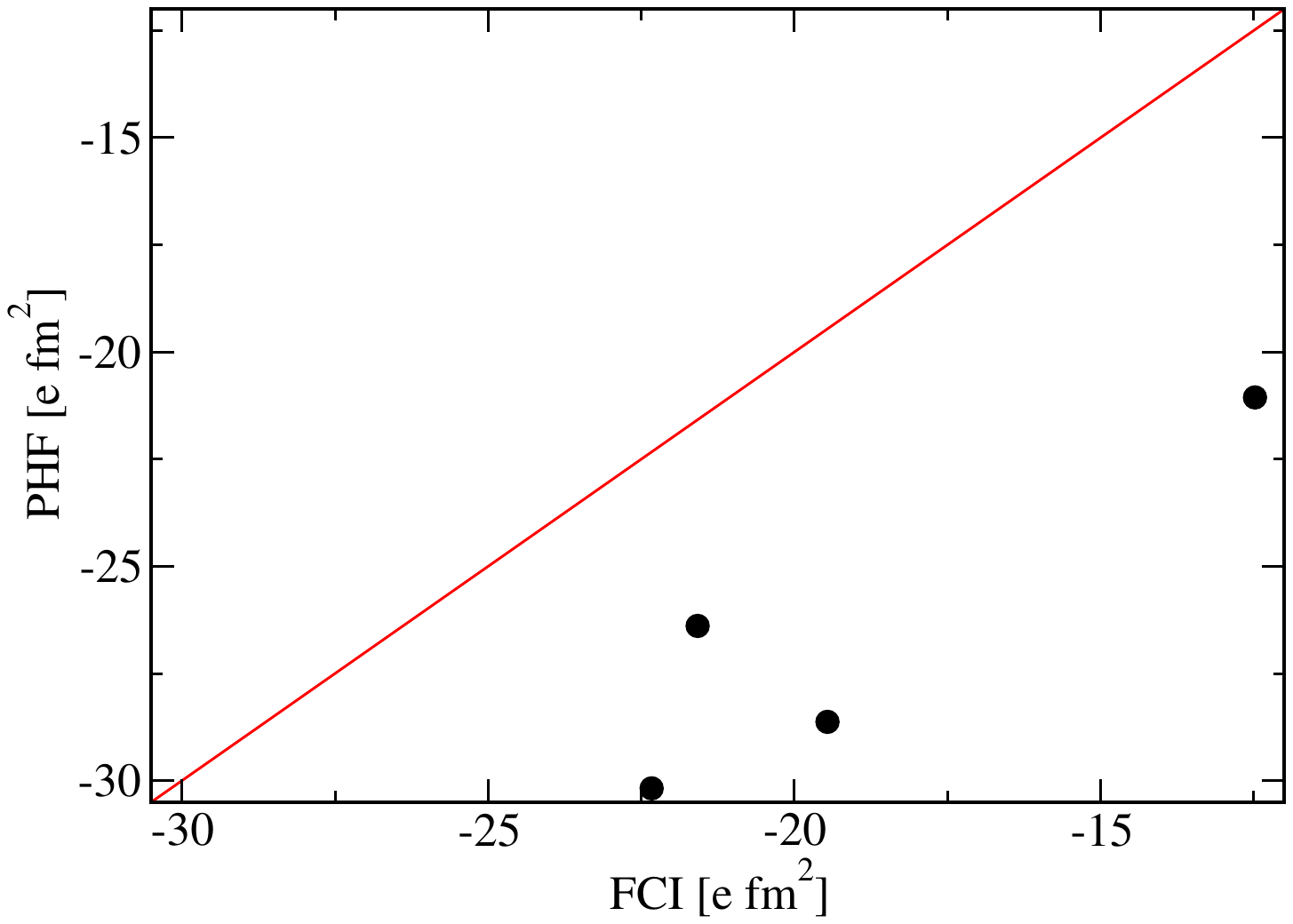}
       \caption{E2 moments or $^{46}$Ti.}
   \end{subfigure}
   \caption{Scatter plot of M1 and E2 moments for $^{46}$Ti. Moments calculated from FCI densities are plotted on the x-axis and from PHF on the y-axis. The $y=x$ line is shown for comparison.}
    \label{fig:ti46mom}
\end{figure}

\begin{figure}[H]
    \centering
    \includegraphics[width=0.5\linewidth]{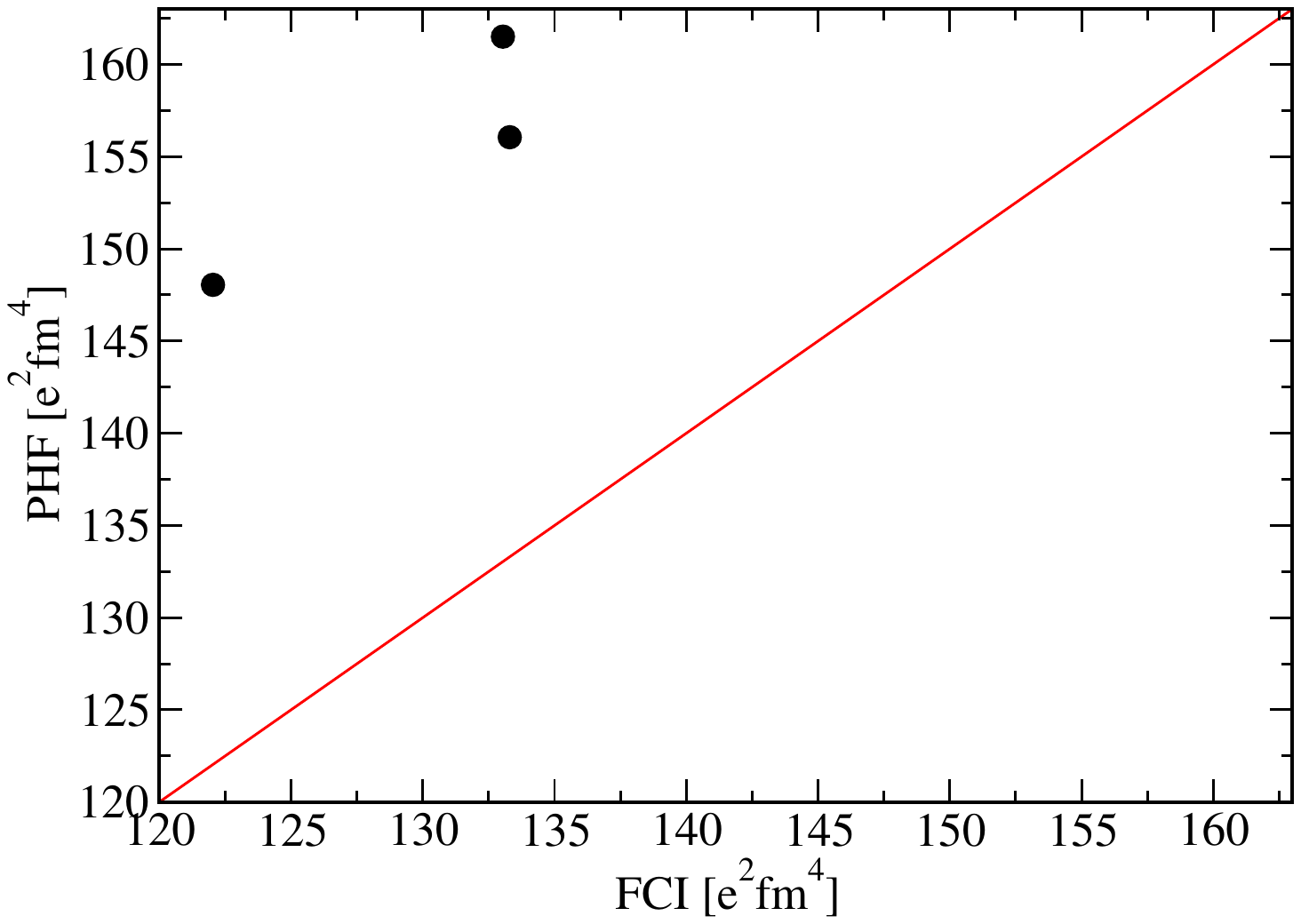}
    \caption{B(E2) transition strengths for $^{46}$Ti}
    \label{fig:ti46trans}
    \end{figure}

\emph{Even-even}. $^{46}$Ti has a single prolate HF minimum at -63.3 MeV, with $\beta=0.325$.
Somewhat surprisingly, while the M1 moments in Fig.~\ref{fig:ti46mom}(a) have reasonably good agreement, the quadrupole observables, both E2 moments in Fig.~\ref{fig:ti46mom}(b) and
B(E2) transition strengths in Fig.~\ref{fig:ti46trans}, have poor agreement.
(Since we have an axially symmetric HF state and thus projected states of only even $J$, 
we present no M1 transition data.)
This is 
contrary to what one would expect for quadrupole observables in even-even nuclides, and 
contrary to most of our other results. Nonetheless, we present it as a cautionary tale, to acknowledge that PHF can fail even in this straightforward case.

\begin{figure}[H]
   \begin{subfigure}{0.45\linewidth}
       \includegraphics[width=\linewidth]{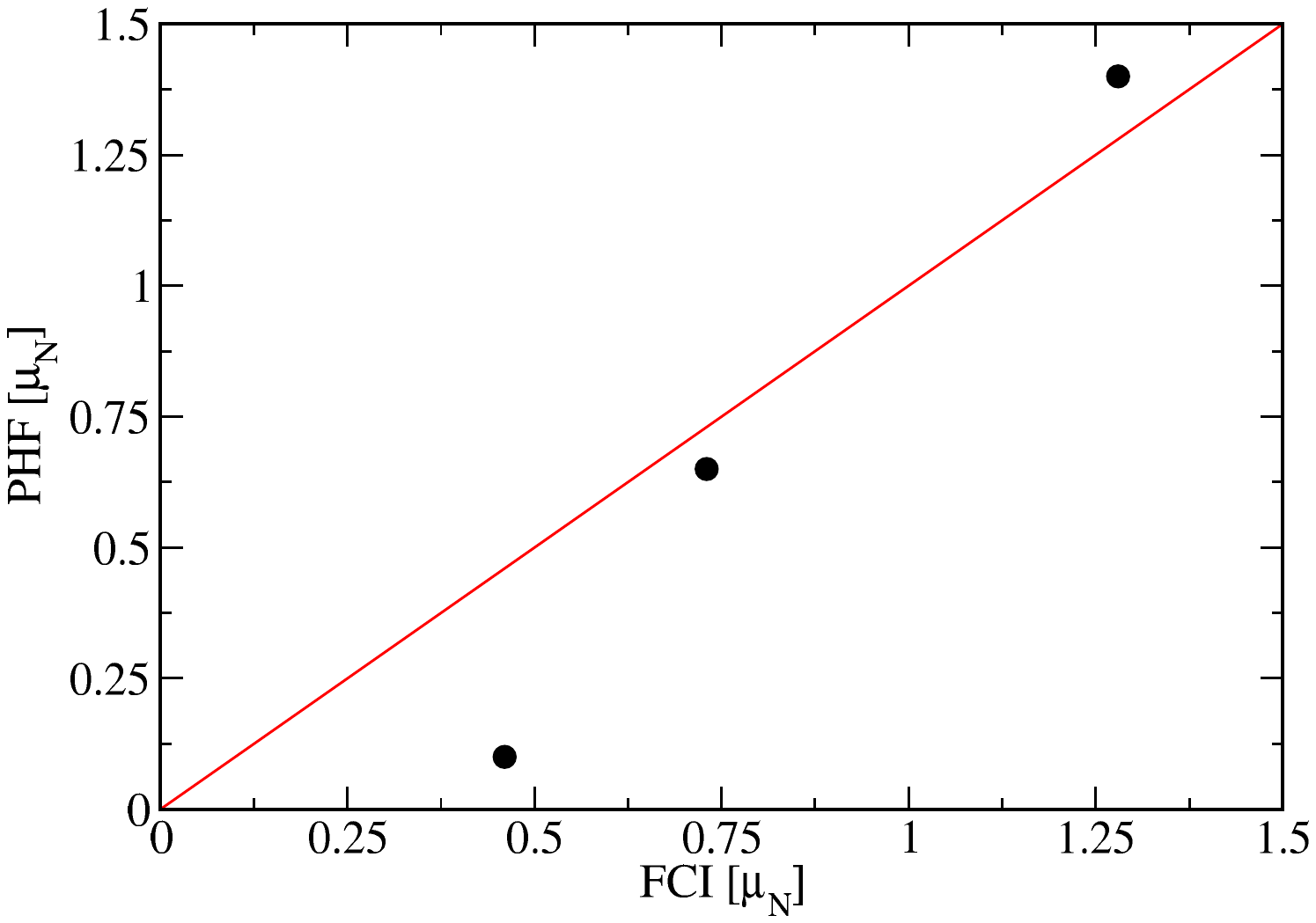}
       \caption{M1 moments for $^{62}$Ni.}
   \end{subfigure}
   \begin{subfigure}{0.45\linewidth}
       \includegraphics[width=\linewidth]{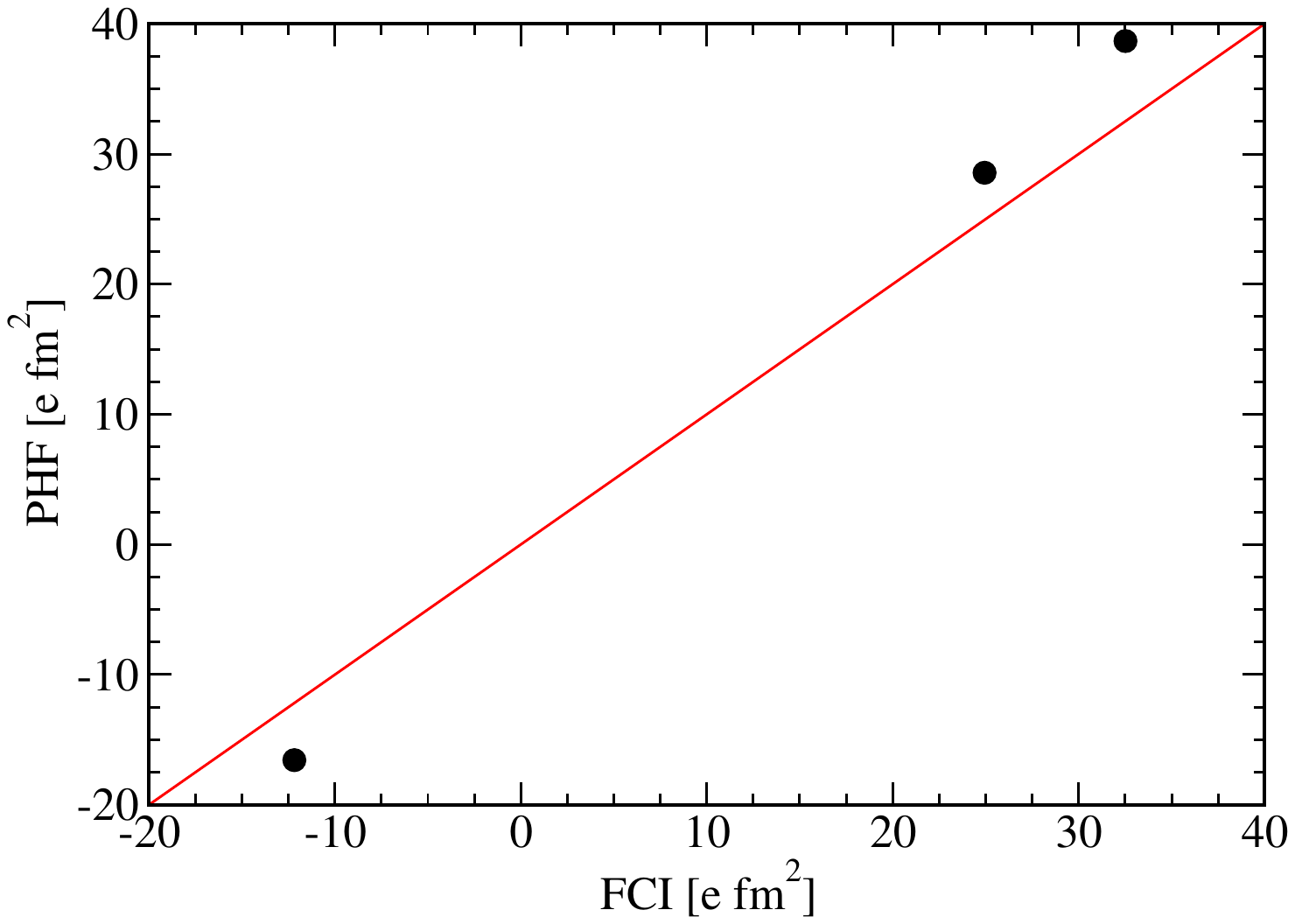}
       \caption{E2 moments for $^{62}$Ni}
   \end{subfigure}
   \caption{Scatter plot of M1 and E2 moments for $^{62}$Ni. Moments calculated from FCI densities are plotted on the x-axis and from PHF on the y-axis. The $y=x$ line is shown for comparison.}
    \label{fig:ni62mom}
\end{figure}

\begin{figure}[H]
    \centering
    \includegraphics[width=0.5\linewidth]{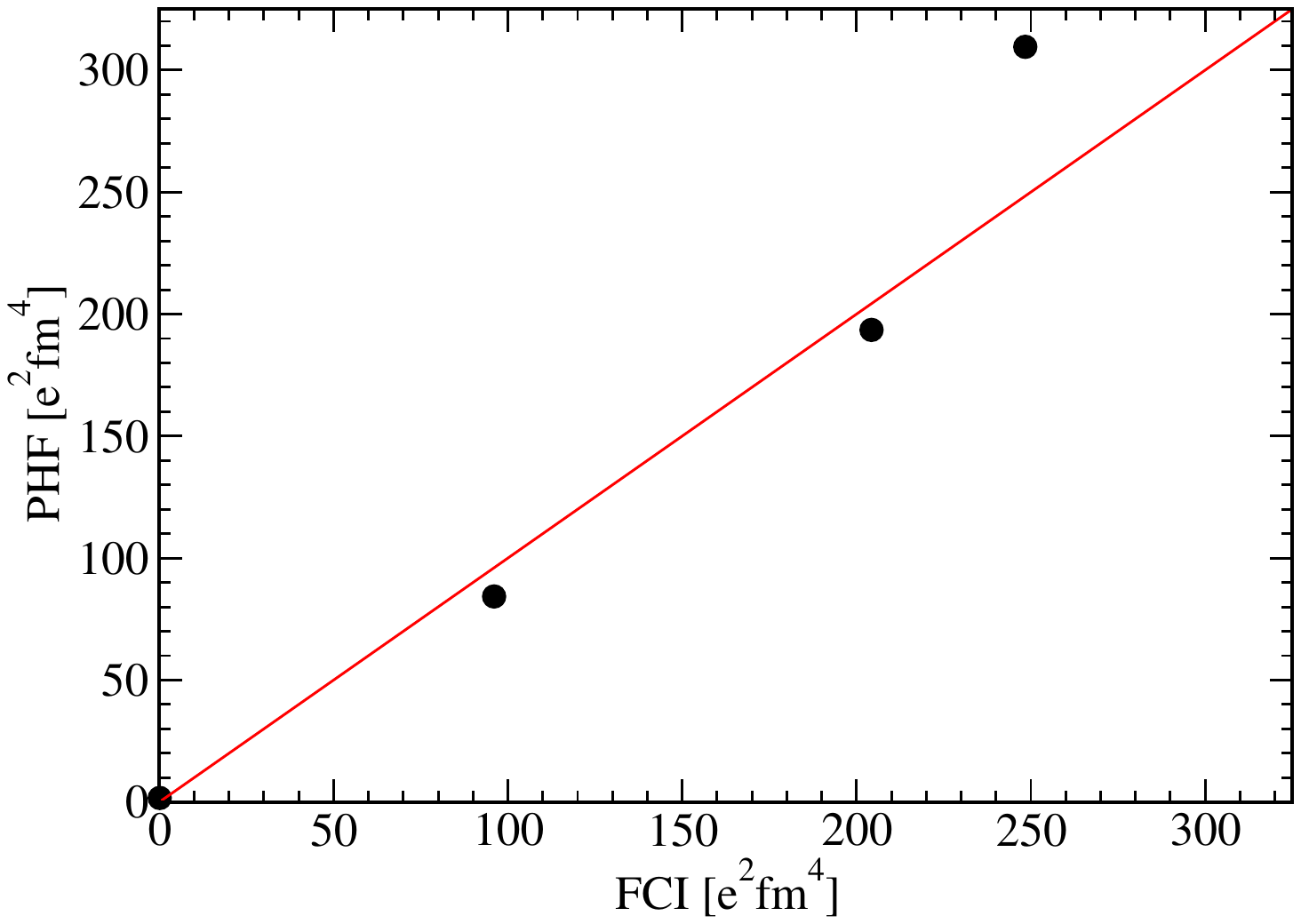}
    \caption{B(E2) transition strengths for $^{62}$Ni}
    \label{fig:ni62trans}
    \end{figure}

$^{62}$Ni has an oblate HF minimum at -261.3 MeV with $\beta = 0.12$ and a second, prolate minimum at -260.5 MeV with $\beta=0.08$. Unlike $^{46}$Ti, here both the moments (M1 in Fig.~\ref{fig:ni62mom}(a) and E2 in Fig.~\ref{fig:ni62mom}), and the B(E2) transition strengths, found in Fig.~\ref{fig:ni62trans}, have good agreement between PHF and FCI. 
Once again, as the projected states have only even $J$, we give no M1 transition data.

\begin{figure}
   \begin{subfigure}{0.45\linewidth}
       \includegraphics[width=\linewidth]{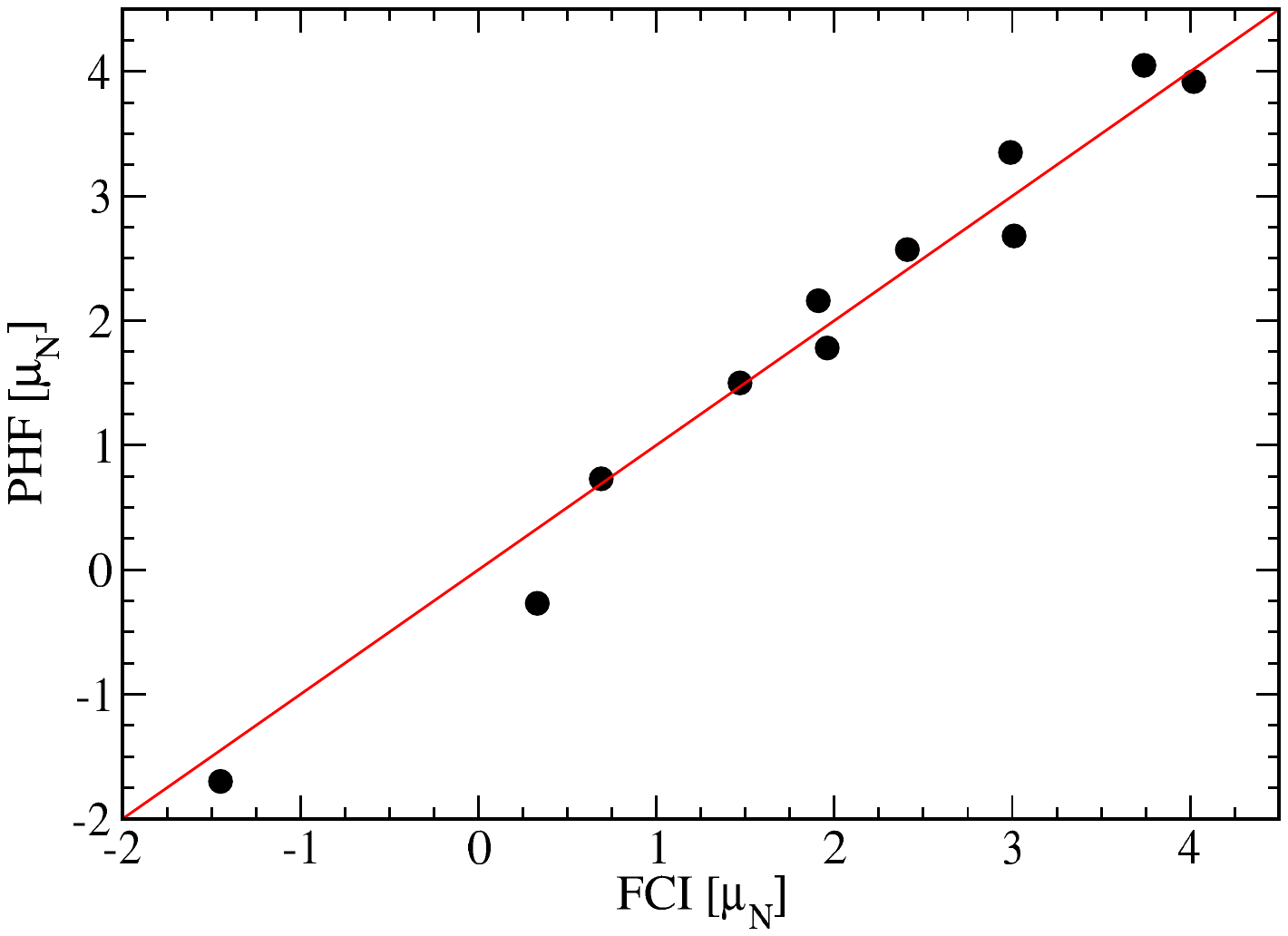}
       \caption{M1 moments for $^{48}$V.}
   \end{subfigure}
   \begin{subfigure}{0.45\linewidth}
       \includegraphics[width=\linewidth]{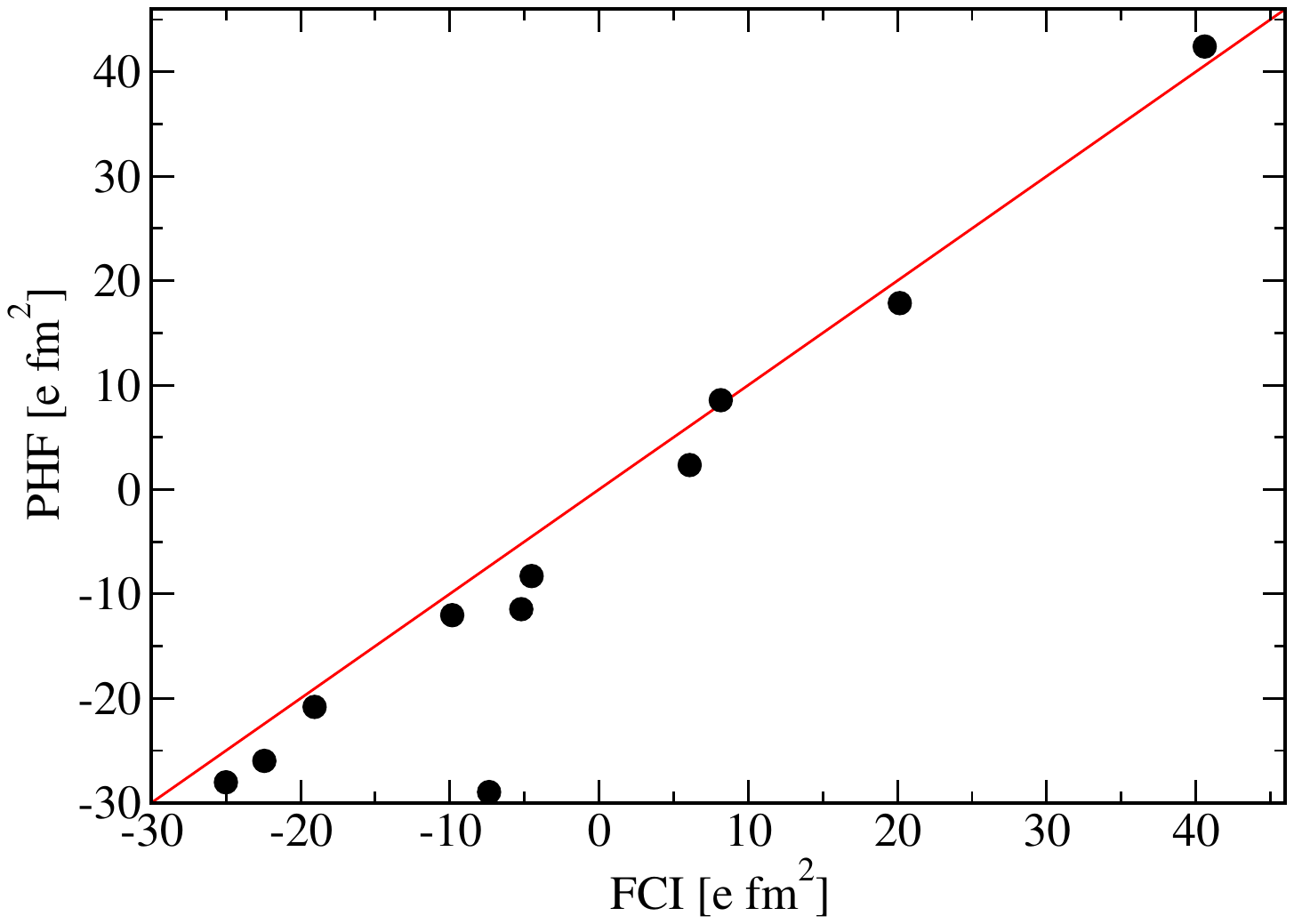}
       \caption{E2 moments for $^{48}$V.}
   \end{subfigure}
   \caption{Scatter plot of M1 and E2 moments for $^{48}$V. Moments calculated from FCI densities are plotted on the x-axis and from PHF on the y-axis. The $y=x$ line is shown for comparison.}
   \label{fig:v48mom}
\end{figure}

\begin{figure}[H]
    \centering
    \begin{subfigure}{0.45\linewidth}
    \includegraphics[width=\linewidth]{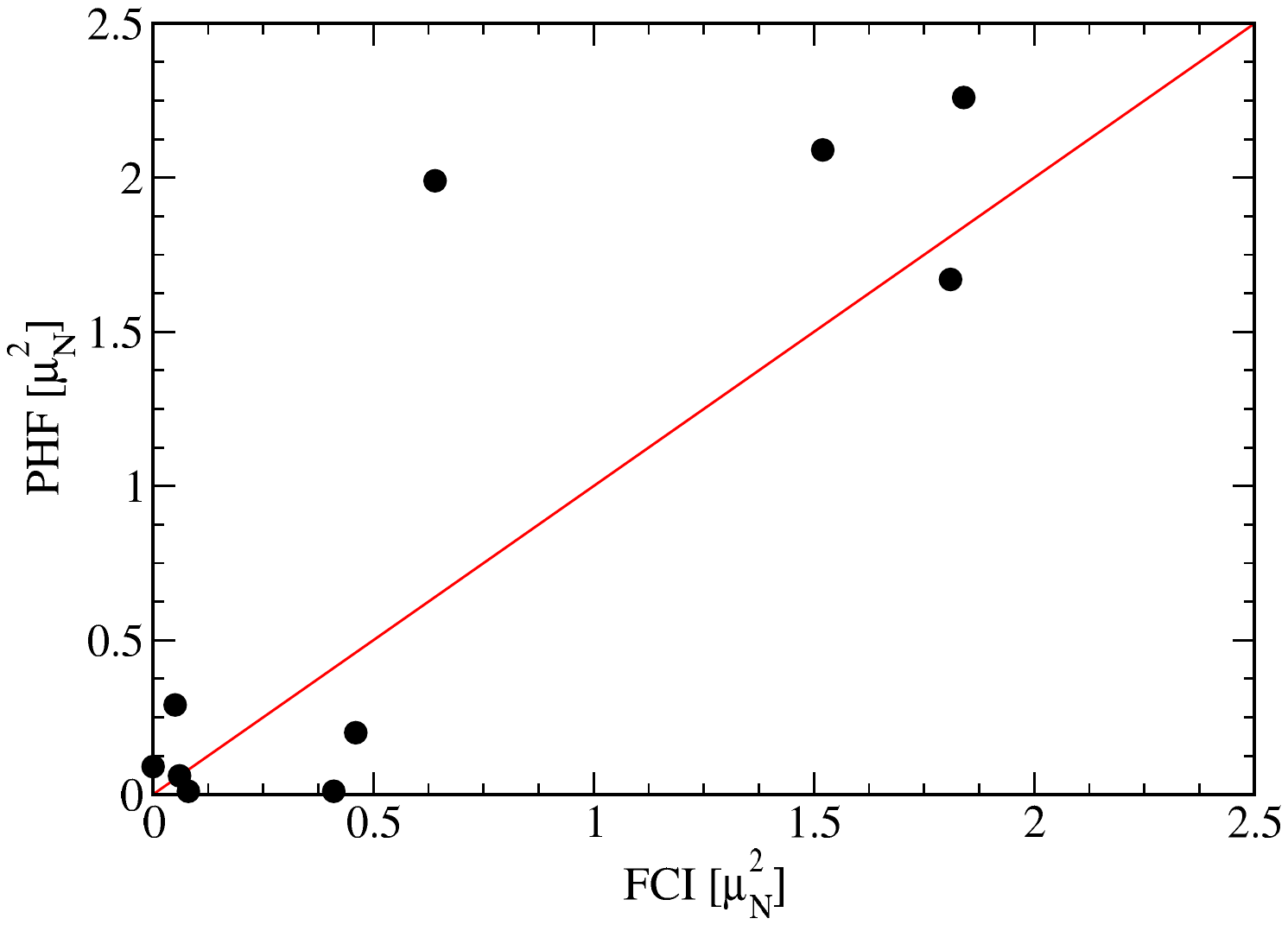}
    \caption{B(M1) transition strengths for  $^{48}$V.}
    \end{subfigure}
    \begin{subfigure}{0.45\linewidth}
         \includegraphics[width=\linewidth]{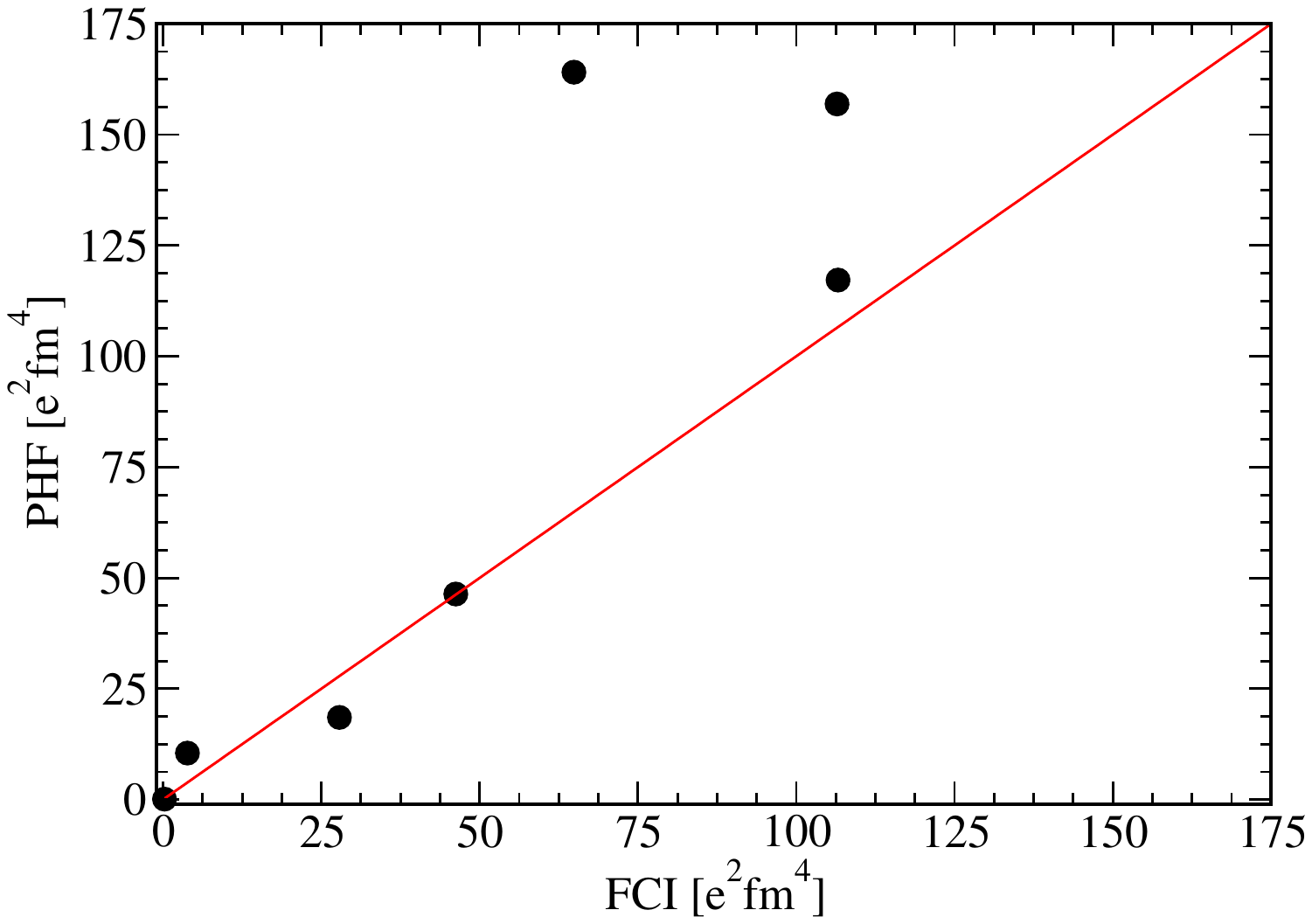}
         \caption{B(E2) transition strengths for $^{48}$V.}
    \end{subfigure}
    \caption{Scatter plot of (a) B(M1) and (b) B(E2) transition strengths for $^{48}$V. Transitions calculated from FCI densities are plotted on the x-axis and from PHF on the y-axis. The $y=x$ line is shown for comparison.}
    \label{fig:v48trans}
\end{figure}

\emph{Odd-odd}. 
$^{48}$V has a prolate HF minimum at -92.25 Mev with $\beta=0.26$.
A second prolate minimum occurs at -92.0 MeV with the same $\beta$, but with a 
very different $\langle J^2 \rangle$ (41.5 and 32.5, respectively). 
Fig.~\ref{fig:pfenergies}(c) show that the PHF excitations energies are generally above those of FCI, though keep in mind the density of states, as typical for odd-odd nuclides, is denser, 
which exaggerates energy differences. The general ordering of states is correct. 
Despite being an odd-odd case, both the M1 and E2 moments, shown in Fig.~\ref{fig:v48mom}(a) 
and (b), respectively, generally agree well between PHF and FCI. The 
B(M1) and B(E2) transition strengths, in Fig.~\ref{fig:v48trans}(a) and (b), respectively, 
have several significant outliers.

\begin{figure}[H]
   \begin{subfigure}{0.45\linewidth}
       \includegraphics[width=\linewidth]{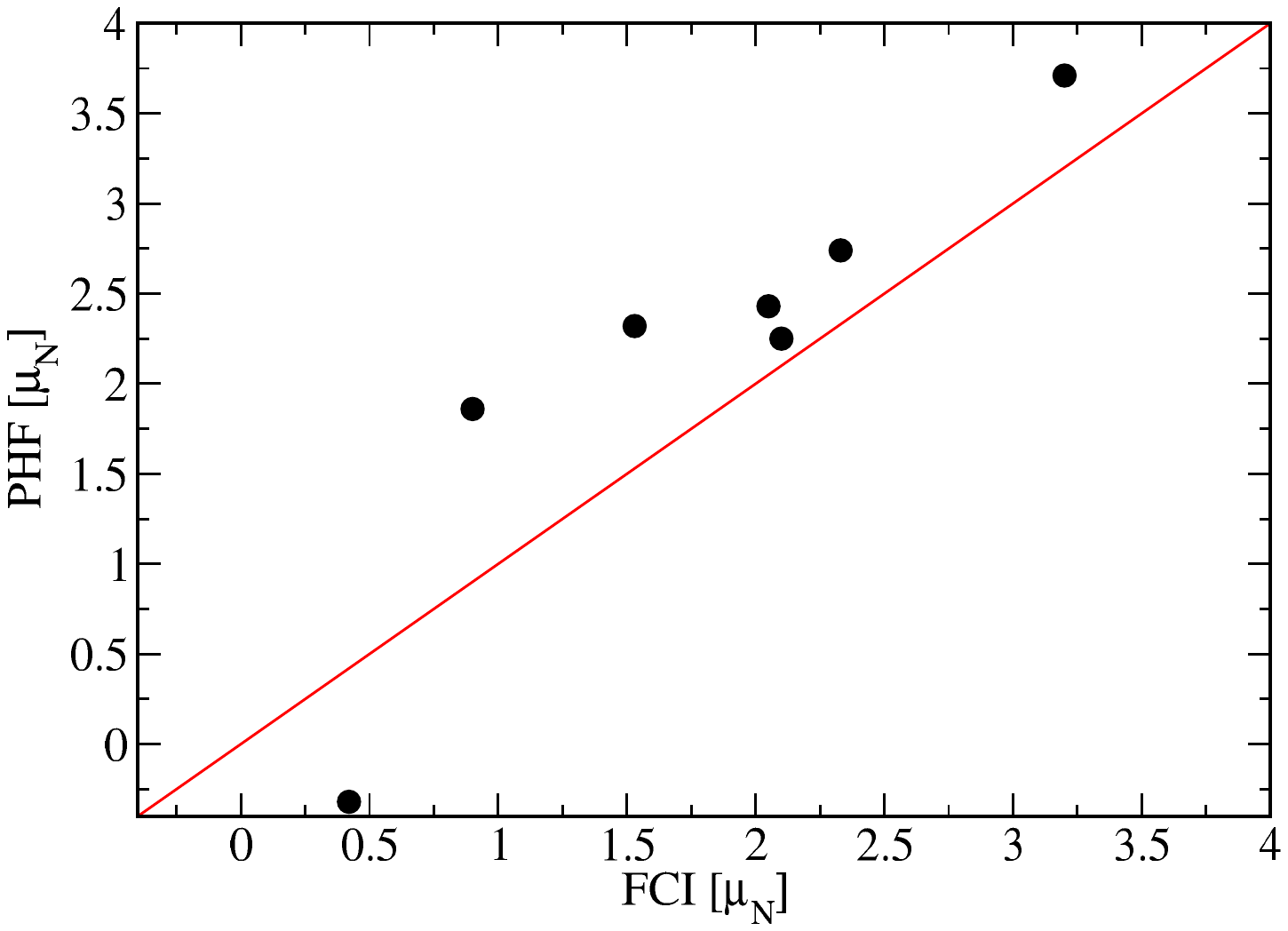}
       \caption{M1 moments for $^{64}$Cu.}
   \end{subfigure}
   \begin{subfigure}{0.45\linewidth}
       \includegraphics[width=\linewidth]{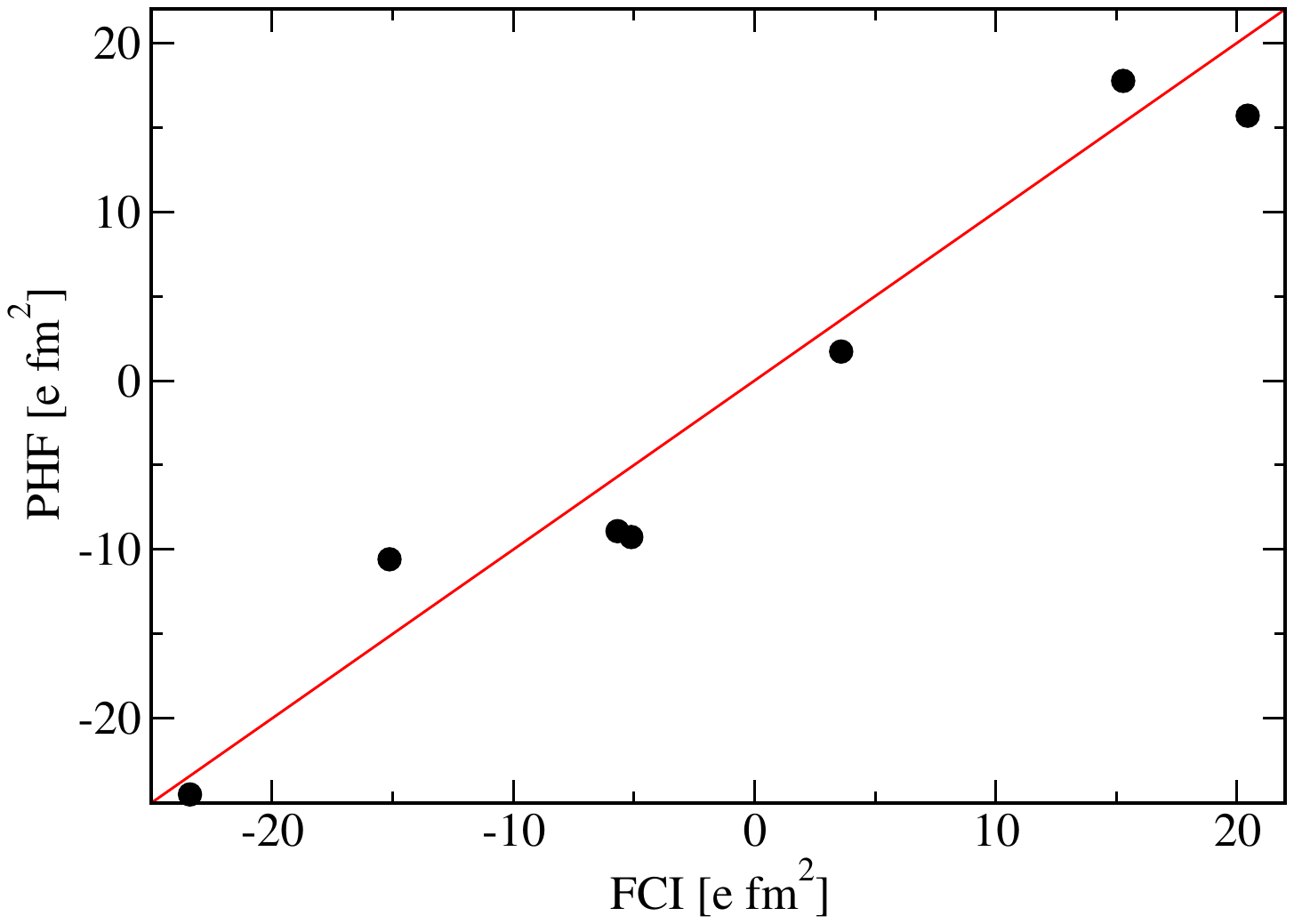}
       \caption{E2 moments for $^{64}$Cu.}
   \end{subfigure}
   \caption{Scatter plot of M1 and E2 moments for $^{64}$Cu. Moments calculated from FCI densities are plotted on the x-axis and from PHF on the y-axis. The $y=x$ line is shown for comparison.}
   \label{fig:cu64mom}
\end{figure}

\begin{figure}[H]
    \centering
    \begin{subfigure}{0.45\linewidth}
         \includegraphics[width=\linewidth]{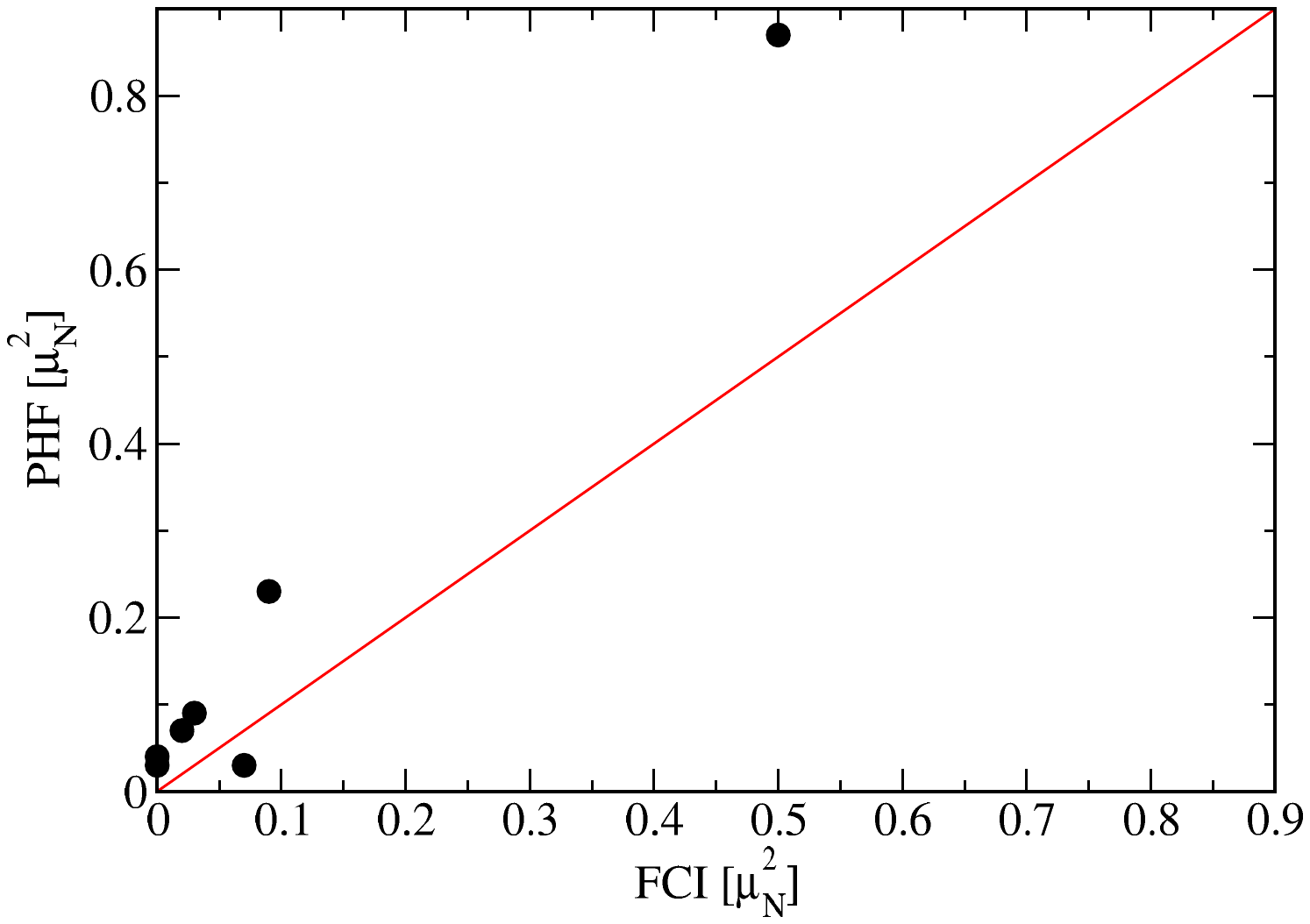}
         \caption{B(M1) transition strengths for $^{64}$Cu.}
    \end{subfigure}
      \begin{subfigure}{0.45\linewidth}
        \includegraphics[width=\linewidth]{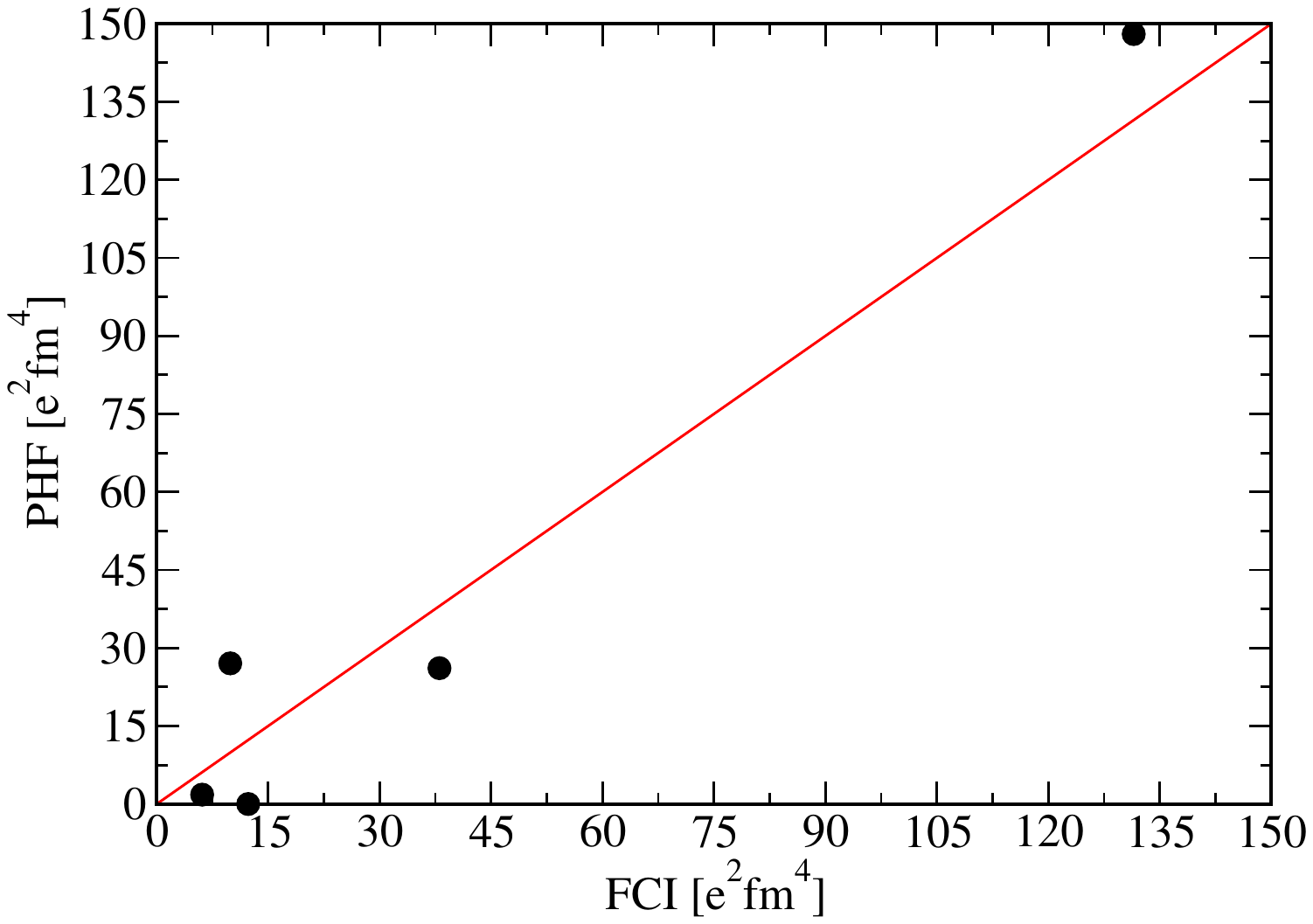}
        \caption{B(E2) transition strengths for $^{64}$Cu.}
    \end{subfigure}
    \caption{Scatter plot of (a) B(M1) and (b) B(M2) transition strengths $^{64}$Cu. Transitions calculated from FCI densities are plotted on the x-axis and from PHF on the y-axis. The $y=x$ line is shown for comparison.}
    \label{fig:cu64trans}
\end{figure}

Our second $pf$-shell odd-odd nuclide,
$^{64}$Cu, 
has a nearly oblate ($\gamma=56^\circ$) HF minimum at -285.1 MeV and 
$\beta = 0.10$, and a secondary triaxial ($\gamma=12^\circ$) minimum at -284.7 MeV 
and $\beta=0.08$. Unlike $^{48}$V, PHF tends to underestimate slightly excitation energies 
relative to FCI. The PHF M1 and E2 moments, shown in Fig.~\ref{fig:cu64mom}(a) and (b), respectively, 
mostly agree with FICI results. The B(M1) and B(E2) transition strengths, 
shown in Fig.~\ref{fig:cu64trans}(a) and (b), respectively, give approximately good agreement.

\begin{figure}[H]
    \centering
    \includegraphics[width=0.5\linewidth]{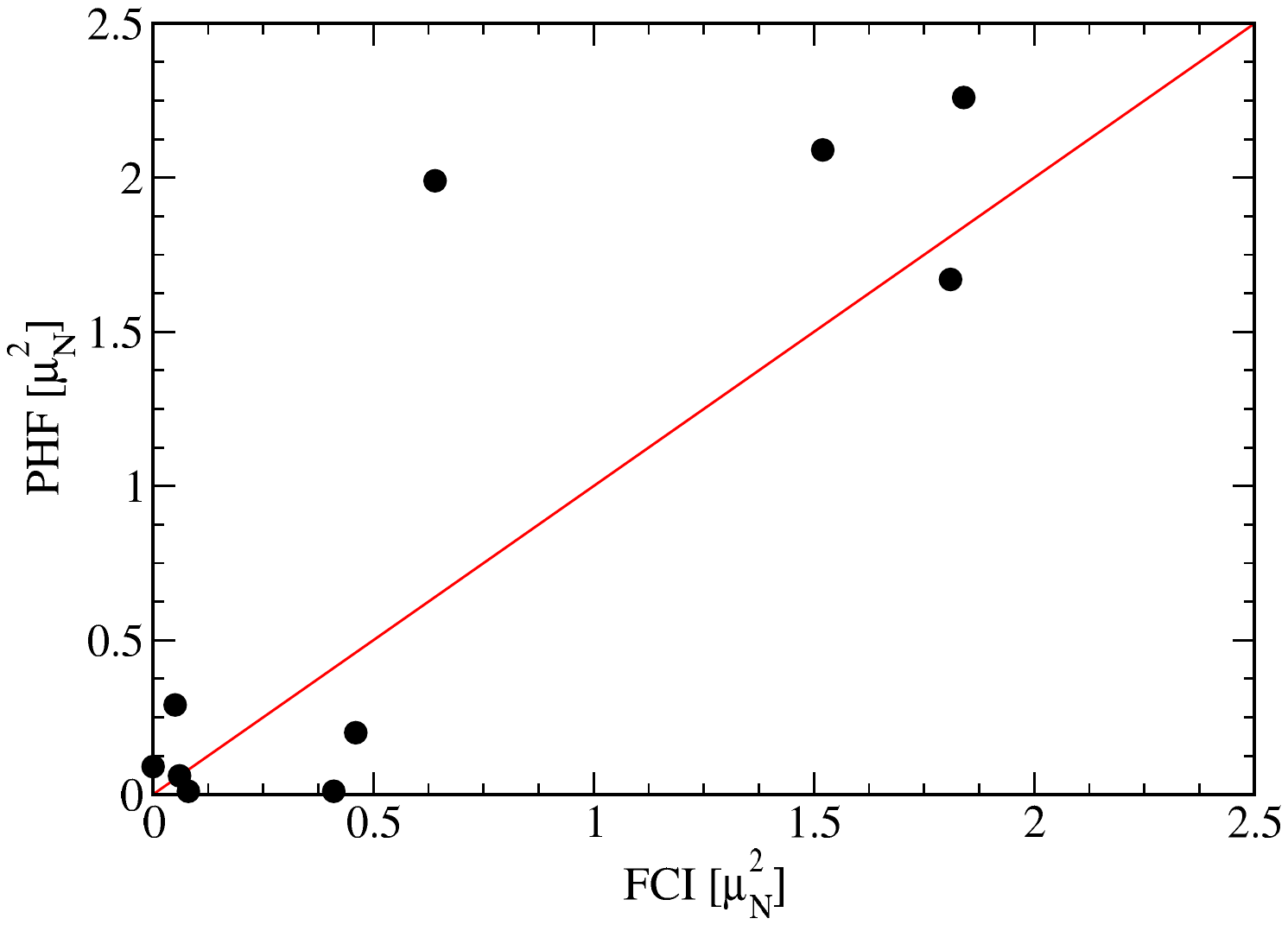}
    \caption{Scatter plot of B(M1) transition strengths for $^{48}$V. Transitions calculated from FCI densities are plotted on the x-axis and from PHF on the y-axis. The $y=x$ line is shown for comparison.}
\end{figure}

\begin{figure}[H]
   \begin{subfigure}{0.45\linewidth}
       \includegraphics[width=\linewidth]{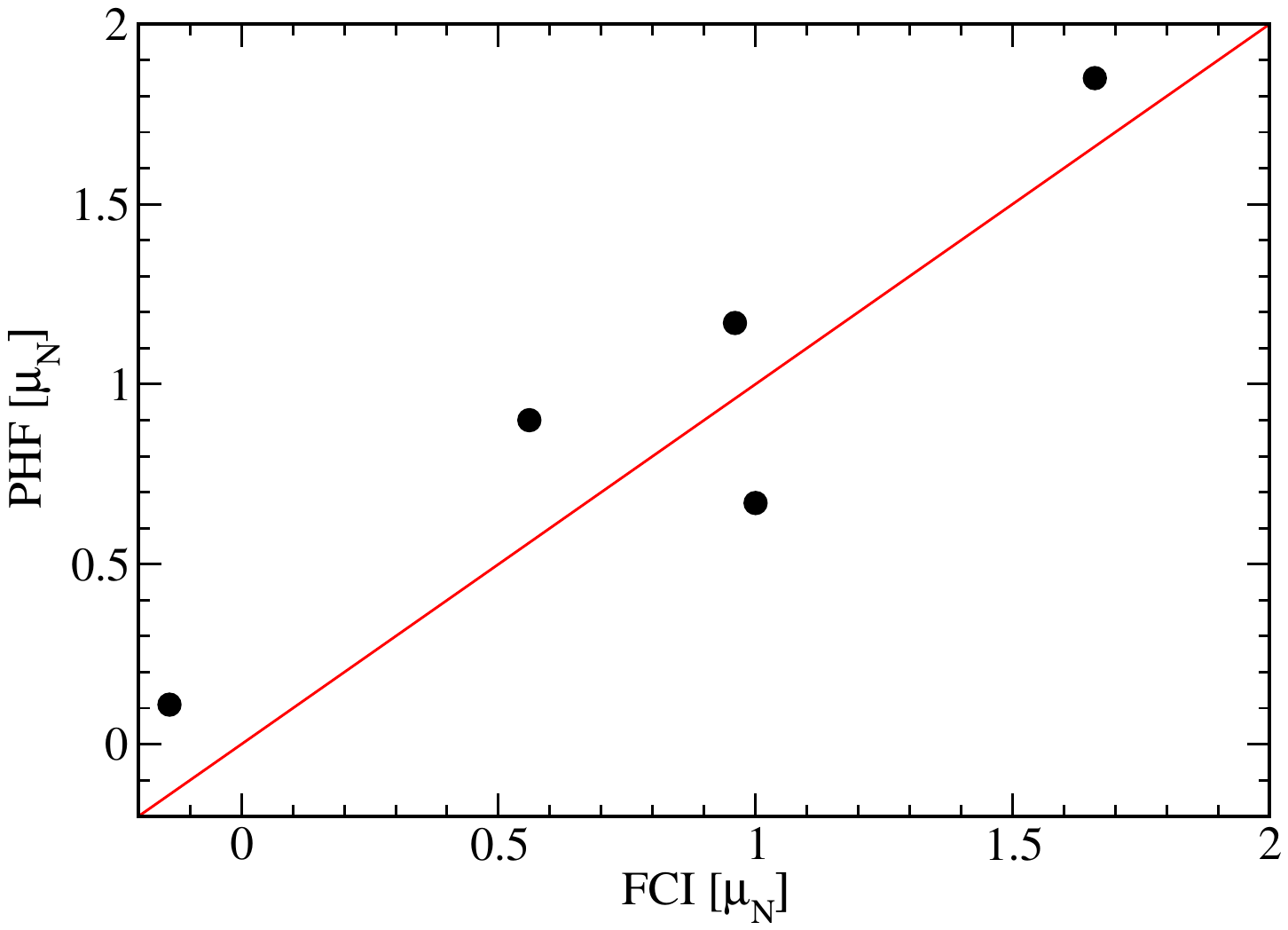}
       \caption{M1 moments for $^{63}$Zn.}
   \end{subfigure}
   \begin{subfigure}{0.45\linewidth}
       \includegraphics[width=\linewidth]{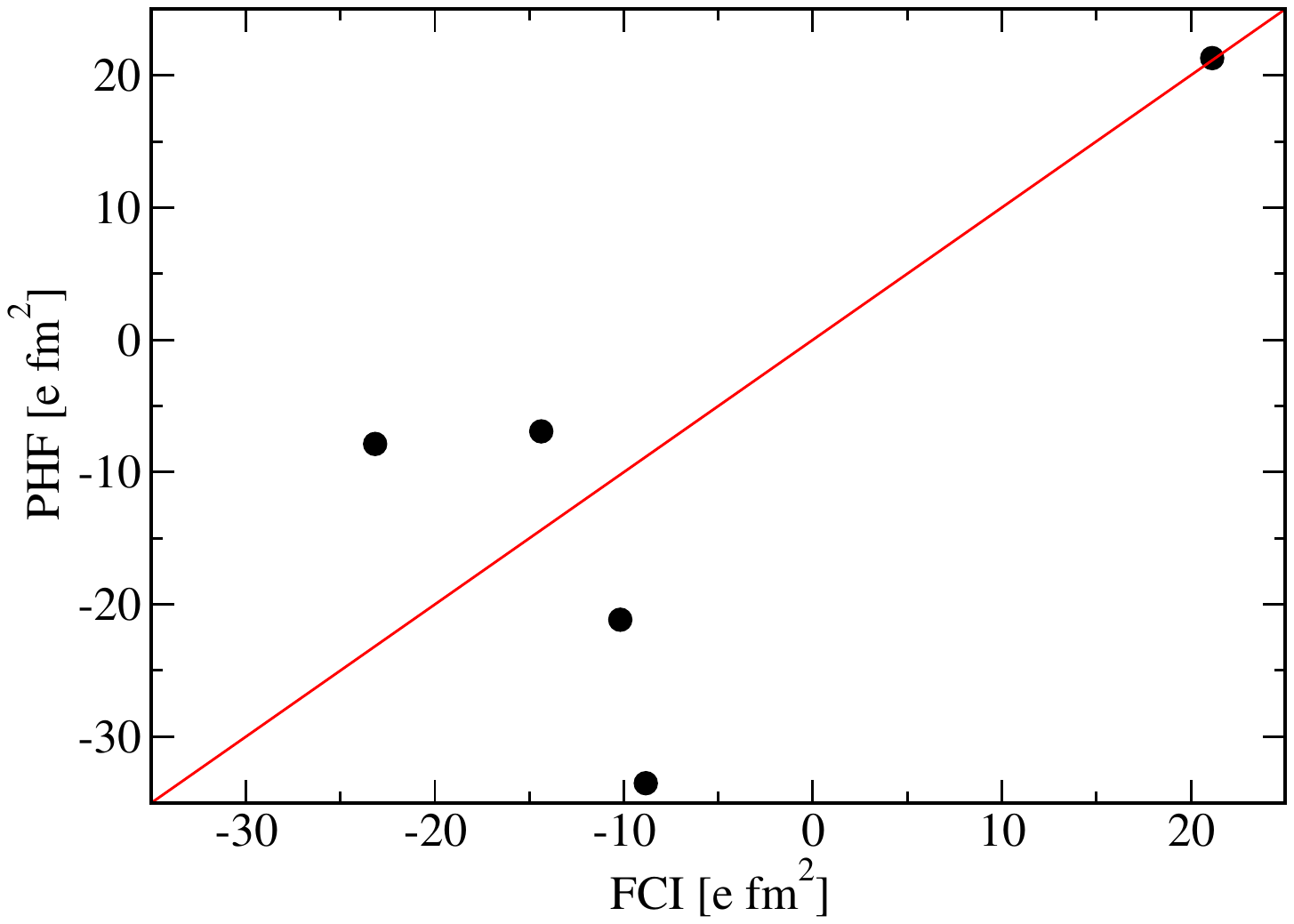}
       \caption{E2 moments for $^{63}$Zn.}
   \end{subfigure}
   \caption{Scatter plot of M1 and E2 moments for $^{63}$Zn. Moments calculated from FCI densities are plotted on the x-axis and from PHF on the y-axis. The $y=x$ line is shown for comparison.}
   \label{fig:zn63mom}
\end{figure}

\begin{figure}[H]
    \centering
    \begin{subfigure}{0.45\linewidth}
        \includegraphics[width=\linewidth]{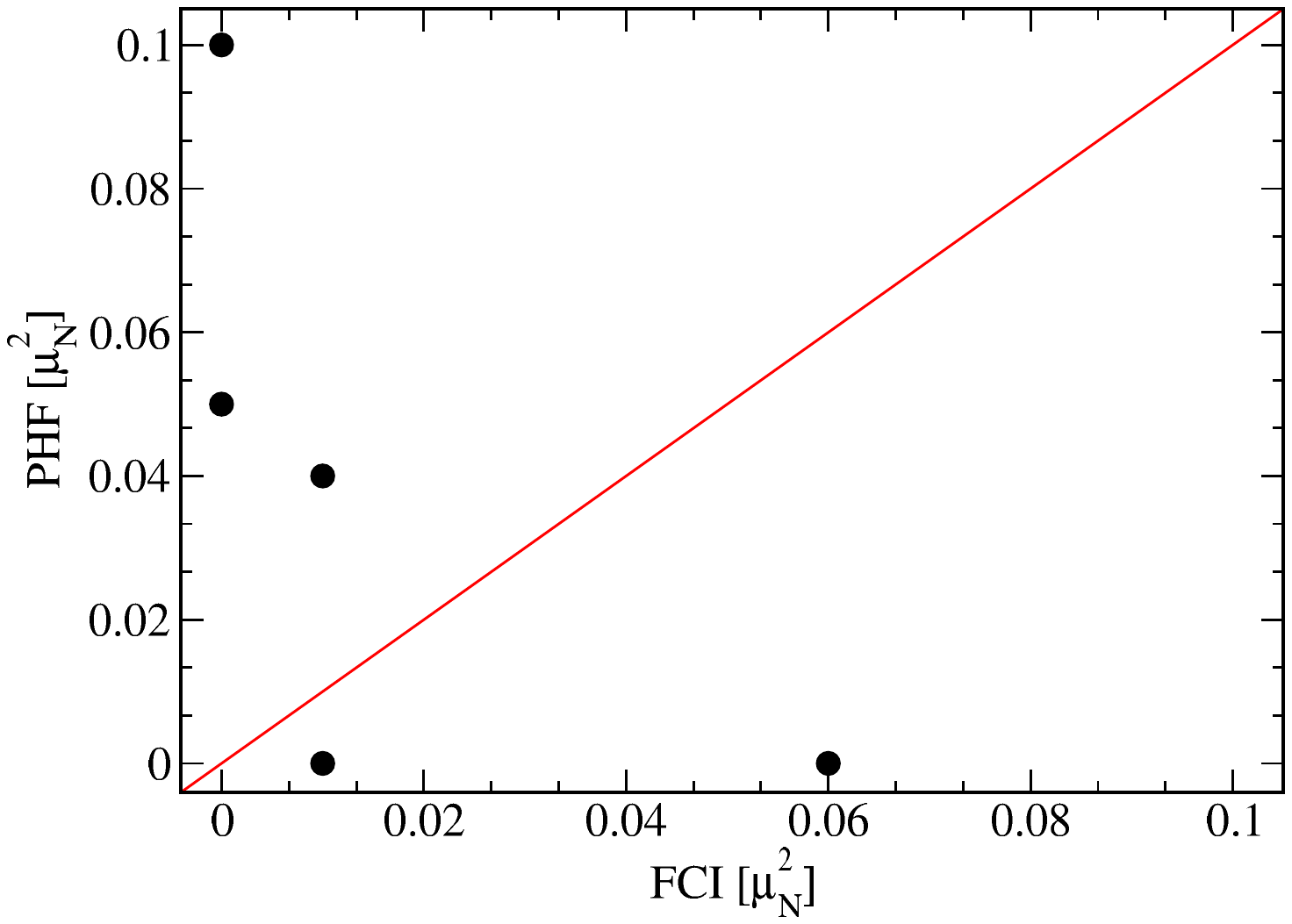}
        \caption{B(M1) transition strengths for $^{63}$Zn.}
    \end{subfigure}
    \begin{subfigure}{0.45\linewidth}
        \includegraphics[width=\linewidth]{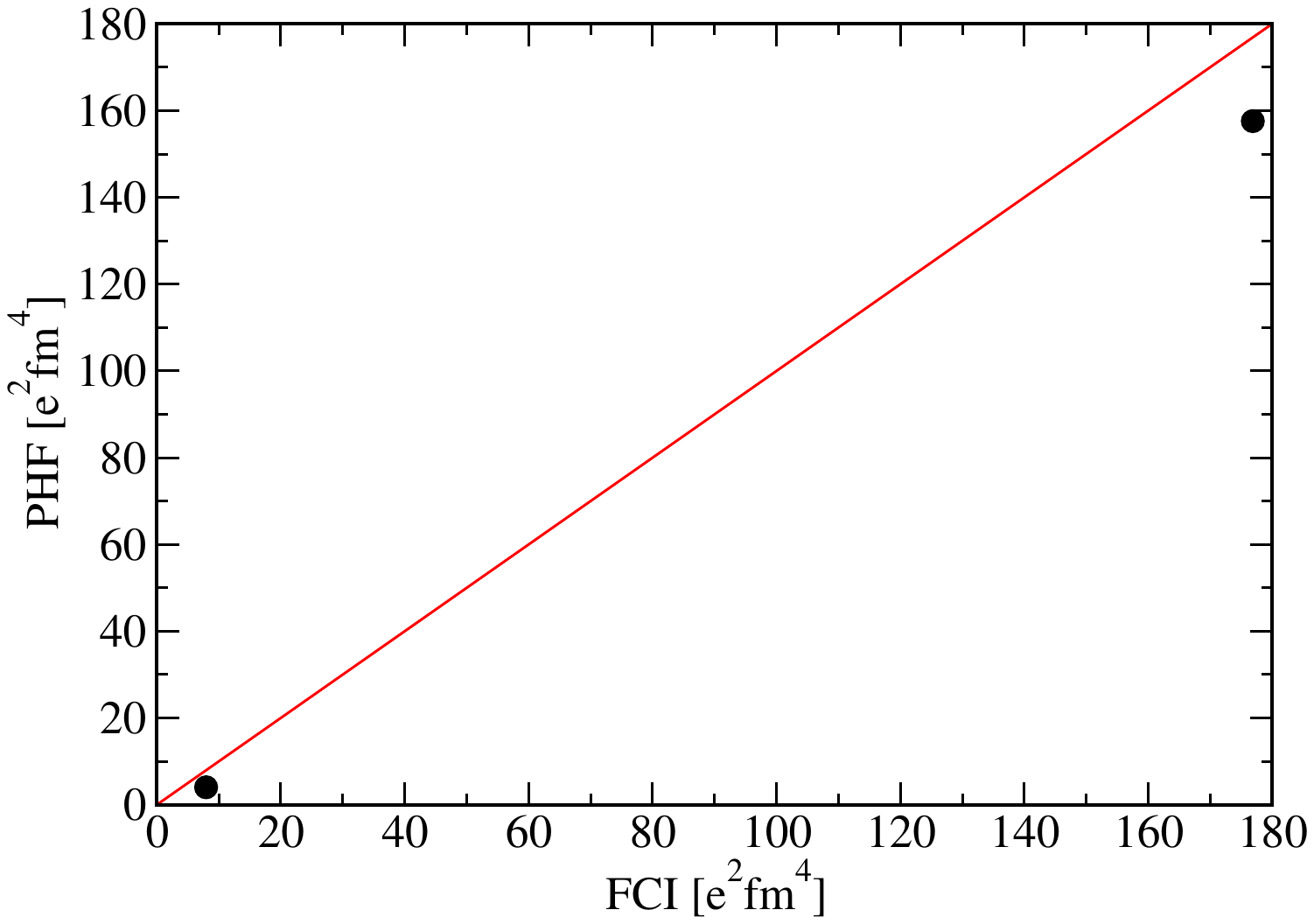}
        \caption{B(E2) transition strengths for $^{63}$Zn.}
    \end{subfigure}
    \caption{Scatter plot of B(M1) and B(E2) transition strengths for $^{63}$Zn. Transitions calculated from FCI densities are plotted on the x-axis and from PHF on the y-axis. The $y=x$ line is shown for comparison.}
    \label{fig:zn63trans}
\end{figure}



\emph{Odd-A}. Finally, we turn to an odd-$A$ nuclide,
$^{63}$Zn, which  has a prolate HF minimum at -282.8 MeV and $\beta=0.11$.
The ordering of low-lying levels, shown in Fig.~\ref{fig:pfenergies}(e), are 
correct, though with deviations between PHF and FCI, which are typical of the 
dense spectrum. The PHF  M1 moments, Fig.~\ref{fig:zn63mom}(a) agree reasonably well with FCI values, while the PHF E2 moments, Fig.~\ref{fig:zn63mom}(b), agree less well. 
The PHF B(M1) transition strengths, displayed in Fig.~\ref{fig:zn63trans}(a), do not agree well with the FCI values, while the two PHF B(E2) values we computed,  see Fig.~\ref{fig:zn63trans}(b), do agree well with FCI values.

\section{Conclusions and acknowledgements}

We have compared select electromagnetic observables from {angular-momentum} projected-after-variation Hartree-Fock spectra to full configuration-interaction calculations, carrying out both in the same shell-model framework, using the same interaction matrix elements. 

In many, though not all, cases we found reasonable agreement between PHF and our FCI benchmark, which may be surprising, especially for M1 observables.

PHF, also in use in quantum chemistry \cite{jimenez2012projected}, is a first step towards more sophisticated approximations, such as (projected) Hartree-Fock-Bogoliubov, 
generator coordinate, and the Monte Carlo shell model. 
This benchmarking will help to identify the value added by such improvemed approaches.

Discussions with C.-F. Jiao helped us understand better how to efficiently extract density matrices from PHF calculations. 
This material is based upon work supported by the U.S. Department of Energy, Office of Science, Office of Nuclear Physics, 
under Award Number  DE-FG02-03ER41272, and by the Office of High Energy Physics, under Award No.~DE-SC0019465.    
Part of this
research was enabled by computational resources supported by a generous gift to
SDSU from John Oldham.

\appendix

\section{Angular momentum projection of density matrices}

\label{sec:details}

In this appendix we give the full details of computing one-body density matrices in the context of angular-momentum projection from multiple reference states. While transition have been computed in previous papers~\cite{PhysRev.159.885,baye1983electromagnetic,PhysRevC.65.024304,PhysRevC.98.054311,PhysRevC.103.064302}, the focus was primarily on electric quadrupole transitions.
Here we give details on extracting one-body density matrices, both static (same initial and final state) and transition (different initial and final states) for general angular momentum rank, with details 
not found in the literature. This would allow, for example,
computation of dark matter cross sections in a general formalism~\cite{gorton2023dmscatter}.
First, though, for full context, we recapitulate 
construction of projected states~\cite{ring2004nuclear,lauber2021benchmarking}.

We start with reference states, in our case, occupation representations of  Slater determinants, labeled by upper case Latin letters $A, B, \ldots $ 
Such reference states are of course not eigenstates of the full many-body 
Hamiltonian, nor in general do they have good angular momentum. 
Each such state can be a linear combination of states with good angular momentum;
\begin{equation}
| B \rangle = \sum_{J, \mu} c^B_{J, \mu} | \psi_B: J,\mu, K_\mu \rangle \label{eq:refexpand}
\end{equation}
Here  use lower-case Greek letters $\lambda, \mu $ to distinguish 
states with the same total angular momentum $J$ but different $z$-projections
 $K_\lambda, K_\mu$. 
For a given $J$, there can be up to $2J+1$ values of $\lambda$ and thus of $K_\lambda$; in Eq.~(\ref{eq:refexpand}) we assume 
that if $\mu \neq \lambda$ then initially $K_\mu \neq K_\lambda$. 
Then normalization $\langle \psi_A | \psi_A \rangle =1$ 
means
\begin{equation}
\sum_{A,J, \lambda} |c^A_{J, \lambda} |^2 = 1.
\end{equation}
(While components of different $J$ or $M/K$ are of course orthogonal,
we do not assume that the 
components of $| \Psi_A \rangle$ need be orthogonal to those of $| \Psi_B \rangle.$ )
Note that we will use $ | A \rangle, | B \rangle $ for unprojected states, either the original reference state or a rotated reference state, and 
$| \psi_A: J, M \rangle $ for projected states. In the code, one 
only directly handles unprojected states; projected states are 
implicitly a linear combination of rotated reference states.

Most projection schemes, whether through quadrature~\cite{ring2004nuclear,sheikh2021symmetry} or linear algebra~\cite{PHF1,PHF2} begins with the rotation operator~\cite{edmonds1996angular},
\begin{equation}
\hat{R}(\alpha, \beta,\gamma)=\hat{R}(\Omega) 
= \exp( i \gamma \hat{J}_z /\hbar)  \exp( i \beta \hat{J}_y/\hbar )   \exp( i \alpha \hat{J}_z /\hbar) 
\end{equation}
where $\alpha,\beta,\gamma = \Omega$ are the Euler angles and 
$\hat{J}_y, \hat{J}_z$ are the generators of rotation about the 
$y$- and $z$-axes. This operator
rotates the 
orientation, or $z$-projection, from the intrinsic value $K_\mu$ to a lab value of $M$:
\begin{equation}
   \hat{R}(\Omega) | J, K_\mu \rangle = \sum_M  
   {\cal D}^J_{M, K_\mu}(\Omega) | J, K_\mu \rightarrow M \rangle, \label{rotate}
\end{equation}
where ${\cal D}^J_{M, K_\mu}(\Omega)$ 
is the Wigner D-matrix~\cite{edmonds1996angular}, and where we use the notation $K \rightarrow M$ to denote a state that has been 
rotated from an initial $z$-projection $K$ to a new $z$-projection $M$.
Note that if we rotate a component in Eq.~(\ref{eq:refexpand}) to the same $M$ as another component with the same $J$, they need not be orthogonal, that is, 
\begin{equation}
    \langle \psi_A: J_A ,\lambda, K_\lambda | 
    \psi_A: J_A, \mu, K_\mu \rightarrow K_\lambda \rangle
\end{equation}
need not be zero, even for $\lambda \neq \mu.$

If we  implement the projection operator $\hat{P}^J_{MK}$ \cite{ring2004nuclear,sheikh2021symmetry,PHF1},  then 
applying Eq.~(\ref{rotate}) to Eq.~(\ref{eq:refexpand}) 
projects out a state of angular momentum $J$ and third component $M$ 
from a state initially with third component $K$:
\begin{equation}
\hat{P}^J_{MK} | B \rangle = \sum_\mu c^B_{J, \mu} \delta_{K, K_\mu}  | \psi_B: J \mu, K_\mu \rightarrow  M \rangle.
\end{equation}
Again, although we suppress the details, whether one is projecting by quadrature or by linear algebra, $\hat{P}^J_{MK} | B \rangle $ is a linear combination of rotated states.

The first quantity one wants to project is 
 the norm or overlap matrix, also called the norm kernel: 
\begin{eqnarray}
N^J_{MK}(A,B) = \langle A| \hat{P}^J_{MK}  | B \rangle  \nonumber \\
= \sum_{\lambda, \mu}  c^{A *}_{J, \lambda}  c^B_{J, \mu} \delta_{M, K_\lambda} \delta_{K, K_\mu} 
\langle \psi_A: J \lambda, K_\lambda   | \psi_B: J \mu, K_\mu \rightarrow  M \rangle.
\end{eqnarray}
The right-hand side is zero except when $M = K_\lambda.$ 
The trace of the norm kernel should equal the number of reference states.

Similarly, the Hamiltonian kernel  is 
 \begin{eqnarray}
 H^J_{MK}(A,B)  \equiv \langle  A |\hat{H}  \hat{P}^J_{MK} | B \rangle \\
 \nonumber  =   \sum_{\lambda \mu} c^{A*}_{J,\lambda} c^B_{J,\mu} \delta_{M, K_{\lambda} } \delta_{K, K_\mu}    \langle \psi_A: J \lambda, K_\lambda | 
 \hat{H} | \psi_B: J \mu, K_\mu \rightarrow  M\rangle.\label{hamdefn}
 \end{eqnarray}

We  solve the generalized eigenvalue problem,
\begin{equation}
\sum_{B, K} H^J_{MK} (A,B)  g^{J,r}_{K}(B) = E^{J,r} \sum_{B, K} N^J_{MK} (A,B)  g^{J,r}_{K}(B) .
\end{equation}
Here $r$ labels a solution for given $J$. 
We can do this by transforming the Hamiltonian kernel, 
\begin{equation}
\tilde{H}^J_{MK}(A,B) = \sum_{ m, k, C,D} 
\left ( N^{J} (A,C)\right)^{-1/2}_{M,m}  
H^J_{m,k}(C,D) 
\left ( N^{J}(D,B) \right)^{-1/2}_{k,K}  ,
\end{equation}
as well as the coefficients,
\begin{equation}
    g_K^{J,r}(A) = \sum_{C,m}  \left (N^J (A,C)\right)^{-1/2}_{K,m} 
    \tilde{g}^{J,r}_{m}(C),
\end{equation}
which allows us to solve as an ordinary eigenvalue problem, 
\begin{equation}
\sum_{K, \beta} \tilde{H}^J_{MK}(A,B) \tilde{g}^{J,r}_{K}(B) 
= E^J_r \tilde{g}^{J,r}_M(A). \label{transformedeigenvalueeqn}
\end{equation}
These matrices are all of small dimension, and in our model spaces, 
 the HF calculations takes a few seconds and the PHF under a minute on a modest laptop.

The orthonormal projected states are
\begin{equation}
    | \psi_r: J  M \rangle = \sum_{A,K} 
    {g}^{J, r }_{K}(A) \hat{P}^J_{M,K} | A \rangle =\sum_{A,C,K,m} 
     \left (N^J \right)^{-1/2}_{K,m} 
    \tilde{g}^{J,r}_{m}(C)
    \hat{P}^J_{K,M} | A \rangle
    \label{projectedstates}
\end{equation}
although these states are only defined implicitly and never constructed explicitly. 

The rest of this appendix is devoted to computing reduced one-body density 
matrix elements between the states defined in Eq.~(\ref{projectedstates}):
\begin{eqnarray}
\rho^{J_f, r; J_i, s}_{J_t}(ab) = 
 \frac{1}{\sqrt{2J_t+1}} 
 \left \langle \psi_r : J_f \left | \left | \left [ \hat{a}^\dagger \otimes \tilde{b} \right ] \right | \right | \psi_s: J_i \right \rangle.
 \label{reducedDM}
\end{eqnarray}
Here $a,b$ label single-particle orbitals, which means the single-particle 
quantum numbers $j$ (angular momentum), $n$ (radial nodes), $l$ (orbital angular momentum, and $t_z$ (proton or neutron); it does not include 
the third component of angular momentum $m$.  For a less cluttered notation we  let $\hat{a}^\dagger, \hat{b}^\dagger$ represent the fermion creation operators 
for orbitals $a,b$ respectively. 

 There are several steps involved in computing Eq~(\ref{reducedDM}):
\begin{enumerate}
    \item Compute uncoupled density matrix elements between two 
    reference states (one of which may be rotated), Eq.~(\ref{rawrho}) below;

    \item Extract angular-momentum coupled densities using Clebsh-Gordan coeffients, Eq.~(\ref{coupledrho});

    \item By quadrature or linear algebra, project out good angular momentum from the reference state $ | B \rangle,$ Eq.~\ref{projectedcoupledrho}).

\item (A useful check, however, is to reverse the order of steps (ii) and (iii), see Eq.~(\ref{sumrule}).)

    \item By further coupling with Clebsch-Gordan coefficients extract 
    the desired coupled density matrix element, Eq.~(\ref{reducedDM}).
    
\end{enumerate}

We start by noting that a coupled one-body operator with good angular 
momentum $J_t$ and third component $M_t$ is
 \begin{eqnarray}
  \left [ \hat{a}^\dagger \otimes \tilde{b} \right ]_{J_t, M_t}
  = \sum_{m_i  m_j} \hat{a}^\dagger_{m_i} \tilde{b}_{-m_j}  (j_a m_i, j_b -m_j | J_t M_t ) \nonumber \\
  = \sum_{m_i  m_j} \hat{a}^\dagger_{m_i} \hat{b}_{m_j}  (-1)^{j_b - m_j}  (j_a m_i, j_b -m_j | J_t M_t ) ,
  \end{eqnarray}
where we use $\hat{a}^\dagger_m$
as the creation operator for a single particle \textit{state} in orbital $a$ 
with third component $m$,  $\tilde{b}_m = \hat{b}_{-m} (-1)^{j_b + m}$ is the annihilation operator for the time-reversed states of $\hat{b}_m$, and 
$(j_a m_i, j_b -m_j | J_t M_t )$ is a Clebsch-Gordan coefficient~\cite{edmonds1996angular}.

It is straightforward to compute $\langle A | \hat{a}^\dagger_{m_i} \hat{b}_{m_j}  | C \rangle$ between any two Slater determinants, included 
rotated Slater determinants~\cite{PhysRevC.48.1518}.
Because the one-body operator can have non-zero angular momentum rank, we allow the 
initial and final states $\psi_A, \psi_B$ to have different angular momentum 
$J_A, J_B$, respectively.
We compute
\begin{eqnarray}
\langle A | \hat{a}^\dagger_{m_i} \hat{b}_{m_j}  \hat{R}(\Omega) | B \rangle
 = \sum_{J_A, \lambda} \sum_{J_B, \mu} 
 c^{A *}_{J_A, \lambda} c^{B}_{J_B, \mu} 
 \sum_M {\cal D}^{J_\beta} _{M, K_\mu} (\hat{\Omega}) \nonumber \\
 \times 
 \langle \psi_A: J_A \lambda, K_\lambda | \hat{a}^\dagger_{m_i} \hat{b}_{m_j}  | \psi_B: J_B \mu, K_\mu \rightarrow M \rangle,
\label{rawrho}
 \end{eqnarray}
 which is nonzero only if $M- m_j = K_\lambda -m_i$.

Next define the coupled densities, on unprojected states,
 \begin{eqnarray}
\left  \langle A  \left |  \left [ \hat{a}^\dagger \otimes \tilde{b} \right ]_{J_t, M_t} \hat{R}(\Omega) \right | B \right \rangle \nonumber \\
= \sum_{m_i  m_j}  (-1)^{j_b - m_j}  (j_a m_i, j_b -m_j | J_t M_t )
\langle A | \hat{a}^\dagger_{m_i} \hat{b}_{m_j}  \hat{R}(\Omega) | B \rangle,  \label{coupledrho}
\end{eqnarray}
which, by Eq.~(\ref{rawrho}), 
\begin{eqnarray}
= \sum_{J_\alpha, \lambda} \sum_{J_\beta, \mu} 
 c^{A *}_{J_A, \lambda} c^{B}_{J_B, \mu} 
 \sum_M {\cal D}^{J_B} _{M, K_\mu} (\hat{\Omega})  \nonumber \\
\left  \langle \psi_A: J_A \lambda, K_\lambda \left  | \left [ \hat{a}^\dagger \otimes \tilde{b} \right ]_{J_t, M_t} \right | \psi_B: J_B \mu, K_\mu \rightarrow M \right \rangle  .
 \end{eqnarray}

 Then, by quadrature/solving the linear equations, one obtains coupled densities 
with a  projected initial state (with angular momentum $J_B$),
 \begin{eqnarray}
\left  \langle A  \left |  \left [ \hat{a}^\dagger \otimes \tilde{b} \right ]_{J_t, M_t} \hat{P}^{J_B}_{MK} \right | B \right \rangle 
\label{projectedcoupledrho}
 = \sum_{J_A, \lambda} \sum_{\mu} 
 c^{A *}_{J_A, \lambda} c^{B}_{J_B, \mu} \delta_{K, K_\mu} \times   \\
 \left \langle \psi_A: J_A \lambda, K_\lambda=M+M_t \left | \left [ \hat{a}^\dagger \otimes \tilde{b} \right ]_{J_t, M_t}  \right | \psi_B: J_B \mu, K_\mu \rightarrow M \right \rangle. \nonumber
 \end{eqnarray}

(At this point, a useful check on calculations is the following. 
Instead of coupling the density operator, project the initial state 
and compute the uncoupled densities,
  \begin{eqnarray}
\left  \langle A  \left |   \hat{a}^\dagger_{m_i} \hat{b}_{m_j}   \hat{P}^J_{MK} \right | B \right \rangle 
 = \sum_{J_A, \lambda,\mu}
 c^{A *}_{J_A, \lambda} c^{B}_{J_B, \mu} \delta_{K, K_\mu}  \delta_{K_\lambda, M+m_i-m_j}  \\
\times \langle \psi_A: J_A \lambda, K_\lambda |   \hat{a}^\dagger_i \hat{b}_j  | \psi_B: J_B \mu, K_\mu \rightarrow M \rangle.
 \nonumber
 \end{eqnarray}
The trace over $i=j$  yields the (angular momentum scalar) number operator,
\begin{eqnarray}
\sum_i \left  \langle A   \left |   \hat{a}^\dagger_i \hat{b}_i   \hat{P}^J_{MK} \right | B \right \rangle =
\left   \langle A   \left |  \hat{n}  \hat{P}^J_{MK} \right | B \right \rangle  \nonumber
\\ 
 = n_p \left   \langle A   \left |    \hat{P}^J_{MK} \right | B \right \rangle  = n_p {\cal N}^J_{MK}(A, B) \label{sumrule},
\end{eqnarray}
 where $n_p$ is the number of particles and ${\cal N}$ is  previously computed projected norm kernel. We found this relation a  helpful test for 
 debugging.)

  Returning to the coupled densities with partial projection,
  Eq.~(\ref{projectedcoupledrho}),
 we appeal to  the Wigner-Eckart theorem to get 
 \begin{eqnarray}
 \left  \langle \psi_A: J_A \lambda, K_\lambda   \left | \left [ \hat{a}^\dagger \otimes \tilde{b} \right ]_{J_t, M_t}  \right | \psi_B: J_B \mu, M \right \rangle
=   \nonumber \\
\frac{  
( J_B M, J_t M_t | J_A K_\lambda ) }{\sqrt{2J_A+1}}
\left \langle \psi_A: J_A \lambda 
 \left | \left | 
 \left [ \hat{a}^\dagger \otimes \tilde{b} \right ]_{J_t} 
 \right | \right | 
 \psi_B: J_B \mu \right \rangle,
 \end{eqnarray}
 so that Eq.~(\ref{projectedcoupledrho}) now becomes
 \begin{eqnarray}
\left  \langle A \left |  \left [ \hat{a}^\dagger \otimes \tilde{b} \right ]_{J_t, M_t} \hat{P}^{J_B}_{MK} \right | B \right \rangle  =  \nonumber \\
\label{eq:intermed1}
\sum_{J_A}  ( J_B M, J_t M_t | J_A K_\lambda ) 
\times \sum_ {\lambda, \mu} 
 c^{A *}_{J_A, \lambda} c^{B}_{J_B, \mu} \delta_{K, K_\mu}  \delta_{K_\lambda, M+M_t}  \\
 \times \frac{1}{\sqrt{2J_A+1}}
\left \langle \psi_A: J_A \lambda 
 \left | \left | 
 \left [ \hat{a}^\dagger \otimes \tilde{b} \right ]_{J_t} 
 \right | \right | 
\psi_B: J_B \mu \right \rangle.  \nonumber 
\end{eqnarray}
Multiplying Eq.~(\ref{eq:intermed1}) by a Clebsch-Gordan 
coefficient, and using  the orthonormality of Clebsch-Gordan coefficients, one can extract
 \begin{eqnarray}
 \sum_{M_t}  ( J_B M=K_\lambda-M_t, J_t M_t | J_A \, K_\lambda ) 
 \left  \langle A \left |  \left [ \hat{a}^\dagger \otimes \tilde{b} \right ]_{J_t, M_t} \hat{P}^{J_B}_{K_\lambda -M_t, \, K_\mu} \right | B \right \rangle \nonumber \\
  =   
 c^{A *}_{J_A, \lambda} c^{B}_{J_B, \mu} 
\frac{1}{\sqrt{2J_A+1}}
\left \langle \psi_A: J_A \lambda 
 \left | \left | 
 \left [ \hat{a}^\dagger \otimes \tilde{b} \right ]_{J_t} 
 \right | \right | 
 \psi_B : J_B \mu \right \rangle .
 \end{eqnarray}
  This is close to our desired result.  
  Define the 
intermediate quantity
\begin{eqnarray}
\sigma^{A: J_A, \lambda; B: J_B, \mu}_{J_t } (ab) \equiv \nonumber \\
 \frac{c^{A *}_{J_A, \lambda} c^{B}_{J_B, \mu}}{\sqrt{2J_t+1}}
\left \langle \psi_A: J_A \lambda 
 \left | \left | 
 \left [ \hat{a}^\dagger \otimes \tilde{b} \right ]_{J_t} 
 \right | \right | 
 \psi_B : J_B \mu \right \rangle = \\
 \nonumber 
 \sqrt{\frac{2J_A+1}{2J_t+1}}\sum_{M_t}  ( J_B M=K_\lambda-M_t, J_t M_t | J_A \, K_\lambda ) \\
 \nonumber \times
 \left  \langle A \left |  \left [ \hat{a}^\dagger \otimes \tilde{b} \right ]_{J_t, M_t} \hat{P}^{J_B}_{K_\lambda -M_t, \, K_\mu} \right | B \right \rangle. \nonumber  
\end{eqnarray}
Note that our new intermediate quantity $\sigma$ has indices 
$\lambda,\mu$ which imply original values of $K_\lambda, K_\mu$, respectively. This is important because the states $| \psi_A: J_A \lambda, K_\lambda \rightarrow M \rangle $ and $ | \psi_B : J_B \mu, 
K_\mu \rightarrow M \rangle $, $A \neq B$, are 
not necessarily orthogonal. Instead, we need to 
transform an orthogonal basis, using the same transformation 
used to solve the Hamiltonian eigenvalue problem:
\begin{eqnarray}
\tilde{\sigma}^{C: J_f K_f; D : J_i K_i}_{J_t} (ab) 
=  \\ 
\nonumber \sum_{M_f M_i A B}
\left ( N^{J}( CA ) \right)^{-1/2}_{K_f, K_\lambda}  
 \sigma^{A: J_A, \lambda; B: J_B \mu}_{J_t } (ab) 
\left ( N^{J} (BD)\right)^{-1/2}_{K_\mu,K_i}  . 
\end{eqnarray}
The final step is to use the solutions to Eq.~(\ref{transformedeigenvalueeqn}), to extract
\begin{eqnarray}
\rho^{J_f, r; J_i, s}_{J_t}(ab) = 
\sum_{C D M_f M_i} \tilde{g}^{J_f, r *}_{M_f}(C) 
\tilde{\sigma}^{C: J_f K_f; D : J_i K_i}_{J_t} (ab) 
 \tilde{g}^{J_i, s}_{M_i}(D).
\end{eqnarray}
Another useful check is the symmetry relation,
\begin{equation}
 \rho^{A: J_A \lambda; B: J_B \mu}_{J_t } (ab)
 = 
 (-1)^{J_A - J_B + j_a -j_b} 
\left [  \rho^{B: J_B \mu; A: J_A \lambda }_{J_t } (ba) \right  ]^*.
 \end{equation}
Extracting the density matrices from one or two reference states in our model spaces takes only a few minutes, even on a laptop computer. 
With the density matrices in hand, one can calculate the 
static or transition matrix element, Eq.~(\ref{eq:operatorme}), 
as long as one knows the  matrix elements of the operator between single-particle orbitals.

\section{Data}

\label{data}

We present selected data for reference. 

\subsection{$^{20}$Ne}

$^{20}$Ne has two HF minima, a prolate minimum at -36.40 MeV, and a second oblate local minimum at -31.83 MeV.

\begin{table}[h!]
    \centering
    \begin{tabular}{|c|c|c|r|r|r|r|r|r|}
    \hline
       & \multicolumn{2}{ c } {$E$ (Mev)} & \multicolumn{2}{|c|} {$E_x$ (MeV)} & \multicolumn{2}{|c|} {$Q$ ($e$-fm$^2$)} & \multicolumn{2}{|c|} {$\mu$ ($\mu_N$)}\\
    $J_n^\pi$ & FCI & PHF & FCI & PHF   & FCI & PHF & FCI & PHF\\
    \hline
       $0_1^+$  &  -40.47 & -39.90 & 0.0 & 0.0 & -- & -- & -- & -- \\
        $2_1^+$  &  -38.72 & -38.41 & 1.75 & 1.50 & -13.63 & -13.65 & 1.02 & 1.02 \\
        $4_1^+$  &   -36.30 & -36.01 & 4.18 & 3.90 & -17.06 & -17.34 & 2.05 & 2.03  \\
        $0_2^+$  &   -33.77 & -32.66 & 6.70 & 7.24 & -- & -- & -- & --  \\
        $2_2^+$  &   -32.93  & -31.70 & 7.54 &  8.20  & +8.40  & +7.61  & 1.09 & 1.03  \\
        $6_1^+$  &   -31.92 & -31.34 & 8.55 & 8.56 & -16.44 & -19.42 & 3.18 & 3.05 \\
        $4_2^+$  &   -30.52 &  -29.92 & 10.71 & 9.99  & -8.14 & -14.49 & 2.20 & 2.04\\
        $8_1^+$  &    -28.96 & -28.62 &  11.51 & 11.28 & -15.10 & -17.21 & 4.31 & 4.08 \\
        $4_3^+$  &   -28.75 &   -27.61 &  11.72  &  12.29 & -7.71 & +5.97 &  2.30 & 2.06  \\
        $6_2^+$  &   -27.77 & -25.76 & 12.70 & 14.68 & -8.05 & -5.22 & 3.47 & 3.07 \\
        $8_2^+$  &  -24.58   & -22.00 & 15.89  &  17.90 & -12.37 & -16.49 & 4.44 &   2.06 \\

       \hline
    \end{tabular}
    \caption{Absolute ($E$) and excitation ($E_x$) energies, and electric quadrupole ($Q$) and magnetic 
    dipole ($\mu$) moments, for $^{20}$Ne, in the $sd$-shell valence space. Here we compare full configuration-interaction (FCI) versus angular-momentum projected Hartree-Fock (PHF). $\mu_N$ is the nuclear magneton.}
    \label{tab:ne20energy}
\end{table}

\begin{table}[H]
    \centering

    \begin{tabular}{|c|c|c|}
    \hline & \multicolumn{2}{ c |} {B(E2) ($e^2$-fm$^4$)}\\
    $i\rightarrow f$ & FCI & PHF \\
    \hline
     $2^+_1 \rightarrow 0^+_1$    &  44.73 &   44.78 \\
     $4^+_1 \rightarrow 2^+_1$    &  52.80  &   53.26 \\
 $6^+_1 \rightarrow 4^+_1$    &   39.25 &  43.82  \\
 $8^+_1 \rightarrow 6^+_1$    &  26.75 &   24.72  \\
 \hline
     $2^+_2 \rightarrow 0^+_2$    &  12.79  & 16.00 \\
     $4^+_2 \rightarrow 2^+_2$    &   3.73 &   0.012 \\
 $6^+_2 \rightarrow 4^+_2$    &  3.58  &   0.70  \\
 $8^+_2 \rightarrow 6^+_2$    &   16.64 &   2.79   \\
 \hline
      $2^+_2 \rightarrow 0^+_1$    & 0.028   &   0.039  \\
    $0^+_2 \rightarrow 2^+_1$    & 7.99  &   9.22  \\
              $4^+_2 \rightarrow 2^+_1$    &   2.12  &   5.57 \\
 $6^+_2 \rightarrow 4^+_1$    &  3.88  &   0.38   \\
  $6^+_2 \rightarrow 4^+_3$    &  15.55  &    7.30  \\
 $8^+_2 \rightarrow 6^+_1$    & 2.39  &    4.68  \\
\hline
    \end{tabular}
    \caption{Selected electric quadrupole transition strengths for $^{20}$Ne, in the $sd$-shell valence space; we separate out intraband transitions for the ground state band (upper block) and for the excited band (middle block), 
    as well as selected interband transitions (lower block). We compare full configuration-interaction (FCI) versus angular-momentum projected Hartree-Fock (PHF).}
    \label{tab:ne20e2}
\end{table}

\subsection{$^{24}$Mg}

    $^{24}$Mg has a  prolate and weakly triaxial ($\gamma = 11.94^\circ$) HF minima at -80.97 MeV
    
\begin{table}[H]
    \centering
    \begin{tabular}{|c|c|c|r|r|r|r|r|r|}
    \hline
       & \multicolumn{2}{ c } {$E$ (MeV)} & \multicolumn{2}{|c|} {$E_x$ (MeV)} & \multicolumn{2}{|c|} {$Q$ ($e$-fm$^2$)} & \multicolumn{2}{|c|} {$\mu$ ($\mu_N$)}\\
    $J_n^\pi$ & FCI & PHF & FCI & PHF & FCI & PHF & FCI & PHF\\
    \hline
       $0_1^+$  &  -87.10 & -85.32 & 0.00 & 0.00 & -- & -- & -- & -- \\
       $2_1^+$  &  -85.60 & -84.05 & 1.50 & 1.28 & -16.94 & -17.14 & 1.03 & 1.02 \\
       $2_2^+$  &  -82.99 & -81.37 & 4.12 & 3.95 & +17.21 & +17.45 & 1.04 & 1.04 \\
       $4_1^+$  &  -82.73 & -81.21 & 4.37 & 4.11 & -21.00 & -21.10 & 2.07 & 2.04 \\
       $3_1^+$  &  -82.03 & -80.42 & 5.07 & 4.91 & -0.08 & -0.13 & 1.55 & 1.54 \\
       $4_2^+$  &  -81.22 & -79.25 & 5.88 & 6.08 & -8.96 & -9.90 & 2.05 & 2.04 \\
       $5_1^+$  &  -79.31 & -77.51 & 7.80 & 7.82 & -14.09 & -14.66 & 2.58 & 2.55 \\
       $6_1^+$  &  -78.83 & -77.10 & 8.27 & 8.23 & -16.74 & -20.78 & 3.13 & 3.06 \\
       $6_2^+$  &  -77.57 & -75.86 & 9.53 & 9.47 & -21.79 & -21.15 & 3.11 & 3.05 \\
    \hline
    \end{tabular}
    \caption{Absolute ($E$) and excitation ($E_x$) energies, and electric quadrupole ($Q$) and magnetic 
    dipole ($\mu$) moments, for $^{24}$Mg in the $sd$-shell valence space. Here we compare full configuration-interaction (FCI) versus angular-momentum projected Hartree-Fock (PHF). $\mu_N$ is the nuclear magneton.}
    \label{tab:mg24energy}
\end{table}

\begin{table}[H]
    \centering
    \begin{tabular}{|c|c|c|}
    \hline 
    & \multicolumn{2}{ c |} {B(E2) ($e^2$-fm$^4$)}\\
    $i \rightarrow f$ & FCI & PHF \\
    \hline
     $2^+_1 \rightarrow 0^+_1$  &  73.16 &  71.74 \\
     $4^+_1 \rightarrow 2^+_1$  &  95.83 &  95.58 \\
     $6^+_1 \rightarrow 4^+_1$  &  88.31 &  92.50 \\
     \hline
     $4^+_2 \rightarrow 2^+_2$  &  37.87 &  39.13 \\
     $6^+_2 \rightarrow 4^+_2$  &  47.20 &  59.14 \\
    \hline
     $2^+_2 \rightarrow 0^+_1$  &   6.80 &   4.61 \\
     $4^+_2 \rightarrow 2^+_1$  &   4.36 &   1.68 \\
     $6^+_2 \rightarrow 4^+_1$  &   0.58 &   0.07 \\
     \hline
    \end{tabular}
    \caption{Electric quadrupole transition strengths for $^{24}$Mg in the $sd$-shell valence space. We compare full configuration-interaction (FCI) and angular-momentum projected Hartree-Fock (PHF) results.We seperate the intraband transitions for the ground state band (upper block) and for the excited band (middle block) as well as selected interband transitions (lower block)}
    \label{tab:mg24e2}
\end{table}

\subsection{$^{25}$Mg}

$^{25}$Mg has a triaxial ($\gamma = 19.08^\circ$) HF minima at -89.11 MeV

\begin{table}[h!]
	\centering
	\begin{tabular}{|c|c|c|r|r|r|r|r|r|}
		\hline
		& \multicolumn{2}{ c } {$E$ (MeV)} & \multicolumn{2}{|c|} {$E_x$ (MeV)} & \multicolumn{2}{|c|} {$Q$ ($e$-fm$^2$)} & \multicolumn{2}{|c|} {$\mu$ ($\mu_N$)}\\
		$J_n^\pi$ & FCI & PHF & FCI & PHF & FCI & PHF & FCI & PHF\\
		\hline
		$5/2_1^+$  &  -94.40 & -92.65 & 0.00 & 0.00 & +19.75 & +20.57 & -0.85 & -0.79 \\
		$1/2_1^+$  &  -93.80 & -91.39 & 0.61 & 1.26 & -- & -- & -0.45 & -0.77 \\
		$3/2_1^+$  &  -93.30 & -91.18 & 1.10 & 1.47 & -12.33 & -12.24 & +0.75 & +0.64 \\
		$7/2_1^+$  &  -92.68 & -91.05 & 1.72 & 1.60 & +2.59 & +2.45 & +0.37 & +0.39 \\
		$5/2_2^+$  &  -92.41 & -90.07 & 2.00 & 2.58 & -15.66 & -16.34 & +0.28 & +0.56 \\
		$3/2_2^+$  &  -91.59 & -87.69 & 2.81 & 4.96 & -11.05 & +12.40 & +1.28 & +1.59 \\
		$7/2_2^+$  &  -91.50 & -89.79 & 2.90 & 2.85 & -20.16 & -20.79 & +1.33 & +1.50 \\
		$9/2_1^+$  &  -90.95 & -89.32 & 3.45 & 3.33 & -0.20 & +5.73 & +1.29 & +0.88 \\
		$9/2_2^+$  &  -90.50 & -88.85 & 3.90 & 3.79 & +26.73 & +20.20 & +0.38 & +0.26 \\
		$5/2_3^+$  &  -90.49 & -87.09 & 3.91 & 5.55 & -15.37 & -5.06 & +0.71 & +1.74 \\
		$7/2_3^+$  &  -89.51 & -86.10 & 4.89 & 6.55 & -18.85 & -10.37 & +2.74 & +2.27 \\
		$11/2_1^+$ &  -89.27 & -87.35 & 5.13 & 5.29 & +9.78 & +9.90 & +1.52 & +1.19 \\
		$13/2_1^+$ &  -88.98 & -85.29 & 5.42 & 7.36 & +20.79 & +0.77 & +4.98 & +2.08 \\
		\hline
	\end{tabular}
	\caption{Absolute ($E$) and excitation ($E_x$) energies, and electric quadrupole ($Q$) and magnetic 
		dipole ($\mu$) moments, for $^{25}$Mg in the $sd$-shell valance space. Here we compare full configuration-interaction (FCI) versus angular-momentum projected Hartree-Fock (PHF). $\mu_N$ is the nuclear magneton.}
	\label{tab:mg25energy}
\end{table}

\begin{table}[H]
    \centering
    \begin{tabular}{|c|c|c|}
    \hline 
    & \multicolumn{2}{ c |} {B(E2) ($e^2$-fm$^4$)}\\
    $i \rightarrow f$ & FCI & PHF \\
    \hline
     $9/2^+_1 \rightarrow 5/2^+_1$   &  37.89 &  39.42 \\
     $13/2^+_1 \rightarrow 9/2^+_1$  &  10.65 &  50.83 \\
     \hline
     $9/2^+_2 \rightarrow 5/2^+_2$   &   0.52 &   0.61 \\
     \hline
     $5/2^+_2 \rightarrow 1/2^+_1$   &  72.00 &  67.17 \\
     $7/2^+_3 \rightarrow 3/2^+_2$   &  71.23 &  49.20 \\
    \hline
    \end{tabular}
    \caption{Electric quadrupole transition strengths for $^{25}$Mg in the $sd$-shell valence space. We compare full configuration-interaction (FCI) and angular-momentum projected Hartree-Fock (PHF) results. $^{25}Mg$ did not exhibit a band structure. Blocks distinguish transitions in different excited states.}
    \label{tab:mg25e2}
\end{table}

\begin{table}[h!]
    \centering
    \begin{tabular}{|c|c|c|}
    \hline 
    & \multicolumn{2}{ c |} {B(M1) ($\mu_N^2$)}\\
    $i \rightarrow f$ & FCI & PHF \\
    \hline
     $3/2^+_1 \rightarrow 1/2^+_1$   &  0.02 &  0.18 \\
     $7/2^+_1 \rightarrow 5/2^+_1$   &  0.53 &  0.56 \\
     $9/2^+_1 \rightarrow 7/2^+_1$   &  0.68 &  0.57 \\
     $11/2^+_1 \rightarrow 9/2^+_1$  &  0.15 &  0.48 \\
     $13/2^+_1 \rightarrow 11/2^+_1$ &  0.01 &  0.79 \\
     \hline
     $7/2^+_2 \rightarrow 5/2^+_2$   &  0.01 &  0.17 \\
     $9/2^+_2 \rightarrow 7/2^+_2$   &  0.00 &  0.00 \\
     \hline
     $7/2^+_3 \rightarrow 5/2^+_3$   &  0.04 &  0.16 \\
     \hline
     $5/2^+_2 \rightarrow 3/2^+_1$   &  0.01 &  0.53 \\
     $5/2^+_3 \rightarrow 3/2^+_2$   &  0.30 &  0.07 \\
    \hline
    \end{tabular}
    \caption{Magnetic dipole transition strengths for $^{25}$Mg in the $sd$-shell valence space. We compare full configuration-interaction (FCI) and angular-momentum projected Hartree-Fock (PHF) results. $^{25}Mg$ did not exhibit a band structure. Blocks distinguish transitions in different excited states.}
    \label{tab:mg25m1}
\end{table}

\subsection{$^{32}$Si}

 $^{32}$Si has an oblate HF minima at -166.344 MeV

\begin{table}[H]
		\centering
		\begin{tabular}{|c|c|c|r|r|r|r|r|r|}
			\hline
			& \multicolumn{2}{ c } {$E$ (MeV)} & \multicolumn{2}{|c|} {$E_x$ (MeV)} & \multicolumn{2}{|c|} {$Q$ ($e$-fm$^2$)} & \multicolumn{2}{|c|} {$\mu$ ($\mu_N$)}\\
			$J_n^\pi$ & FCI & PHF & FCI & PHF & FCI & PHF & FCI & PHF\\
			\hline
			$0_1^+$  &  -170.52 & -168.71 & 0.00 & 0.00 & -- & -- & -- & -- \\
			$2_1^+$  &  -168.47 & -167.00 & 2.05 & 1.71 & 13.35 & 12.73 & 0.95 & 0.80 \\
			$4_1^+$  &  -164.63 & -163.12 & 5.89 & 5.59 & 18.25 & 17.24 & 2.27 & 2.12 \\
			$6_1^+$  &  -160.90 & -158.73 & 9.62 & 9.98 & 18.36 & 20.10 & 4.21 & 3.72 \\
			$8_1^+$  &  -155.44 & -153.60 & 15.08 & 15.11 & 20.48 & 22.58 & 6.40 & 5.92 \\
			\hline
		\end{tabular}
		\caption{Absolute ($E$) and excitation ($E_x$) energies, and electric quadrupole ($Q$) and magnetic 
			dipole ($\mu$) moments, for $^{32}$Si in the $sd$-shell valence space.}
		\label{tab:si32energy}
	\end{table}

    \begin{table}[H]
	\centering
	\begin{tabular}{|c|c|c|}
		\hline 
		& \multicolumn{2}{ c |} {B(E2) ($e^2$-fm$^4$)}\\
		$i \rightarrow f$ & FCI & PHF \\
		\hline
		$2^+_1 \rightarrow 0^+_1$  &  43.54 &  38.87 \\
		$4^+_1 \rightarrow 2^+_1$  &  67.14 &  60.56 \\
		$6^+_1 \rightarrow 4^+_1$  &  52.73 &  62.99 \\
		$8^+_1 \rightarrow 6^+_1$  &  47.14 &  55.81 \\
		\hline
	\end{tabular}
	\caption{Selected electric quadrupole transition strengths for $^{32}$Si in the $sd$-shell valence space.}
	\label{tab:si32e2}
\end{table}

\subsection{$^{34}$S} 

$^{34}$S has 2 HF minima. An oblate minima at -198.097 MeV and a prolate ($\gamma = 1.53$) local minima at -197.026 Mev. 

\begin{table}[H]
		\centering
		\begin{tabular}{|c|c|c|r|r|r|r|r|r|}
			\hline
			& \multicolumn{2}{ c } {$E$ (MeV)} & \multicolumn{2}{|c|} {$E_x$ (MeV)} & \multicolumn{2}{|c|} {$Q$ ($e$-fm$^2$)} & \multicolumn{2}{|c|} {$\mu$ ($\mu_N$)}\\
			$J_n^\pi$ & FCI & PHF & FCI & PHF & FCI & PHF & FCI & PHF\\
			\hline
			$0_1^+$  &  -202.50 & -200.17 & 0.00 & 0.00 & -- & -- & -- & -- \\
			$2_1^+$  &  -200.37 & -198.25 & 2.13 & 1.93 & +3.97 & +8.19 & 1.09 & 0.83 \\
			$2_2^+$  &  -199.38 & -195.75 & 3.12 & 4.43 & -7.71 & +0.40 & 1.61 & 1.33 \\
			$4_1^+$  &  -197.67 & -195.79 & 4.84 & 4.39 & +4.91 & +9.56 & 2.32 & 2.17 \\
			$4_2^+$  &  -195.76 & -191.48 & 6.75 & 8.70 & +10.62 & +5.16 & 1.38 & 2.07 \\
			$6_1^+$  &  -192.99 & -191.32 & 9.51 & 8.85 & +8.33 & +10.75 & 4.40 & 3.40 \\
			$8_1^+$  &  -187.55 & -186.61 & 14.95 & 13.56 & +13.95 & +12.01 & 4.18 & 3.44 \\
			\hline
		\end{tabular}
		\caption{Absolute ($E$) and excitation ($E_x$) energies, and electric quadrupole ($Q$) and magnetic 
			dipole ($\mu$) moments, for $^{34}$S in the $sd$-shell valence space. $\mu_N$ is the nuclear magneton.}
		\label{tab:s34energy}
	\end{table}

	\begin{table}[H]
		\centering
		\begin{tabular}{|c|c|c|}
			\hline 
			& \multicolumn{2}{ c |} {B(E2) ($e^2$-fm$^4$)}\\
			$i \rightarrow f$ & FCI & PHF \\
			\hline
			$2^+_1 \rightarrow 0^+_1$  &  37.09 &  20.77 \\
			$4^+_1 \rightarrow 2^+_1$  &  49.73 &  34.46 \\
			$6^+_1 \rightarrow 4^+_1$  &  44.32 &  31.60 \\
			$8^+_1 \rightarrow 6^+_1$  &  24.49 &  12.39 \\
			\hline
			$4^+_2 \rightarrow 2^+_2$  &   5.00 &   2.03 \\
			\hline
		\end{tabular}
		\caption{Selected electric quadrupole transition strengths for $^{34}$S in the $sd$-shell valence space.}
		\label{tab:s34e2}
	\end{table}

\clearpage


\bibliographystyle{unsrt}
\bibliography{johnsonmaster}

\end{document}